%% file: tepper_etal_OVI.tex
\documentclass[useAMS,usenatbib]{mn2e}

\pdfoutput=1

\usepackage{amsmath}
\usepackage{graphicx}
\usepackage[german,british]{babel}
\usepackage[varg]{txfonts}
\usepackage{biblio}
\usepackage{bibentry}
\usepackage{natbib}
\usepackage{color}

\usepackage[draft]{hyperref}
\providecommand{\eprint}[1]{\href{http://arxiv.org/abs/#1}{#1}}

\input{newcommands_mnras}

\title[\OVI\ absorbers at low redshift]{
Absorption signatures of warm-hot gas at low redshift: \OVI}

\author[Tepper-Garc\'\i{}a et al.]{%
Thorsten Tepper-Garc\'\i{}a,$^{1}$\thanks{E-mail: tepper@astro.physik.uni-potsdam.de} 
Philipp Richter,$^{1}$ 
Joop Schaye,$^{2}$ 
C. M. Booth,$^{2}$  \newauthor
Claudio Dalla Vecchia,$^{2,3}$
Tom Theuns$^{4,5}$ and 
Robert P.C. Wiersma$^{2,6}$\\
$^{1}$Universit\"at Potsdam, Karl-Liebknecht-Str. 24/25, 14476 Potsdam, Germany\\
$^{2}$Leiden Observatory, Leiden University, P.O. Box 9513, 2300 RA Leiden, The Netherlands\\
$^{3}$Max Planck Institut f\"ur Extraterrestrische Physik, Giessenbachstra\ss{}e 1, 85748 Garching, Germany\\
$^{4}$Institute for Computational Cosmology, Department of Physics, University of Durham, South Road, Durham, DH1 3LE, UK\\
$^{5}$Department of Physics, University of Antwerp, Groenenborgerlaan 171, B-2020 Antwerpen, Belgium\\
$^{6}$Max Planck Institut f\"ur Astrophysik, Karl-Schwarzschild-Str. 1, 8574, Garching, Germany}
\begin{document}

\date{Accepted ----. Received ----; in original form ----}

\pagerange{\pageref{firstpage}--\pageref{lastpage}} \pubyear{----}

\maketitle

\label{firstpage}

\begin{abstract}

We investigate the origin and physical properties of \OVI\ absorbers at low redshift ($z = 0.25$) using a subset of cosmological, hydrodynamical simulations from the OverWhelmingly Large Simulations (OWLS) project. Intervening \OVI\ absorbers are believed to trace shock-heated gas in the Warm-Hot Intergalactic Medium (WHIM) and may thus play a key role in the search for the missing baryons in the present-day Universe.
When compared to observations, the predicted distributions of the different \OVI\ line parameters (column density $\NOVI$, Doppler parameter $\bovi$, rest equivalent width $W_{\rm r}$) from our simulations exhibit a lack of strong \OVI\ absorbers, a discrepancy that has also been found by \citet{opp09b}. This suggests that physical processes on sub-grid scales (e.g.\ turbulence) may strongly influence the observed properties of \OVI\ systems.
We find that the intervening \OVI\ absorption arises mainly in highly metal-enriched ($10^{-1} \ll Z/Z_\odot \lesssim 1$)  gas at typical overdensities of $1 \ll \rho/\left <\rho\right >  \lesssim 10^2$. One third of the \OVI\ absorbers in our simulation are found to trace gas at temperatures $T < 10^{5} ~\K$, while the rest arises in gas at higher temperatures, most of them around $T =10^{5.3\pm0.5} \K$. These temperatures are much higher than inferred by \citet{opp09b}, probably because that work did not take the suppression of metal-line cooling by the photo-ionising background radiation into account.
While the \OVI\ resides in a similar region of ($\rho,T$)-space as much of the shock-heated baryonic matter, the vast majority of this gas has a lower metal content and does not give rise to detectable \OVI\ absorption.
As a consequence of the patchy metal distribution, \OVI\ absorbers in our simulations trace only a very small fraction of the cosmic baryons ($<2$ percent) and the cosmic metals. Instead, these systems presumably trace previously shock-heated, metal-rich material from galactic winds that is now mixing with the ambient gas and cooling.
The common approach of comparing \OVI\ and \HI\ column densities to estimate the physical conditions in intervening absorbers from QSO observations may be misleading, as most of the \HI\ (and most of the gas mass) is not physically connected with the high-metallicity patches that give rise to the \OVI\ absorption.

\end{abstract}


%--------------------------------------------------------------------------------------------------------------------------------------------------------------------------------
\section{Introduction} \label{sec:intro}

Diffuse ionised gas in the intergalactic medium (IGM) represents the major baryon reservoir in the Universe at any redshift. From observations of the Ly$\alpha$ forest line density at high redshift in quasar (QSO) and active galactic nuclei (AGN) spectra it can be deduced that the photo-ionised IGM contains more than ninety per cent of the baryons at redshift $z=3$ \citep[][]{rau97a,wei97b,sch01a}. However, at $z=0$ the fraction of baryons residing in the Ly$\alpha$ forest is strongly reduced to $\sim 30-40$ per cent \citep[][]{pen04a}, while at the same time the (observable) amount of baryons in condensed structures (i.e., baryons in galaxies and galaxy clusters) has increased to only a few per cent \citep[][]{fuk04a}. These observations thus indicate that a significant fraction of the baryons formerly residing in the photo-ionised IGM have ``disappeared''. Cosmological simulations have predicted that, as a result of the large-scale structure formation in the Universe, most of these ``missing baryons'' have moved into a hot, shock-heated intergalactic gas phase \citep[][]{cen99a,dav01a,ber08a}. This shock-heated intergalactic gas phase is referred to as the {\em Warm-Hot Intergalactic Medium} (WHIM) and is expected to have characteristic temperatures in the range \mbox{$T \sim 10^{\,5}-10^{\,7} \, \K$} and densities of \mbox{$n_{\rm H} \sim10^{\,-4}-10^{-\,6} \, \cm^{\,-3}$}. Constraining the distribution and physical properties of the WHIM is important to understand how the gas (and the metals) are transported from galaxies into the IGM and recycled into galaxies, and what role the shock-heated IGM has for the evolution of galaxies in the local Universe.

Since emission from such a thin plasma is extremely dim \citep{fur04a,ber09a,ber10a} and the hydrogen is almost fully ionised, the analysis of highly ionised heavy elements (in particular the high ions of oxygen, \OVI\, \OVII\ , and \OVIII) in Ultra-Violet (UV) and X-ray absorption against distant extra-galactic background sources has become the leading method to study the properties and baryon content of the WHIM at low redshift \citep[for a recent review see][]{ric08b}. Most promising is the search for five-times ionised oxygen (\OVI) in the UV spectra of low-redshift AGN, as obtained with space-based UV spectrographs such as {\em Hubble Space Telescope} (HST) {\em Space Telescope Imaging Spectrograph} (STIS) and the {\em Far Ultraviolet Spectroscopic Explorer}  (FUSE) \citep[\eg][]{tri00a,ric04a}. Oxygen is a relatively abundant element with two strong \OVI\ transitions at $\lambda1031.9 {\rm \AA}$ and $\lambda1037.6 {\rm \AA}$, and so far, more than 50 intervening \OVI\ absorbers at low redshift have been identified and analysed \citep[\eg][]{tri08b,dan08b}.

In spite of the large \OVI\ samples obtained to date, there is still no general consensus about the physical conditions of the gas giving rise to \OVI\ absorption. While \citet[][]{dan08b} find that \OVI\ (and associated \NV) are reliable tracers of collisionally ionised gas at temperatures $10^{5} \K < T < 10^{6} \K$ (\ie, the low-temperature WHIM), \citet[][]{tho08b} argue that \OVI\  arises mainly in photo-ionised gas at temperatures $T < 10^{5} \K$. Similarly, \citet[][]{tri08b} find that well-aligned \OVI-\HI\ absorbers have line widths that are consistent with photo-ionised gas. Nevertheless, these authors show that more than half of their \OVI\ absorbers are complex, \ie, multi-phase, and could thus trace both cold photo-ionised gas and lower metallicity collisionally ionised gas at $T > 10^{5} \K$.

The interpretation of the abundance and nature of intervening \OVI\ absorbers arising in collisionally ionised gas in terms of the distribution and baryon content of the WHIM is not straight-forward. In collisional ionisation equilibrium (CIE; as usually assumed for a shock-heated plasma like the WHIM), \OVI\ predominantly traces gas at \mbox{$T \sim 3 \times 10^5 \, \K$}, while in the higher temperature regime of the WHIM oxygen is further ionised to \OVII{} and \OVIII{}, observable only in the X-ray band for which high-quality spectral data are sparse. Thus, only a (small) fraction of the WHIM can in principle be traced by intervening \OVI\ absorbers. In addition, several authors \citep[\eg][]{gna07a} have argued that the assumption of CIE is not valid in case of the WHIM (\ie, the gas may be out of an ionisation equilibrium) and more recently, \citet[][]{wie09a} have shown that photo-ionisation of the WHIM strongly influences the cooling efficiency of the gas in the temperature range of interest. These physical arguments, together with the not well established oxygen abundance of the WHIM at low redshift, indicate that a reliable estimate of the baryon budget of the WHIM (i.e., the amount of ionised hydrogen) from \OVI\ observations requires a deep understanding of the physical conditions of the gas.

In this paper, we make use of a set of numerical simulations from the OverWhelmingly Large Simulations (OWLS) project \citep{sch10a} to study the physical properties and baryon content of intervening \OVI\ absorbers at low redshift. The main advantage of these simulations compared to previous ones is the implementation of important physical processes that have been largely ignored in earlier studies. As will be shown here, the influence of the photo-ionisation on the cooling function of the WHIM \citep[][]{wie09a} represents a particularly important aspect for a correct interpretation of the observed properties of the \OVI\ absorbers and their role for our understanding of the WHIM.

Even though we do not attempt to tune our simulations to reproduce any \OVI\ observables in this paper, we will compare our results to observations where possible. Moreover, a thorough analysis of variations with respect to our fiducial model and the effect of post-run physics variations on the resulting physical properties of the simulated \OVI\ absorbers are left for a future paper. Note that we will often refer to \citet[][]{opp09b} and make comparisons between our and their results, since \citet[][]{opp09b} is the most comprehensive study on \OVI\ absorbers in simulations to date \citep[see also][]{cen10a}. As we will show, the fact that \citet{opp09b} assumed CIE when computing the contribution of heavy elements to the radiative cooling rates will likely have affected their conclusions.

This paper is organised as follows: in Sec.~\ref{sec:sims} we briefly introduce the simulation we use, \ie, our fiducial model. We describe how we create our synthetic spectra and present their analysis in Sec.~\ref{sec:spec}. In Sec.~\ref{sec:obs} we discuss a comparison of our results to observations, while in Sec.~\ref{sec:phys} we analyse the physical properties of the \OVI\ gas in our simulation in detail. We present and discuss our conclusions in Sec.~\ref{sec:dis}. A detailed analysis of the effect of the signal-to-noise ratio, simulation box size, and resolution on the \OVI\ line parameter distributions are left for the Appendix.

%--------------------------------------------------------------------------------------------------------------------------------------------------------------------------------
% TABLE: Simulation list: reference model
\input{tables/refsim_list}
%--------------------------------------------------------------------------------------------------------------------------------------------------------------------------------

%--------------------------------------------------------------------------------------------------------------------------------------------------------------------------------
\section[]{Simulations} \label{sec:sims}

The simulations used in this work are part of a large set of cosmological simulations that together comprise the OWLS project, described in detail in \citet[][and references therein]{sch10a}. Briefly, the simulations were performed with a significantly extended version of the $N$-Body, Tree-PM, Smoothed Particle Hydrodynamics (SPH) code \textsc{gadget iii} -- which is a modified version of \textsc{gadget ii} \citep[last described in][]{spr05b} -- a Lagrangian code used to calculate gravitational and hydrodynamical forces on a system of particles. The initial conditions were generated from an initial glass-like state \citep[][]{whi96a} with \textsc{cmbfast} \citep[version 4.1; ][]{sel96a} and evolved to redshift $z = 127$ using the \citet[][]{zel70a} approximation. The reference model of OWLS adopts a flat $\Lambda$CDM cosmology characterised by the set of parameters $\{\Omega_{\,\rm m}, \, \Omega_{\,\rm b}, \, \Omega_{\,\Lambda}, \, \sigma_{\,8}, \, n_{\,\rm s}, \, h\} = \{ 0.238, \, 0.0418, \, 0.762, 0.74, \, 0.95, \, 0.73 \}$ as derived from the Wilkinson Microwave Anisotropy Probe (WMAP) 3-year data\footnote{These parameter values are largely consistent with the WMAP 7-year results \citep[][]{jar10a}, the largest difference being the value of $\sigma_{\,8}$, which is $2 \, \sigma$ lower in the WMAP 3-year data than allowed by the WMAP 7-year data.} \citep[][]{spe07a}. The reference model includes prescriptions for star formation (SF) and for feedback from core collapse supernovae (SNe) described in \citet[][]{sch08e} and \citet[][]{dal08b}, respectively. The implementations of radiative cooling and heating are described in \citet[][]{wie09a} and summarised below (cf. Sec.~\ref{sec:cool}). For a thorough description of the implementation of stellar evolution in the reference model we kindly refer the reader to \citet[][]{wie09b}.

We analyse the physical properties of \OVI\ absorbers identified in simulated spectra, focusing on a high resolution simulation run which assumes the reference model of the OWLS dubbed {\em REF\_L050N512}. This simulation was run down to $z=0$ in a periodic box of size $L = 50$ comoving $\hMpc$, using $512^{\,3}$ dark matter particles and equally many baryonic particles. This run has been shown to have a large enough box size and high enough resolution to obtain a converged prediction for the cosmic star formation history for $z \lesssim 4$, which, however, drops off less rapidly with time than is observed for $z < 1$ \citep[][]{sch10a}. Also, the predicted metal mass distribution for different components (stars, star-forming gas, non-star-forming gas) and gas phases from this particular run is shown to be converged for $z < 2$ and agrees broadly with observations \citep[][see discussion in their Appendix~C]{wie09b}. We refer the reader to \citet[][and references therein]{sch10a} for a more detailed description of our fiducial run (and variations thereof). In Table~\ref{tab:ref_sims} we summarise its basic  numerical properties, together with the corresponding properties of other runs assuming the reference model. These runs are discussed in Appendix~\ref{sec:conv}, where we address the convergence of our fiducial run with respect to box size and resolution, using various quantities derived from synthetic spectra as explained in the following section.

%--------------------------------------------------------------------------------------------------------------------------------------------------------------------------------
\section{Spectral analysis} \label{sec:spec}

In this section we briefly describe our method to compute synthetic spectra using the spectra generating package \textsc{specwizard} written by Schaye, Booth, \& Theuns, which follows the approach described in \citet[][their Appendix A4]{the98b}. A detailed analysis of the effect of the signal-to-noise ratio (S/N) on the distributions of line parameters (column density, Doppler parameter, rest equivalent width) derived from the spectra is presented in Appendix~\ref{sec:spec_quant}, and the main results are only summarised here for brevity (see below).

%--------------------------------------------------------------------------------------------------------------------------------------------------------------------------------
\subsection{Synthetic spectra} \label{sec:syn_spec}

Quite generally, the first step to compute a synthetic spectrum is to draw a random physical \los{} from a simulation box at a given redshift $z$. A physical \los{} is simply defined as the line between a given point on opposite faces of the simulation box, and the collection of SPH particles with projected distances to this line smaller than their corresponding smoothing length. The next step is to calculate the ionisation balance for each SPH particle as a function of redshift, density, and temperature, which we do using precomputed tables obtained with the photoionisation package \textsc{cloudy} \citep[version 07.02 of the code last described by][]{fer98a}, assuming the gas is exposed to the \citet[][]{haa01a} model for the X-Ray/UV background radiation from galaxies and quasars.

A synthetic spectrum is then computed as follows. For a chosen pixel size $\Delta x$ (in proper length units), the \los{} is divided into $N_{\,\rm pix} = \left[a(z) \, L/h \right] / \Delta x$ bins in distance, where $h$ and $a(z)$ are the normalised Hubble constant and the expansion factor at the box's redshift $z$, respectively. Given the positions, peculiar velocities, and temperatures of all relevant SPH particles as well as the densities for each ion species (\eg, \OVI), we compute the ion density-weighted temperature and peculiar velocity for each species at each bin according to the SPH interpolation scheme. Proper distance bins $\Delta x$ along the \los{} are transformed into velocity bins $\Delta v$ via $\Delta v = H(z) ~ \Delta x$, ion number densities $n_{\rm ion}$  into column densities ${\rm N}_{\rm ion} = n_{\rm ion} ~\Delta x$, and temperatures into Doppler widths. The optical depth $\tau(v)$ at each pixel for a particular transition is computed assuming a thermal (\ie, Gaussian) profile, taking peculiar velocities into account. Finally, the optical depth spectrum is transformed into a continuum-normalised flux via $F(v) = \exp [-\tau(v)]$.

We generate 1000 random \loss{} through the simulation box of our fiducial run at 5 different redshifts spanning the range\footnote{The redshift step is $\mbox{d}z = 0.125$.} $z = 0.0 - 0.5$, from which we synthesise a total of 5000 continuum-normalised spectra containing absorption by the strongest transition \OVIstrong{} of the \OVI\ doublet (\OVIstrong{}, \OVIweak{}). In order to mimic (and thus compare our results to) observations performed by HST/STIS as closely as possible, we convolve our spectra with a instrumental Gaussian Line-Spread Function (LSF) with a Full-Width-At-Half-Maximum ${\rm FWHM} = 7 \, \kms$, and resample our spectra onto $3.5 \, \kms$ pixels. We add Gaussian noise to each spectrum\footnote{Unless stated otherwise, the ${\rm S/N}$ values quoted throughout this work denote the ${\rm S/N}$ {\em per pixel} in the continuum.} assuming a flux-dependent root-mean-square (rms) amplitude given by\footnote{We assume a {\em minimum}, \ie\  flux-independent noise level $({\rm S/N})_{\rm min} = 10^2$. Note, however, that our results are not sensitive to the assumed $({\rm S/N})_{\rm min}$ value.} \mbox{$({\rm S/N})^{\,-1} \, F(v)$} for \mbox{${\rm S/N}$ = 10, 30, and 50}. Note that the typical ${\rm S/N}$ of STIS observations average ${\rm S/N} \sim 10$ \citep[\eg\ ][]{tri08b}, and hence we use our synthetic spectra with ${\rm S/N} = 10$ when testing our results against observations in the next section. Spectra with ${\rm S/N} = 30$ and ${\rm S/N} = 50$ are used to assess the effect of the signal-to-noise ratio on the line statistics (see Appendix~\ref{sec:spec_quant}). Also, and even though a \mbox{${\rm S/N}$} of 50 is obviously much higher than the typical ${\rm S/N}$ of STIS observations,  we choose it as a standard value for the spectra used in the analysis of the physical conditions of \OVI\ bearing gas in our simulation -- presented in Sec.~\ref{sec:phys} -- since this results in a larger sample of absorption lines. Besides, the {\em Cosmic Origins Spectrograph} (COS) recently installed on HST is expected to provide data at such high ${\rm S/N}$ for reasonable integration times\footnote{So far, QSO spectra with a mean \mbox{${\rm S/N}$ = 50} have been obtained with COS for a total observing time of $\sim 9 {\rm \, ks}$, compared to \mbox{${\rm S/N}$ = 10} for $\sim 28 {\rm \, ks}$ using STIS.}, although at somewhat lower spectral resolution.

\begin{figure}
\resizebox{1.\colwidth}{!}{\includegraphics{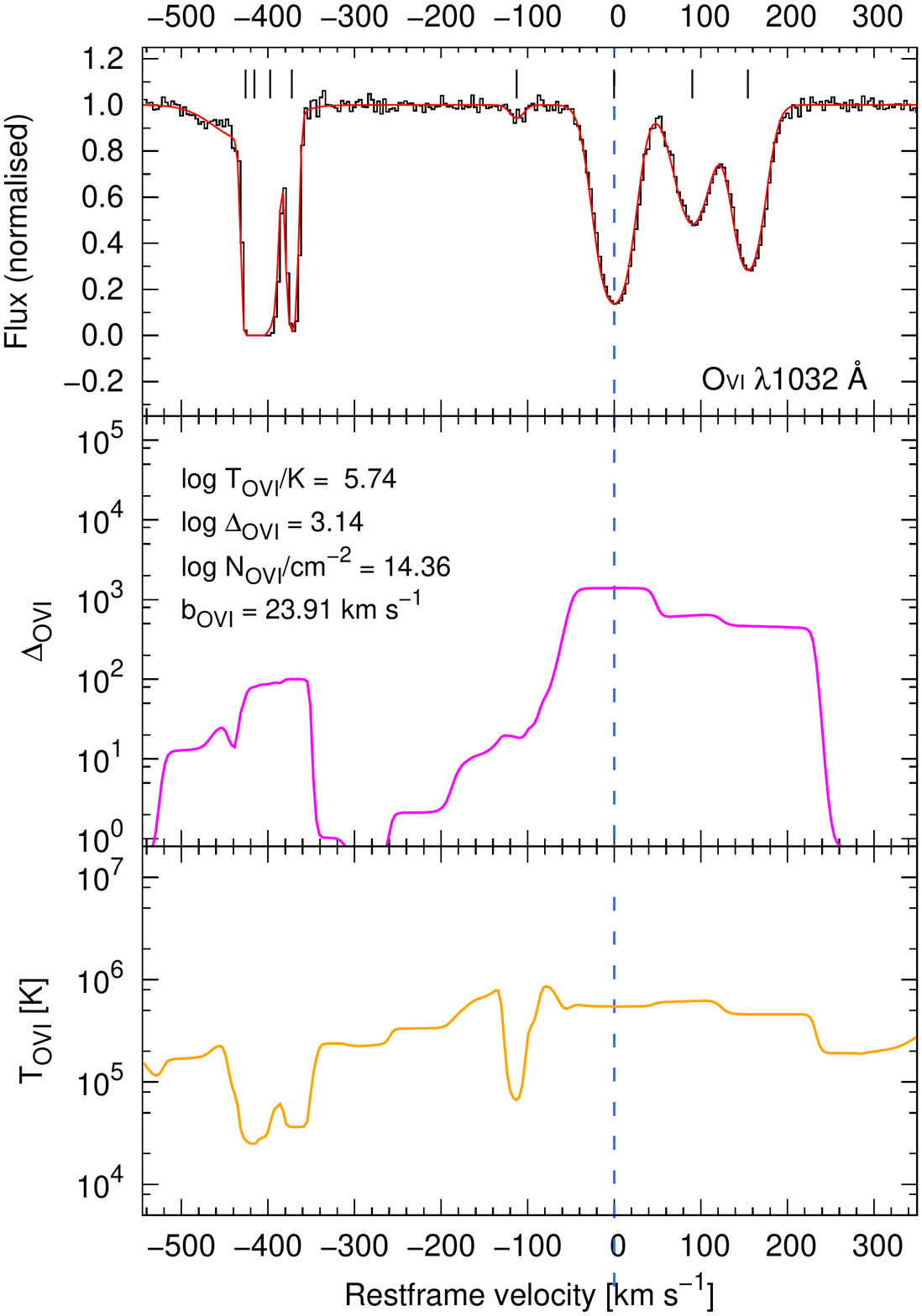}}
%\resizebox{1.\colwidth}{!}{\includegraphics{plots/physical_lsc_spec2615_o6_h1}}
\caption[ ]{The top panel shows a synthetic, normalised spectrum with ${\rm S/N} = 50$ (black) and its corresponding fit (red) along a random \los{} through a simulation box of size $L = 50 \hMpc$ at $z = 0.25$, showing a series of \OVI\ \OVIstrong{} absorption features. The vertical black marks flag the positions of the centroids of individual identified components. Note that only the relevant velocity range is shown.
The middle and bottom panels show, respectively, the optical-depth weighted overdensity (magenta) and temperature (orange) along the \los\ {\em in redshift-space} (see Sec.~\ref{sec:phys}). The values included in the middle panel correspond to the optical-depth weighted gas temperature, optical-depth weighted overdensity, \OVI\ column density, and \OVI\ Doppler parameter for the line flagged by the blue dashed line. That the optical-depth weighted quantities correspond in each case to an optical-depth weighted {\em average over the full line profile}. Note in particular the correspondence between density peaks and absorption features}
\label{fig:spec}
\end{figure}

We fit \OVI\ \OVIstrong{} absorption features identified in our spectra with individual Voigt-profile components using a modified version of the package \textsc{autovp} \citep[][]{dav97a}, which includes the Voigt-profile approximation of \citet[][]{tep06a}. The fit results include the column density $\NOVI$ and Doppler parameter $\bovi$ with corresponding uncertainties, as well as the rest equivalent width $W_{\, r}$ for each component. An example of a synthetic spectrum and its corresponding fit is shown in top panel of Fig.~\ref{fig:spec}.

The analysis of our 5000 spectra in the range \mbox{$ 0 \leq z \leq 0.5$} yields a total surveyed redshift path \mbox{$\Delta z = 98.5$}, along which we identify a total of 1093, 2246, and 3115 \OVI\ absorbers for \mbox{${\rm S/N}$=10, 30, and 50}, respectively, thus obtaining a statistically significant sample of \OVI\ absorption lines. This allows a detailed analysis of the line parameter distributions for each absorption line sample obtained from spectra with different ${\rm S/N}$, discussed in detail in Appendix~\ref{sec:spec_quant}. Briefly, we show that we can reliably identify lines with \OVI\ column densities above $\log (\NOVI/\cmsq) \approx 12.3$ at  ${\rm S/N} = 50$, $\log (\NOVI/\cmsq) \approx 12.6$ at  ${\rm S/N} = 30$, and $\log \NOVI/\cmsq \approx 13.2$ at  ${\rm S/N} = 10$, even though we systematically underestimate the total column density along each \los, indicating that we miss a fraction of the absorption lines present in the spectra. For all identified lines, however, we are able to measure rest equivalent widths quite accurately. Moreover, we find that the Doppler parameter distribution is sensitive to signal-to-noise ratios ${\rm S/N} \lesssim 10$, but fitted Doppler parameters $\bovi \lesssim 40 \kms$ are reliable if ${\rm S/N} \gtrsim 30$. Furthermore, we show in Appendix~\ref{sec:conv} that the line parameter distributions are converged with respect to box size and resolution. These results are of great importance when testing the predictions from simulations against observations using quantities derived from synthetic spectra, since they allow us to bracket the sources of a potential disagreement between predicted and observed quantities more easily. Also, they are important for the analysis of the connection between physical conditions of the absorbing gas and direct observables, \ie, \OVI\ column density and Doppler parameter (see Sec.\ref{sec:phc}).

For ease of discussion, in what follows we shall refer to our \OVI\ absorber sample described above as {\em sample 1}. The particular line sample obtained from spectra with ${\rm S/N} = 10$ is discussed in the next section where we compare the results from our simulation to observations. We generate and analyse another 5000 spectra with ${\rm S/N} = 50$ for our fiducial run at $z = 0.25$. In these spectra we identify a total of 3034 \OVI\ absorbers over a total redshift path $\Delta z = 97.7$, and we shall refer to this \OVI\ line sample as {\em sample 2}. The latter is used and analysed in detail in Sec.~\ref{sec:phys}. Note that the number of identified components and the total surveyed redshift path in sample 2 are very similar to the corresponding quantities of sample 1. Also, the distribution of column density, Doppler parameter, and rest equivalent width resulting from sample 2 are very similar to those of sample 1, but not identical, since the latter averages the evolution of the \OVI\ absorbers from $z = 0.5 \to 0$, while the former is restricted to $z = 0.25$.

%--------------------------------------------------------------------------------------------------------------------------------------------------------------------------------
% FIGURE:  random sightline with OVI spectrum and optical-depth weighted quantities
%--------------------------------------------------------------------------------------------------------------------------------------------------------------------------------

%--------------------------------------------------------------------------------------------------------------------------------------------------------------------------------
\section{Comparison to observations} \label{sec:obs}

The \OVI\ line parameter distributions (presented in Appendix~\ref{sec:spec_quant}) from our sample 1 for an adopted ${\rm S/N} = 10$ can readily be compared to three well measured \OVI\ observables, namely, the column density distribution function (CDDF for short), the equivalent width distribution, and the correlation between Doppler parameter and column density.

\begin{figure*}
\resizebox{1.\colwidth}{!}{\includegraphics{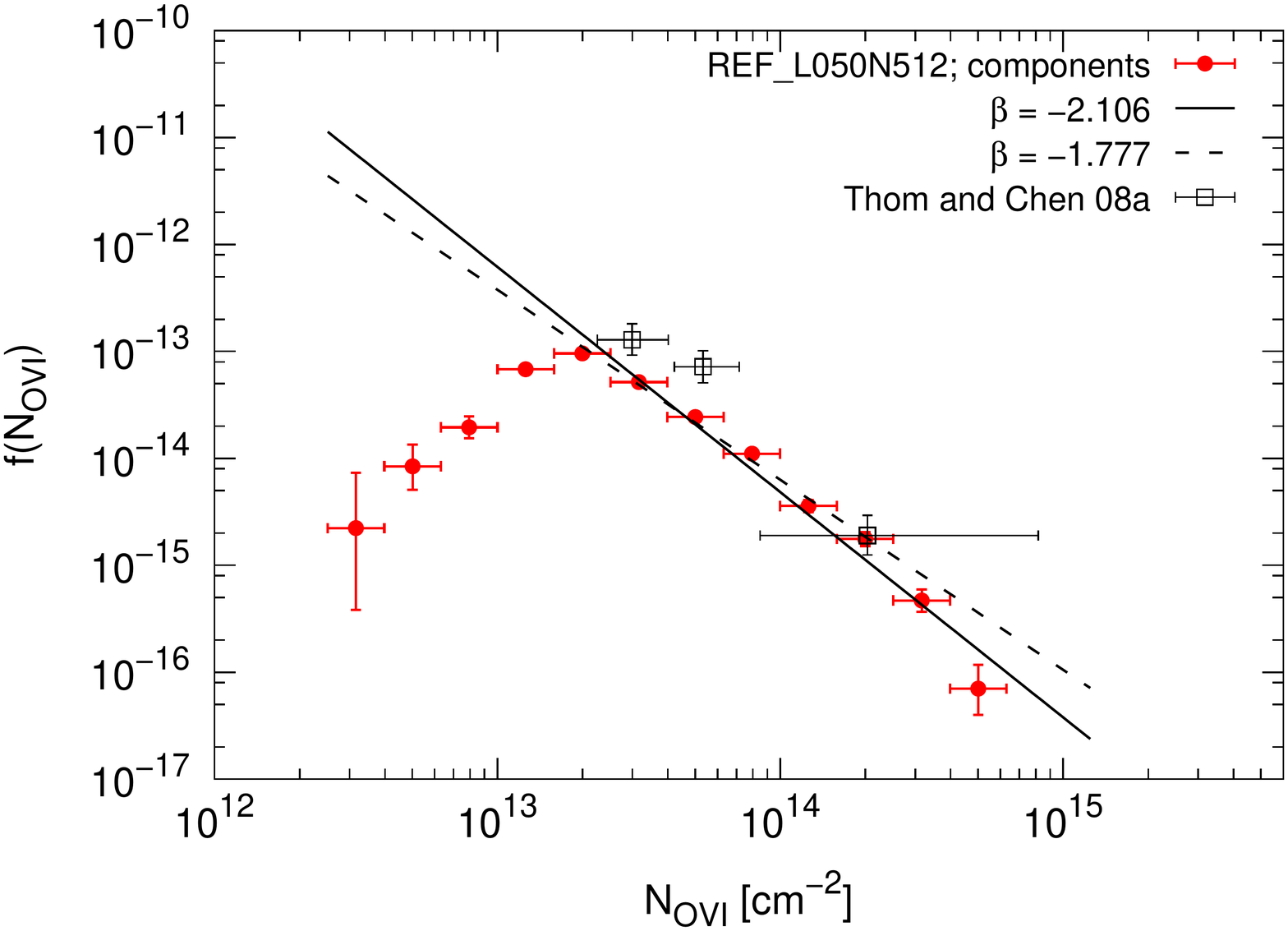}}
\resizebox{1.\colwidth}{!}{\includegraphics{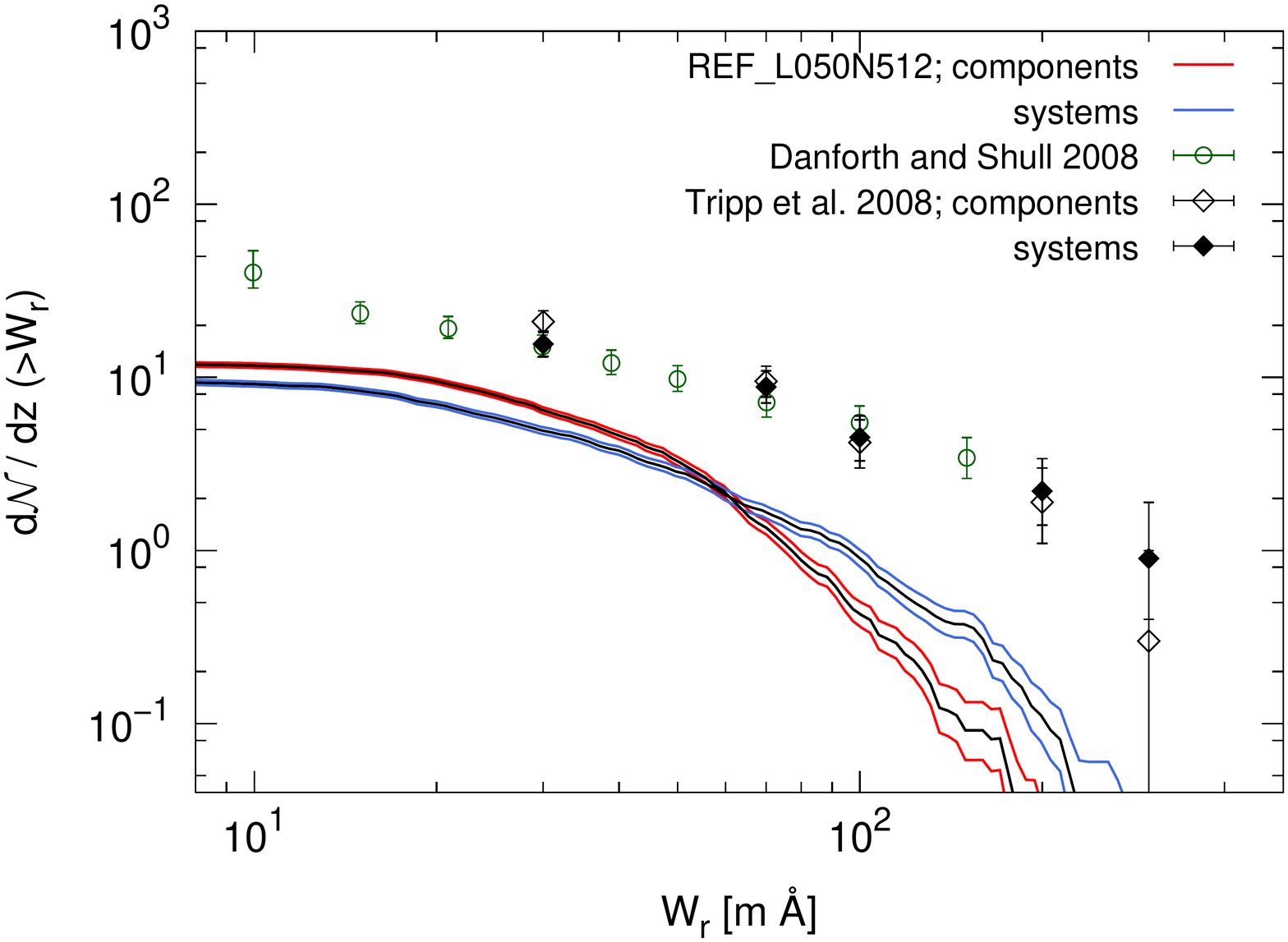}}
\resizebox{1.\colwidth}{!}{\includegraphics{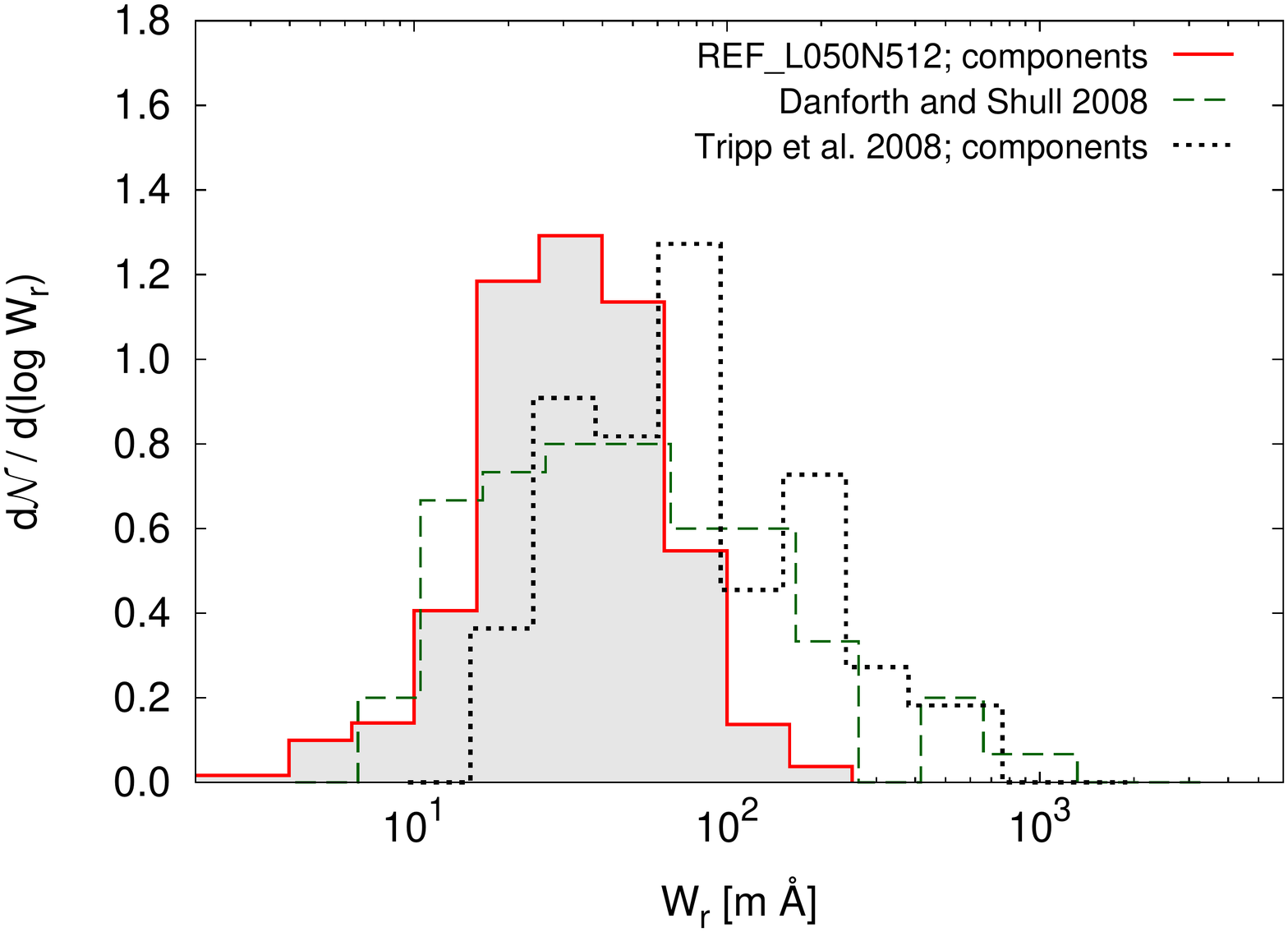}}
\resizebox{1.\colwidth}{!}{\includegraphics{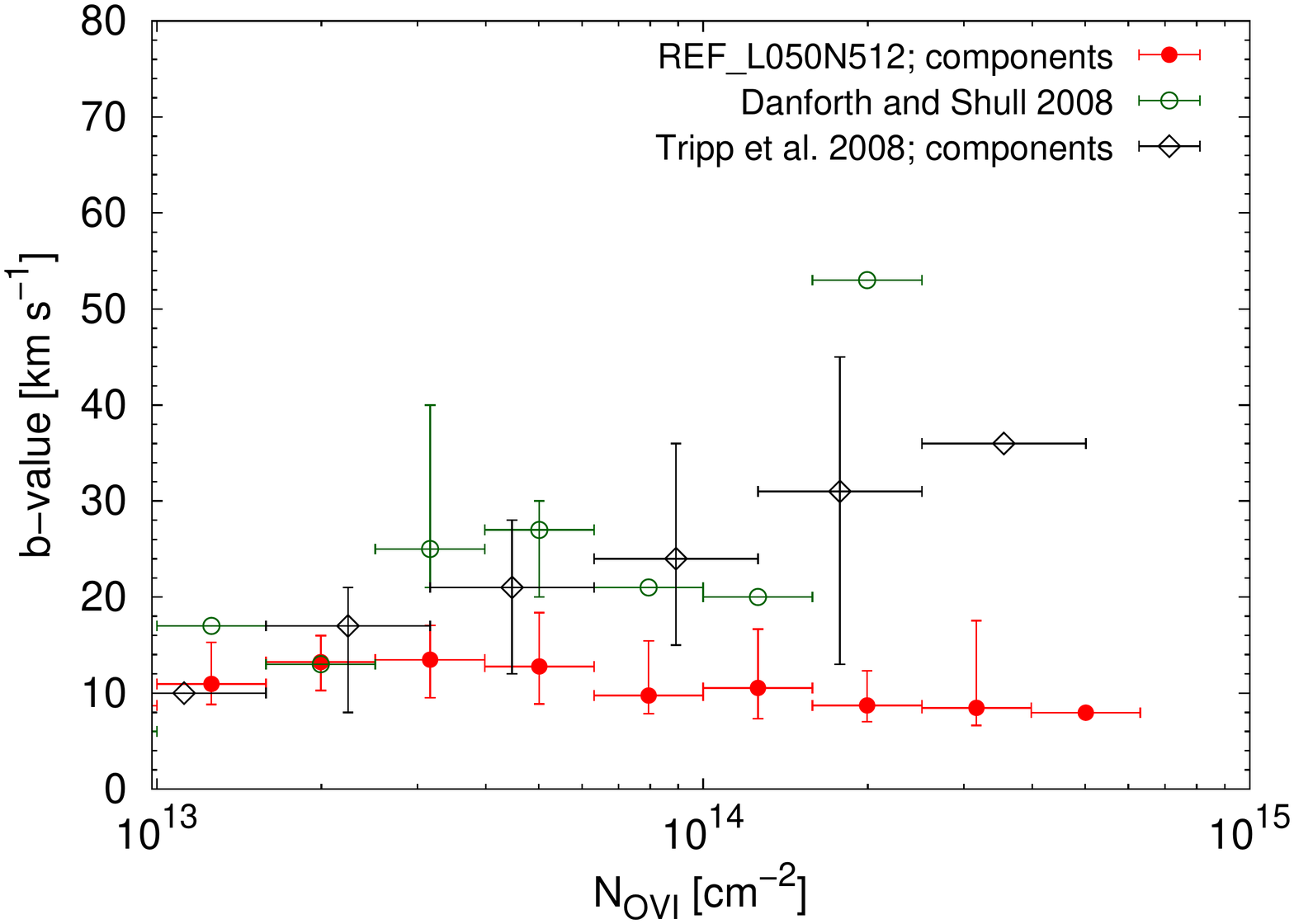}}
\caption[ ]{Comparison between the observed \OVI\ line parameter distributions and results from our fiducial run using spectra with ${\rm S/N} = 10$ (our sample 1), which roughly corresponds to the average signal-to-noise of the data.
{\em Top left:} \OVI\ column density distribution function. The $x$-bars indicate the bin size, while the $y$-error bars show Poisson single-sided $1\sigma$ confidence limits based on the tables by \citet[][]{geh86a}. The dashed and solid lines show, respectively, a fit in form of a power-law to the CDDF in the column density ranges $[1.6\times10^{\,13}, \, 6.3\times10^{\,14}] \, \cm^{\,-2}$ and \mbox{$[1.6\times10^{\,13}, \, 2\times10^{\,14}] \, \cm^{\,-2}$}. 
{\em Top right:} Cumulative line number density from our simulation for {\em single} components (red) and {\em systems} (blue), and observations (symbols). In each case, the black solid line shows the actual value at each given limiting equivalent width, while the blue and red lines display Poisson noise.
{\em Bottom left:} Differential column density distribution for single components. The result for our simulation is highlighted by the shaded area.
{\em Bottom right:} Line Doppler width {\em vs.} \OVI\ column density from our synthetic spectra (red) and observations by \citet[][black]{tri08b} and \citet[][green]{dan08b}. For ease of comparison, our predictions and the data have been binned in column density; in each bin, the dots indicate the median value, and the $x$- and $y$-bars indicate the bin size, and the 25th and 75th percentiles, respectively (see text for details).}
\label{fig:obs}
\end{figure*}

The top-left panel of Fig.~\ref{fig:obs} displays the CDDF \mbox{$f(\NOVI) \equiv \partial^{\,2} \mathcal{N} / \partial \NOVI \, \partial {\rm \chi}$}, \ie, the number of lines per logarithmic  column density interval per unit absorption path length $\chi$ (see Sec.~\ref{sec:bar}). Comparison of the result from our fiducial run to observations by \citet[][black symbols]{tho08a} shows that the amplitude of the predicted CDDF is slightly lower at column densities \mbox{$[10^{\,13}, \, 10^{\,14}] \, \cm^{\,-2}$}, but it agrees well with the high-column density data point, which has, however, a large associated uncertainty. Assuming the CDDF can be described in terms of a single power-law, \citet[][]{dan08b}\footnote{We note at this point that several studies \citep[\eg, ][]{wak09a,coo10a,wil10a} have disputed a subset of the line identifications of  \citet[][]{dan06a} and \citet[][]{dan08b}, of which we make use in this section for comparison to our results.} find that a fit in the column density range \mbox{$[10^{\,13}, \, 6.3\times10^{\,14}] \, \cm^{\,-2}$} to the CDDF obtained from their \OVI\ absorber sample results in a power-law index $\beta  = -1.98 \pm 0.11$. For the column density range \mbox{$[1.6\times10^{\,13}, \, 6.3\times10^{\,14}] \, \cm^{\,-2}$}, we find $\beta  = -2.11 \pm 0.15$ (solid line), which is consistent with the observed value. Note that this slope (with the appropriate normalisation) is also consistent with the data from \citet[][]{tho08a}. If instead we consider only lines with column densities in the narrower range \mbox{$[1.6\times10^{\,13}, \, 2\times10^{\,14}] \, \cm^{\,-2}$}, we find $\beta  = -1.78 \pm 0.08$ (dashed line), which fits the distribution better as can be judged by the smaller uncertainty in $\beta$. We conclude that the slope of the \OVI\ CDDF resulting from our fiducial run is consistent with current observational constraints up to column densities $\NOVI = 5\times10^{14} \, \cmsq$, but the amplitude is lower (by a factor 2) than observed. It is worth noting at this point that a perfect match between predicted and observed CDDF would be surprising, since the stellar yields used in our simulation are uncertain at the factor of two level \citep[see][their Appendix A3]{wie09b}.

%--------------------------------------------------------------------------------------------------------------------------------------------------------------------------------
% FIGURE: Cumulative line number density and line width vs. column density: simulations and observations
%--------------------------------------------------------------------------------------------------------------------------------------------------------------------------------

Given that most of our identified lines lie on the linear part of the curve-of-growth (see Fig.~\ref{fig:bdf}), there is a (nearly) one-to-one correspondence between the predicted CDDF and the predicted equivalent width distribution, and we thus do not expect the latter to agree well with the observed one. The comparison between the {\em cumulative} equivalent width distribution both from observations and our sample 1 is shown in the top-right panel of Fig.~\ref{fig:obs}. The observations are taken from \citet[][their Table~4, containing 75 confirmed equivalent width measurements]{dan08b} and from \citet[][their Table~2 with 77 single components]{tri08b}. While \citet[][]{tri08b} explicitly distinguish between {\em components} and {\em systems}, \citet[][]{dan08b} do not make a clear distinction between both types of absorbers. We follow the approach by \citet[][]{tri08b} for a better comparison with their results. We define an absorption system as a simply connected region along the normalised, fitted spectrum with an \OVI{} optical depth at each pixel above a given threshold $\tau_{\, {\rm th}}$. We adopt\footnote{For comparison, \citet[][]{tri08b} estimate a detection threshold given by $\tau_{{\rm th}} = 0.1$ for their data with an average $S/N \approx 10$.} $\tau_{{\rm th}} = 0.097$, which corresponds to the central optical depth of a \OVI{} \OVIstrong{} absorption line with a column density \mbox{$\log (\NOVI / \cm^{-2}) = 12.3$} -- corresponding to the {\em smallest} detectable column density in our spectra with S/N = 10; see. Fig.~\ref{fig:cddf} -- and a Doppler parameter at the STIS resolution limit $b_{\rm min} = 4.2 ~\kms$. Note that the resulting rest equivalent distribution for systems is not completely insensitive  to the adopted value of  $\tau_{\, {\rm th}}$. As can be seen in the top-right panel of Fig.~\ref{fig:obs}, our fiducial run is not able to reproduce the amplitude of the observed distribution, neither for components nor for systems, as anticipated given the disagreement between the amplitude of the observed and predicted CDDF. Also, our predicted turn-over at \mbox{$W_{\,r} \sim 100 \, {\rm m\AA}$} in the distribution of components is stronger than observed.

To gain a better insight into the reason behind the discrepancy between the predicted and observed equivalent width distributions for single components, we re-plot the predicted and observed distributions shown in the top-right panel of Fig.~\ref{fig:obs} in differential form, as shown in the bottom-left panel of the same figure. Note that the data and the results from our spectra have been binned using \mbox{$ \Delta \log W_{\rm r} = 0.2 ~{\rm dex}$}, and that the resulting distributions have been normalised to unit area. It is apparent that the observed distributions are broadly consistent with each other, even though the \citet[][]{tri08b} distribution peaks at higher \mbox{$W_{\,r}$}. In contrast, the distribution resulting from our fiducial run shows an excess of absorption lines at \mbox{$W_{\,r} \approx 30 \, {\rm m\AA}$}, as well as a lack of absorption lines with \mbox{$W_{\,r} > 100 \, {\rm m\AA}$}, which explains why our predicted turn-over at \mbox{$W_{\,r} \sim 100 \, {\rm m\AA}$} in the cumulative equivalent width distribution is stronger than observed. As is shown in Appendix~\ref{sec:spec_quant} these mismatches are not due to fitting inaccuracies but are rather intrinsic to our simulation. Note further that, while \citet[][]{tri08b}'s measurements are restricted to  \mbox{$W_{\,r} \geq 30 \, {\rm m\AA}$}, \citet[][]{dan08b} estimate their \OVI\ system sample to be complete down to \mbox{$W_{\,r} \approx 10 \, {\rm m\AA}$}.

In principle, there are two plausible reasons for the lack of strong (\ie, higher equivalent width) lines. On the one hand, it could be due to the apparent deficit of absorption components with \mbox{$\NOVI > 10^{14.5} \cmsq$} in the predicted CDDF (top-left panel of Fig.~\ref{fig:obs}), when compared to the extrapolated power-law, assuming of course that the latter gives a correct description of the data. On the other hand, this lack could also arise due to the absence of lines with \mbox{$\NOVI \sim10^{14.5} \cmsq$} {\em and} Doppler parameters much larger than the predicted median $b$-value ($\bovi = 12.9 ~\kms$ at S/N = 50). Indeed, since an \OVI\ line  saturates at these high column densities, the equivalent width is particular sensitive to the Doppler parameter. An \OVI\ line with \mbox{$\NOVI \sim10^{\,14.5} \cm^{\,-2}$} can, for example, have an equivalent width as high as \mbox{$W_{\,r} \sim 400 \, {\rm m\AA}$} if $\bovi \sim 50 \, \kms$. Lines with such high $b$-values are, however, almost absent from our spectra (see Fig.~\ref{fig:bdf}). Since there is no reason to assume that the observed power-law can be extrapolated to high column density values, and given the good agreement between our predicted and the observed CDDF slope, we consider the absence of lines with large $b$-values at a given column density as the most plausible explanation for the lack of high equivalent width lines in our simulation as compared to the data.

To further investigate this possibility, let us now consider the correlation between Doppler parameters and \OVI\ column densities. In the framework of a unifying model, \citet[][]{hec02a} have shown that a $\bovi - \NOVI$ correlation, as is observed in \OVI\ absorption systems in a variety of environments (Milky Way disk and halo, High-Velocity Clouds, the Magellanic Clouds, IGM), naturally arises in diffuse, radiative cooling gas at temperatures \mbox{$T \sim 10^{5} - 10^{6} \, \K$}. Nevertheless, there is still some controversy about the validity of such a model for IGM \OVI\ absorbers. Indeed, while \citet[][]{dan06a} find no such correlation in their data, \citet[][]{leh06a} show that their (own and compiled) data follow the predicted relation rather well. Similarly, \citet[][]{tri08b} find a correlation between the line widths and column densities  in their \OVI\ sample, but these authors note that the significance is not enough to support the model by \citet[][]{hec02a}. 

Bearing this controversy in mind, we compare the $\bovi-\NOVI$ correlation resulting from our simulation to observations. For this purpose, we use the results by \citet[][their 40 \OVI\ single component sample]{dan06a} and \citet[][their high quality sample of 77 intervening \OVI\ absorbing components]{tri08b} only, since both these samples are large enough, since each of them provides evidence either against or in favour of the predicted correlation, and because we use one of these data sets for the comparison between our predicted and the observed cumulative equivalent width distribution. To facilitate the comparison between our results and observations, we bin the \OVI\ Doppler parameters from our simulation and the data in column density bins of size \mbox{$\Delta \log \NOVI = 0.3$ dex}, and compute the median, and the 25th and 75th percentiles in each bin. The result is shown in the bottom-right panel of Fig.~\ref{fig:obs}. Apparently, the predicted correlation between Doppler parameters and column densities matches the observations at the lowest column densities \mbox{$\log (\NOVI/\cmsq) \lesssim 13.5$}, but does not so at higher column densities. While the observations show an increase in the Doppler width with increasing column density, the result from our simulation shows no clear trend. We warn, however, that the trend suggested by the binned data is to be taken with caution, given the enormous scatter in the observations.

The bottom-right panel of Fig.~\ref{fig:obs} clearly shows that the observed $b$-values are much larger at column densities \mbox{$\NOVI > 5\times10^{\,13}$} than the corresponding {\em median} $b$-values of our identified \OVI\ absorbers. This is consistent with our suspicion that the mismatch between the predicted and observed equivalent width distribution is due to a lack of \OVI\ lines with high column densities {\em and} large Doppler parameters. Furthermore, this result also indicates that the observed \OVI\ absorption systems are subject to a substantial broadening mechanism, either in the form of (small-scale) turbulence (which is not captured/properly modelled by the SPH scheme in our simulation) or Hubble broadening. The alternative that the observed \OVI\ absorbers arise in gas at higher temperatures, thus resulting in lines with larger Doppler parameters, can be ruled out, since a Doppler parameter of, \eg, $\bovi = 40 \, \kms$ corresponds to a gas temperature $T \sim 1.5\times10^{6} \, \K$ for oxygen, which would result in a too low \OVI\ ion fraction $(n_{\ionsubscript{O}{VI}}/n_{\rm O})  < 10^{-2}$, either in CIE \citep[][]{sut93a} or non-equilibrium conditions \citep[][]{gna07a}.

It is noteworthy that a similar disagreement between the predicted and the observed \OVI\ cumulative line-number density and $\bovi -\NOVI$ correlation is reported by \citet[][]{opp09b}. These authors argue that turbulence is the crucial mechanism leading to the large $b$-values measured in IGM \OVI\ systems, and solve the discrepancy between their simulations and observations by adding sub-resolution turbulent broadening as a function of hydrogen density -- partly constrained by observations-- to the Doppler parameter of their simulated absorption lines. Quite remarkably, they find that this approach, besides reproducing the $b-\NOVI$ correlation by construction, simultaneously brings their predicted equivalent width distribution into better agreement with observations. Since the CDDF slope resulting from our simulation is consistent with the data, while the $b$-value distribution is not, it is quite plausible that broadening our \OVI\ lines following the approach by \citet[][]{opp09b} would help to reconcile our results with observations. Although this is tempting, we will not consider it here. Instead, we will present and discuss a series of modifications to our reference run (\ie\  simulation runs with different parameters) and {\em post-run variations} thereof -- including the addition of sub-resolution turbulence-- and their implications for \OVI\ absorbers statistics in a future paper. For now, we will proceed with the analysis of the physical properties of the \OVI\ bearing gas in our reference run as is.

%--------------------------------------------------------------------------------------------------------------------------------------------------------------------------------
\section{Physical Properties of \OVI\ absorbers} \label{sec:phys}

In this section we analyse the physical properties (density, temperature, metallicity, ionisation state, baryon content) of the \OVI\ bearing gas traced by the absorption features identified in our synthetic spectra. For the sake of simplicity, we restrict ourselves to the analysis of our fiducial run at $z = 0.25$, which roughly corresponds to the median redshift of the \OVI\ observations found in the literature. Hence, in what follows, we will use our \OVI\ line sample 2 introduced at the end of Sec.~\ref{sec:syn_spec}.

\subsection{Optical depth-weighted physical quantities} \label{sec:odwq}

Defining physical properties such as the density or temperature of the gas responsible for the absorption features identified in simulated spectra is not a trivial task, and different methods have been described in the literature. \citet[][their equation 7]{opp09b}, for example, ascribe physical properties to simulated \HI{} and \OVI\ absorption systems by weighting the desired quantity by the product of the SPH particle mass, the mass fraction of the corresponding element (\eg\, oxygen), and the ionisation fraction of the corresponding species (\eg\, \OVI\ ) {\em at the line centre}, taking both peculiar and thermal velocities into account. \citet[][]{sch99a}, on the other hand, define the density (temperature) of the absorbing gas at the line centre as the sum of the density (temperature) of all gas elements that contribute to the absorption in that pixel, weighted by their contribution to that pixel's {\em optical depth}. A similar approach is followed by  \citet[][]{ric06a} to ascribe a density and a temperature to the absorption features identified as Broad \lya{} Absorbers (BLAs) in their simulations.

By definition, optical-depth weighted physical quantities directly relate an absorption feature identified in a synthetic spectrum to the absorbing gas in the simulation, and this gives insight about the physical state of the gas in which absorption actually takes place. This in turn allows for a meaningful comparison between quantities obtained from a spectral analysis (\eg\ , column density, Doppler parameter) and the `true' physical properties as given in the simulations (\eg, volume density, temperature). For example, for a uniform density, uniform temperature gas cloud, the optical-depth weighted properties of the cloud would be simply its given density and temperature. In case there is, \eg, a density (temperature) gradient towards the centre, then naturally the denser gas will be more important in determining the absorption line properties. The optical-depth weighted quantity properly reflects this, and provides the properly weighted average density or temperature across the absorbing cloud. We therefore choose to use optical-depth weighted physical quantities for our analysis of simulated spectra. In contrast to the approaches mentioned above, however, we do not only consider the line centre, but compute a weighted average over the line profile. So, to compute a physical quantity, say the density, associated with a certain line, we first compute the optical-depth weighted density in redshift space along the \los\ as in \citet[][]{sch99a}. Next we compute the average of the optical-depth weighted density over the line profile, weighted by the optical depth of each pixel and assign this last weighted average to the line. We have compared the physical quantities resulting from this approach with using just the value of the optical-depth weighted quantity at the line centre and found that the difference is small, with averaging over the full profile typically giving slightly smaller values.

In what follows, quantities weighted by \OVI\ optical depth will be denoted by adding a corresponding subscript; thus, for example, the optical-depth weighted gas overdensity (temperature) is denoted by $\deltaw$ ($\tempw$). Note that we define $\Delta \equiv \rho_{b}  /  \left <\rho_{b}\right > $, where \mbox{$\left <\rho_{b} \right >  = 4.18\times10^{-31} ~\left( h / 0.73 \right)^{2}~{\rm g ~ cm^{-3}}$} is the mean cosmic baryon density. An example of the optical-depth weighted temperature and overdensity along a random \los\ through a simulation box at $z=0.25$, as well as the corresponding quantities for a given \OVI\ absorption line are shown in middle and bottom panels of Fig.\ref{fig:spec}.

%--------------------------------------------------------------------------------------------------------------------------------------------------------------------------------
\subsubsection{Temperatures and overdensities} \label{sec:temp_od}

In a recent study, \citet[][]{wie09b} investigated the distribution of metals using a simulation run which assumed the same reference model as our fiducial run but at 8x lower mass resolution and twice as large a box size. Their results, which were shown to be insensitive to box size and nearly converged for both their and our fiducial runs, show that the diffuse, photo-ionised IGM, \ie\ , gas at overdensities $\Delta \lesssim 10^{1.5}$ and temperatures  $10^{3} \K \lesssim T \lesssim 10^{4.5} \K$, harbours a large fraction of the mass in the simulation, but contains a negligible fraction of the metals. They find that the metals are mostly spread over low-density structures at temperatures $T \gtrsim 10^{\,5} \K$, {\em despite} the inclusion of metal-line cooling, indicating that the Warm-Hot Intergalactic Medium (WHIM) at $10^{5} \K \lesssim T \lesssim 10^{7} \K$ and $\Delta \lesssim 10^{3}$ contains a significant amount of metals \citep[][their Figure 10]{wie09b}. As discussed by these authors, the reason that a very high fraction of the metals are in low-density gas at $T \gtrsim 10^{5} \K$ is that high-velocity winds transport metals from galaxies out to large distances, and these winds shock-heat the gas to such high temperatures. 

Motivated by their findings, here we want to investigate if and how the baryon and metal distributions, as well as different gas phases in our simulation are traced by \OVI\ detected in absorption. To this end, we compute an optical-depth weighted temperature, $\tempw$, and overdensity, $\deltaw$, for each of our \OVI\ absorbers in sample 2, and compare the resulting $(\tempw,~\deltaw)$-distribution to the intrinsic mass and metal distributions in our simulation.

\begin{figure}
\resizebox{1.1\colwidth}{!}{\includegraphics{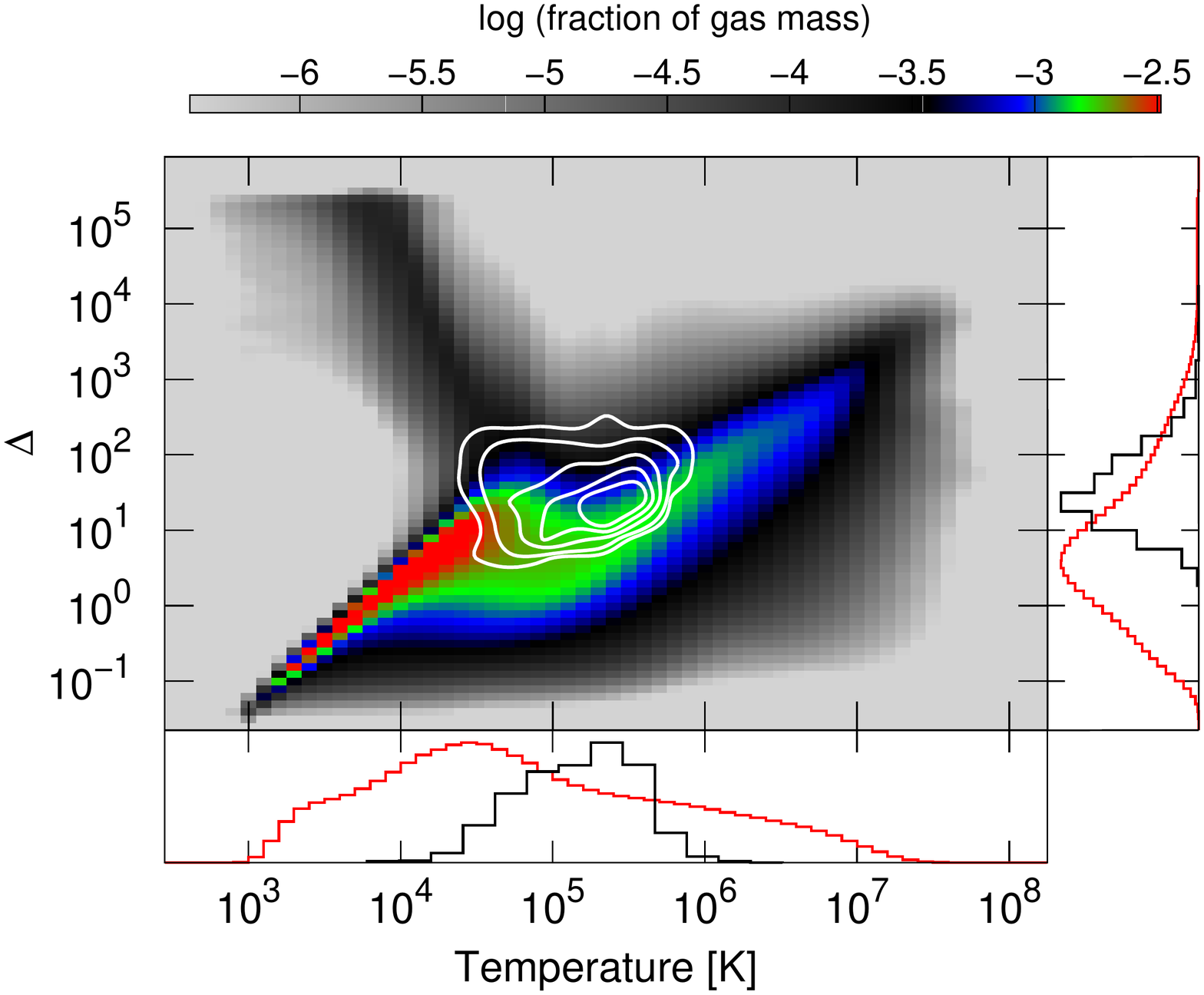}}\\
\resizebox{1.1\colwidth}{!}{\includegraphics{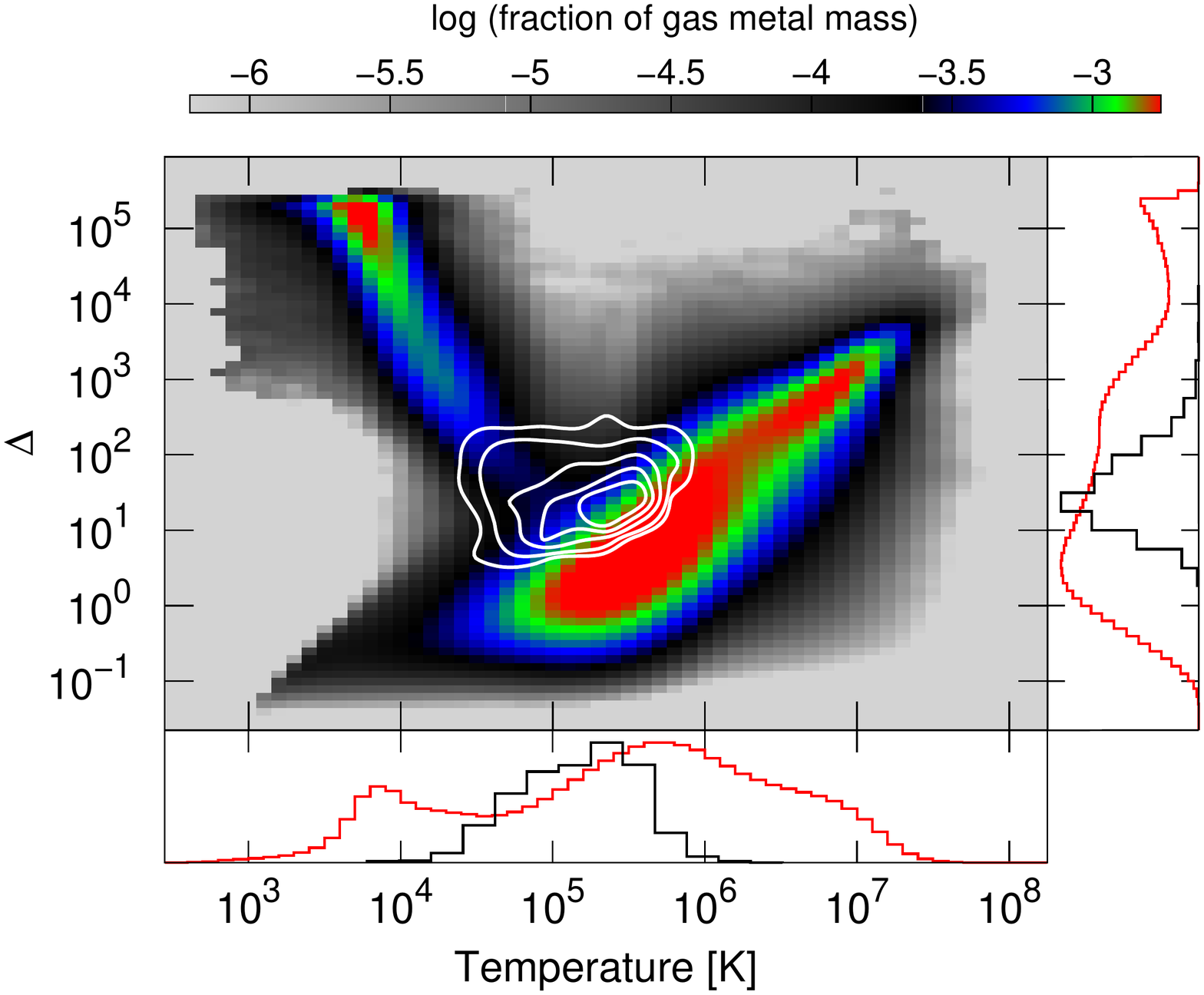}}\\
\resizebox{1.1\colwidth}{!}{\includegraphics{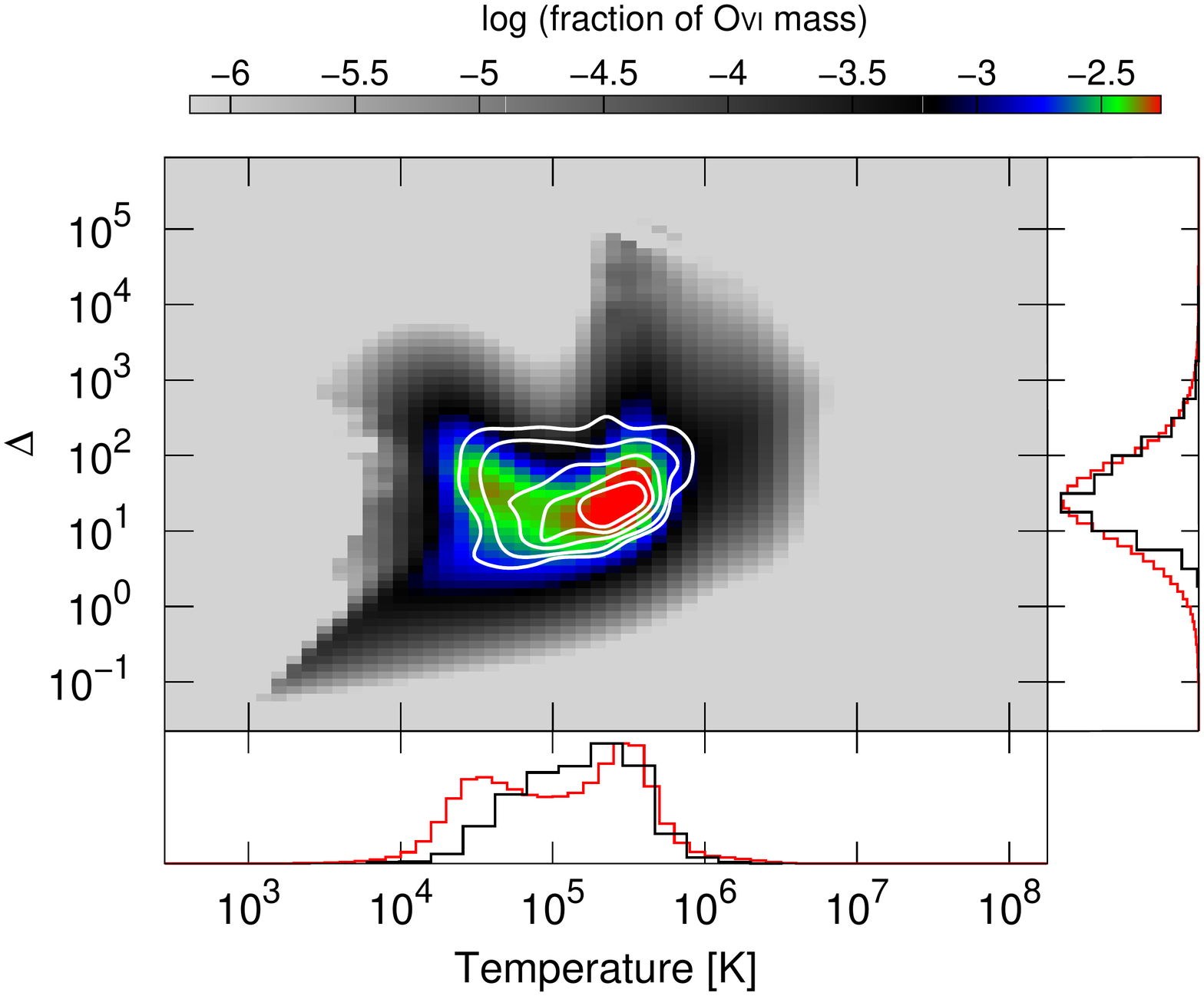}}
\caption[]{Distribution (coloured areas) of gas mass (top), gas metal mass (middle), and \OVI\ mass (bottom) in the temperature-overdensity plane at $z = 0.25$, together with the distribution of \OVI\ absorbers (white contours) of sample 2. The histograms at the bottom and to the right of each panel show the corresponding gas (red) and $\OVI$ absorbers (black) distributions marginalised over $T$ and $\Delta$, respectively, normalised to their peak values. The colour coding has been chosen in such a way that the red/green/blue/black areas contain 50/75/90/100 per cent of the total gas mass (top), total gas metal mass (middle), total \OVI\ mass (bottom), respectively. The white contours enclose, starting from the innermost, 25, 50, 75, 90, and 99 per cent of the total number of \OVI\ absorbers. Note that the white contours and the black histograms are the same in all three panels. 
}
\label{fig:ODvsT}
\end{figure}

The top panel of Fig.~\ref{fig:ODvsT} shows the gas mass distribution $\partial^{2} \, M_{\rm gas} / (M_{\rm gas, tot} \, \partial \log \Delta \, \partial \log T )$ (coloured areas) at $z = 0.25$ and the distribution of \OVI\ absorbers (white contours) in the temperature-density plane. For the case of the absorbers (contours) the axes correspond to the \OVI\ optical-depth weighted temperature and density. The colour coding shows the amplitude of the gas mass distribution. The red/green/blue/black areas enclose 50/75/90/100 per cent of the total gas mass, with the gas mass fraction in each bin given by the value indicated in the colour bar. The white contours contain, starting from the innermost, 25, 50, 75, 90, and 99  per cent of the total number (3034) of identified \OVI\ absorbers. For reference and for the subsequent analysis, we include the sub-panels below and to the right which show, respectively, the distributions (normalised to their peak value) marginalised over $\Delta$ (lower sub-panel) and $T$ (right sub-panel), of gas mass (red), and \OVI\ absorbers  (black). Note that we have excluded from this and all subsequent phase diagrams the star-forming gas (i.e., the interstellar medium) -- defined as gas with densities exceeding our adopted star-formation threshold $\nHs = 0.1 \, \cm^{-3}$ which corresponds to\footnote{The relation between hydrogen number density $\nH$ and (baryonic) overdensity \mbox{$\Delta$} is given by
\beq
	\nH = \frac{\left<\rho_{\rm b}\right>}{\mH}~X_{\rm H}~(1+z)^{\,3} \Delta \approx 1.9 \times 10^{\,-7} \cm^{\,-3}~\left( \frac{X_{\rm H}}{0.752} \right)~(1+z)^{\,3} \Delta \notag
\eeq}
 $\Delta \approx 3\times10^{5}$ at $z = 0.25$
-- since the temperature of this gas simply reflects the pressure, imposed via an equation of state $P \propto \rho^{4/3}$, of the unresolved multiphase ISM \citep[see][for details]{sch08e}.

%--------------------------------------------------------------------------------------------------------------------------------------------------------------------------------
% FIGURE: Temperature-overdensity phase diagram and projected histograms and optical-depth vs. mass-weighted
%--------------------------------------------------------------------------------------------------------------------------------------------------------------------------------

Focusing first on the gas mass distribution, we can see that approximately half of the baryonic mass in our simulation is contained in a phase corresponding to the diffuse IGM (red region), which shows a tight correlation between temperature and (over)density governed by the balance between adiabatic cooling due to the cosmic expansion and photo-heating by the meta-galactic UV background \citep[][]{hui97a}. We also find a significant amount of gas at low overdensities ($1 \lesssim \Delta \lesssim 10^2$) and much higher temperatures ($10^{5} \K \lesssim T \lesssim 10^{7} \K$), corresponding to shock-heated material. The plume of the distribution at $\Delta \gtrsim 10^{3}$ and $T \gtrsim 10^{7} \K$ , represents hot gas in galaxy clusters, \ie\  the so-called intra-cluster medium (ICM). Note that there is no well-defined demarcation between gas phases at different temperatures and overdensities (see red histograms); rather, the transition between these phases is smooth.

In contrast to the extended temperature and overdensity distributions of the gas mass in our simulation, the distribution of $\tempw$ and $\deltaw$ shown by the white contours is constrained to a relatively narrow temperature range \mbox{$10^{4.5} \K \lesssim T \lesssim 10^{6} \K$} and overdensities \mbox{$1 \ll \Delta < 10^{3}$}. Our simulation thus indicate that \OVI\ traces shock-heated material with temperatures around $T \sim 10^{\,5.3 \pm 0.5} \K$, and at slightly higher overdensities than the diffuse IGM, around $\Delta \sim 10 - 10^2$ typically. The comparison to the overall gas mass distribution also suggests that \OVI\ traces a significant amount of the baryons in the simulation. We will discuss the baryon content of \OVI\ bearing gas in more detail in Sec.~\ref{sec:bar}, and will show that, contrary to our expectation, the \OVI\ bearing gas contains only a small fraction of the warm-hot baryons. 

The middle panel of Fig.~\ref{fig:ODvsT} shows the gas metal mass distribution $\partial^{2} \, M_{\rm Z} / (M_{\rm Z, tot} \, \partial \log \Delta \, \partial \log T )$. Each particle's metal mass is defined by $m_{\rm Z} \equiv Z_{\rm sm} \times m_{\rm g}$, where $Z_{\rm sm}$ is the {\em smoothed} metallicity \citep[see Sec.~\ref{sec:mmf} and][for a detailed discussion on smoothed metallicities; we note that this plot would look nearly identical if particle metallicities had been used]{wie09b}. Again, the colour scheme shows the amplitude of the distribution with the colour cuts chosen in such a way that the red/green/blue/black areas enclose 50/75/90/100 per cent of the total gas {\em metal} mass, with the gas metal mass fraction in each bin given by the value indicated in the colour bar. The white contours are the same as in the top panel. Clearly, we can distinguish two phases that harbour a substantial fraction of the metals in the simulation at $z = 0.25$: shock heated material at temperatures $T \gtrsim 10^{5} \K$ and overdensities $\Delta \lesssim 10^{3}$, and cold-warm gas at $T \lesssim 10^{4.5} \K$ and $\Delta \gtrsim 10^{2}$. In agreement with \citet[][their Fig.~7]{wie09b} we estimate that these phases together contain $\sim 30$ per cent of the total metals in the simulation, while the rest is contained in star-forming gas ($\sim 10$ per cent) or locked up in stars ($\sim 60$ per cent). We find that nearly 20 per cent of the metals in our simulations are contained in WHIM gas at $10^{5} \K \leq T \leq 10^{7} \K$, but the \OVI\ absorbers together contain only $< 0.1$ per cent of the metal budget. Hence, \OVI\ absorbers do not trace the bulk of metals, even though the majority arise in gas with a high metal fraction, as can be judged from the ($\tempw, \, \deltaw$) distribution. 

Presumably, there is a continuous exchange of material between the cold-warm, diffuse IGM and these metal-rich gas phases. Intergalactic gas is accreted onto the potential wells of galaxies and thereby shock-heated. At the same time, galactic winds shock-heat and transport metal-rich gas from the high-density, star-forming regions into the low-density IGM, some of which may then cool down and may eventually be re-accreted onto galaxies to fuel star formation (``intergalactic fountain''). In this scenario, and given the location of the \OVI\ distribution with respect to the gas metal mass distribution (middle panel), \OVI\ absorbers would mainly arise in  enriched, shock-heated galactic-wind material that has probably started to mix with the ambient gas and cool down, wandering from the high-temperature, low-density region to the low-temperature, high-density regime on phase space. As such, many of the \OVI\ absorbers tracing warm-hot gas may be short-lived. We will provide evidence for this later on when we discuss the metallicity of \OVI\ bearing gas below and in Sec.~\ref{sec:mmf}.

To better understand the nature of these \OVI\ absorbers, we investigate if (and how much of) the \OVI\ bearing gas in our simulation is traced by the \OVI\ we see in absorption. To this end, we compare in the bottom panel of Fig.~\ref{fig:ODvsT} the \OVI\ mass distribution $\partial^{2} \, M_{\ionsubscript{O}{VI}} / (M_{\ionsubscript{O}{VI}, {\rm tot}} \, \partial \log \Delta \, \partial \log T )$ to the distribution of \OVI\ absorbers in the ($\tempw, \, \deltaw$) plane. A particle's oxygen mass is given by ${\rm m_{\ionsubscript{O}{VI}}  \equiv (n_{\ionsubscript{O}{VI}}/n_{\rm O})\times X_{\rm O} \times m_{\rm g}}$, where $\rm X_{\rm O}$ is the {\em smoothed} oxygen mass fraction. Colours and contours have the same meaning as in the middle and top panels. We find that the bulk of the \OVI\ is found in gas at moderate overdensities $\Delta \sim 10^{0.5} - 10^{2.5}$, and temperatures between $T = 10^{4} \K$ and $10^{\,6} \K$, which is consistent with the overall distribution of our \OVI\ absorbers. Also, the marginalised temperature distribution (lower sub-panel, red histogram) is clearly bi-modal, showing that \OVI\ is distributed among two phases with different temperatures: $T \sim 10^{4.5} \K$, corresponding to photo-ionised gas, and $T \sim 10^{5.5} \K$, corresponding to collisionally ionised gas. We find that more than 65 per cent of the \OVI\ seen in absorption traces mainly the hot gas phase at temperatures $T \gtrsim 10^{5} \K$. This means that approximately one third of the \OVI\ absorbers arise in cooler, photo-ionised gas. This number agrees with the result of \citet[][]{tri08b} who find compelling evidence that  34 per cent of the \OVI\ absorbers in their sample arises in gas at $T < 10^5 \K$. 

Our finding that most of the \OVI\ arises in hot gas is not caused by the limited ${\rm S/N}$ of our spectra, nor by the limited resolution, since qualitatively the same result is found in noise-free spectra with very high resolution. As can be judged by the amplitude of the (two-dimensional) distribution around  $T \sim 10^{4.5} \K$ and $10 < \Delta < 10^{2}$, the \OVI\ content in this gas phase is low, which could mean that also the \OVI\ column density is too low for all of this gas to be detectable in absorption. Our assumption is reinforced by the fact, going back to the middle panel, the overall metal content of gas at temperatures and overdensities comparable to the low temperature \OVI\ bearing gas is quite low. The fact that the peak temperature of the hotter phase is slightly shifted to lower values as compared to the high-temperature peak of the intrinsic temperature distribution, suggests that \OVI\ absorption arising in cooler gas phases overlaps in redshift space with \OVI\ absorption from hotter gas, thus leading to slightly lower {\em optical-depth weighted} gas temperatures. To a lesser extent, it is a consequence of averaging optical-depth weighted quantities over the full line profile, as discussed in Sec.~\ref{sec:odwq}.

An interesting fact is that the low temperature \OVI\ bearing gas closes the gap between the high-temperature, low-density and the cold, high-density metal-rich gas phases shown in the middle panel, which suggests that some of the gas traced by \OVI\ has already cooled down to an equilibrium temperature $T \sim 10^{\,4.5} \K$, which is typical for enriched gas at $z = 0.25$ (see Sec.~\ref{sec:cool}, Fig.~\ref{fig:cool_1}). 

Summarising, we find that, while the mass in our simulation is more or less equally distributed between two phases with temperatures \mbox{$T \lesssim 10^{4.5} \K$} and \mbox{$T \gtrsim 10^{4.5} \K$}, and overdensities $0.1 \lesssim \Delta \lesssim 10^{3}$, \OVI\ traces gas at high temperatures $T > 10^{5} \K$ and moderate overdensities $10 < \Delta < 10^{2}$, which are typical for the warm-hot intergalactic medium (WHIM) in our \citep[][]{wie09b} and other simulations \citep[][]{cen99a,dav01a,ber08a,tor09a}. This gas phase typically contains $\sim40$ of the baryons in our simulation, but the \OVI\ bearing gas turns out to contain only a small fraction of the cosmic baryons at low redshift (see Sec.~\ref{sec:bar}). Furthermore, we find that roughly 40 per cent of the metals in our simulation are distributed between warm-hot gas at moderate to high overdensities $1 < \Delta < 10^{3}$ ($\sim30$ per cent), and cold star-forming gas at very high densities ($\sim10$ per cent). While \OVI\ certainly arises in gas containing metals, it is not a tracer of the main metal reservoirs in our simulation.

Our result that absorption by \OVI\ in our simulations is preferentially related to gas at temperatures \mbox{$10^{4.5} \K  < T < 10^{6} \K$}, with a high fraction (more than 65 per cent) having temperatures falling in the range \mbox{$10^{5} \K < T < 10^{6} \K$}, \ie\ , in the low temperature regime of the WHIM, is quite interesting, since \citet[\eg\ ][]{opp09b} find exactly the opposite trend, namely that the overwhelmingly majority of the \OVI\ in their simulations is found in gas a temperatures \mbox{$T \sim 10^{4.2 \pm 0.2} \K$}, typical of photo-ionised gas. It is important to note that our simulation does show a gas phase containing \OVI\ at photoionisation temperatures (see Fig.~\ref{fig:ODvsT}, bottom panel), which is however not detected in absorption, most probably due to its low overall metal content. In Sec.\ref{sec:cool}  we will present a deeper analysis of the disagreement between the findings by \citet[][]{opp09b} and our results, which we will attribute to their neglect of the effect of photo-ionisation on metal-line cooling.

%--------------------------------------------------------------------------------------------------------------------------------------------------------------------------------
% FIGURE: Distribution of metal mass fractions (vs. temperature overdensity)
\begin{figure*}
{\resizebox{1.\colwidth}{!}{\includegraphics{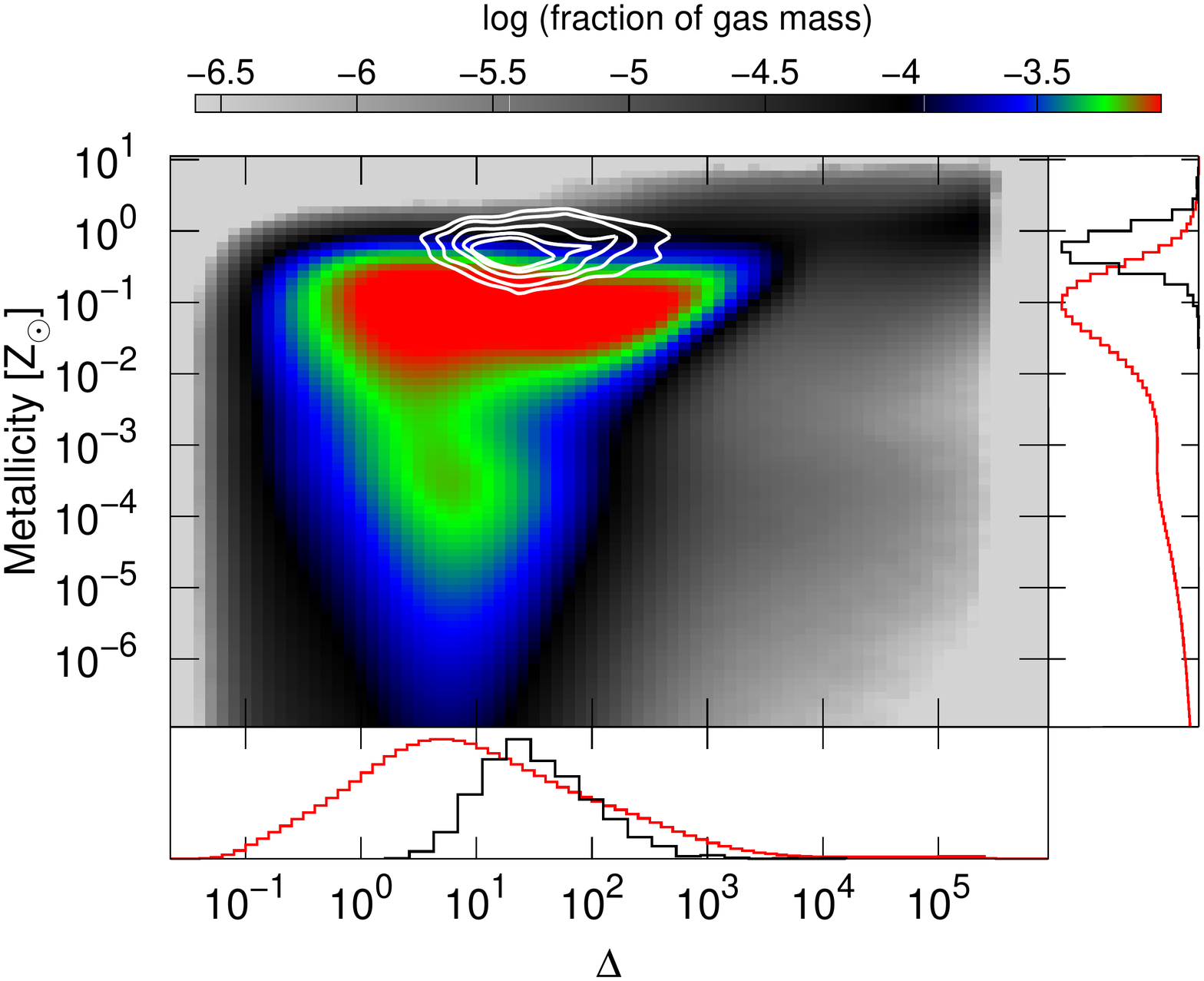}}}{\resizebox{1.\colwidth}{!}{\includegraphics{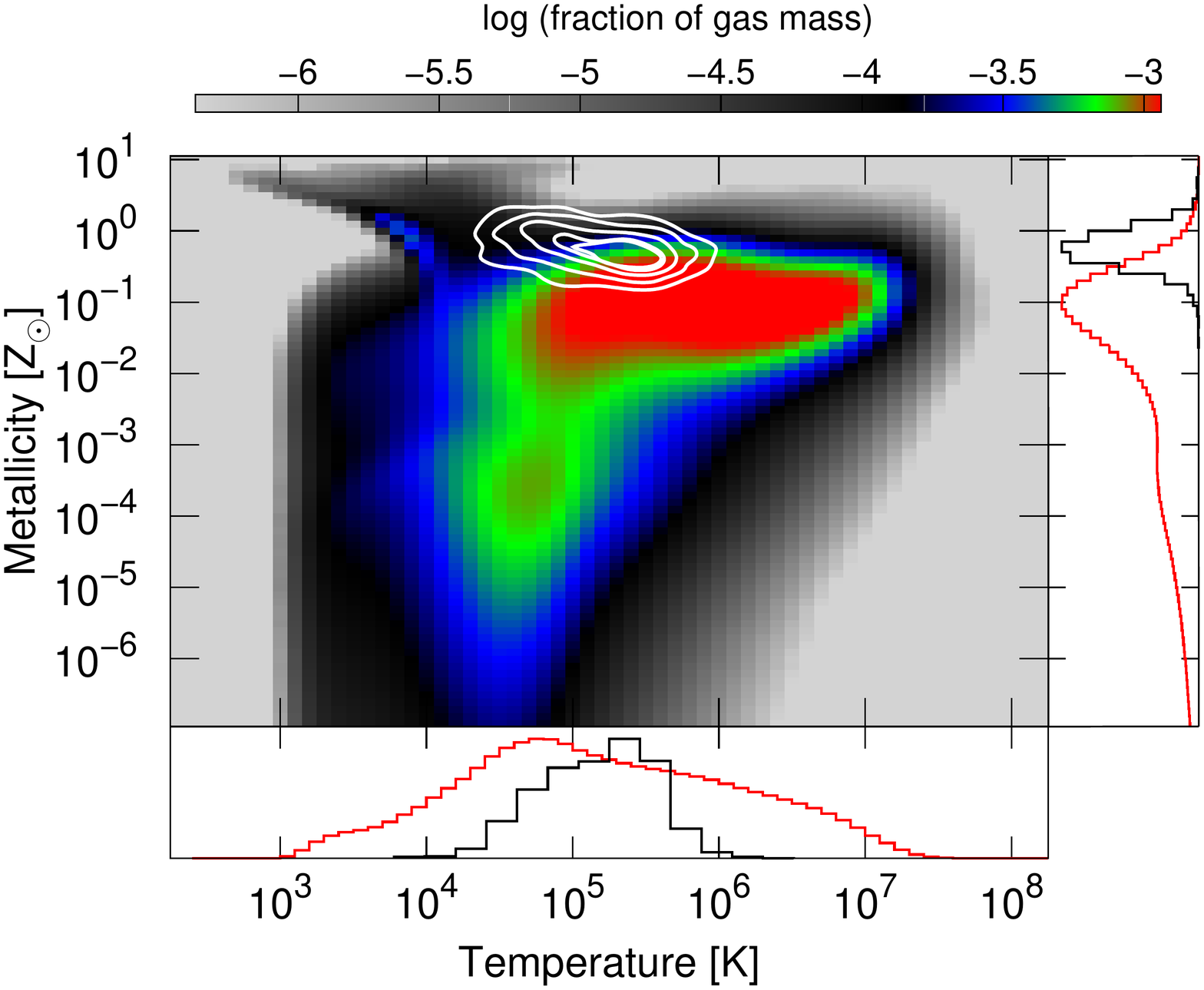}}}
\\
{\resizebox{1.\colwidth}{!}{\includegraphics{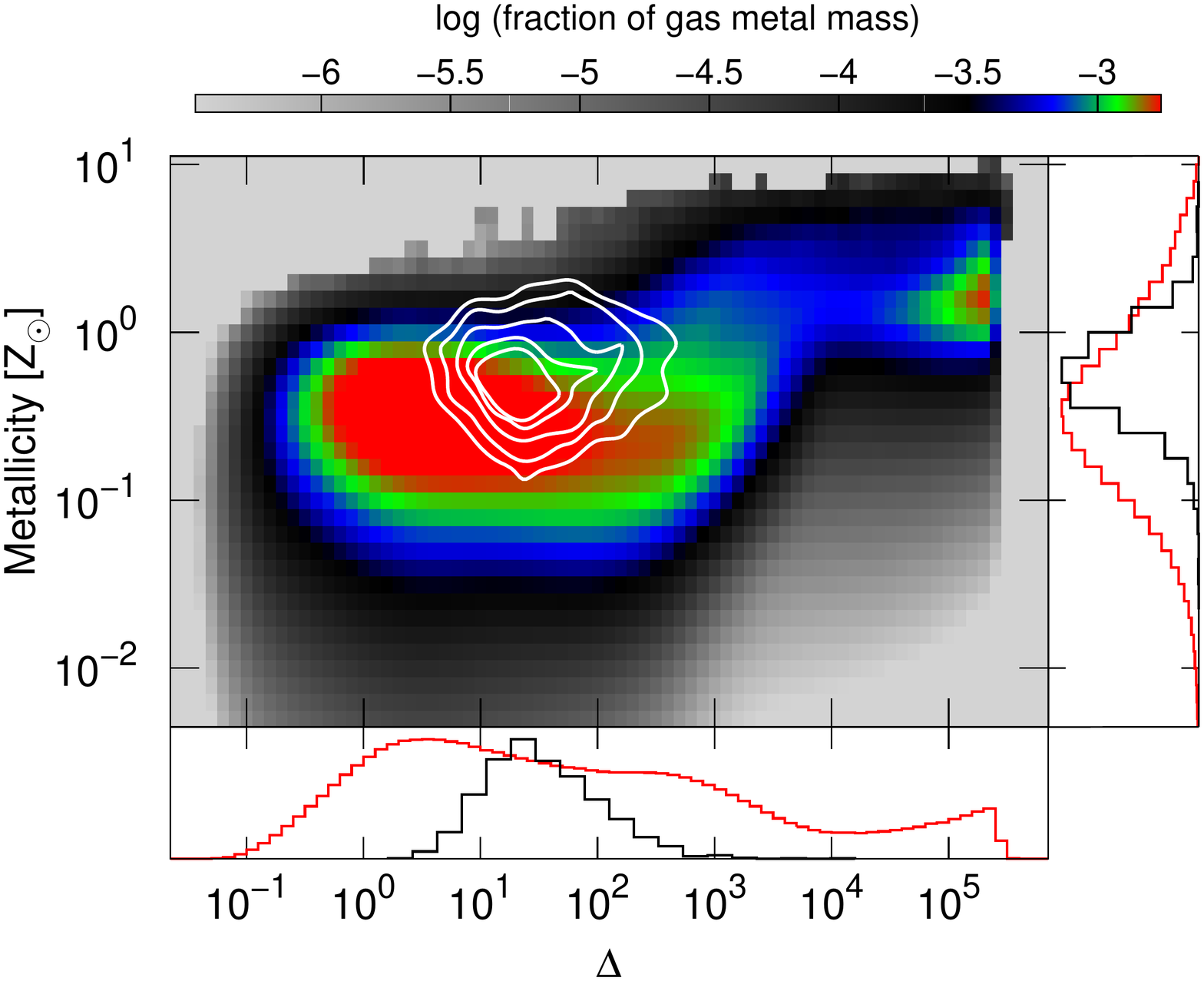}}}
{\resizebox{1.\colwidth}{!}{\includegraphics{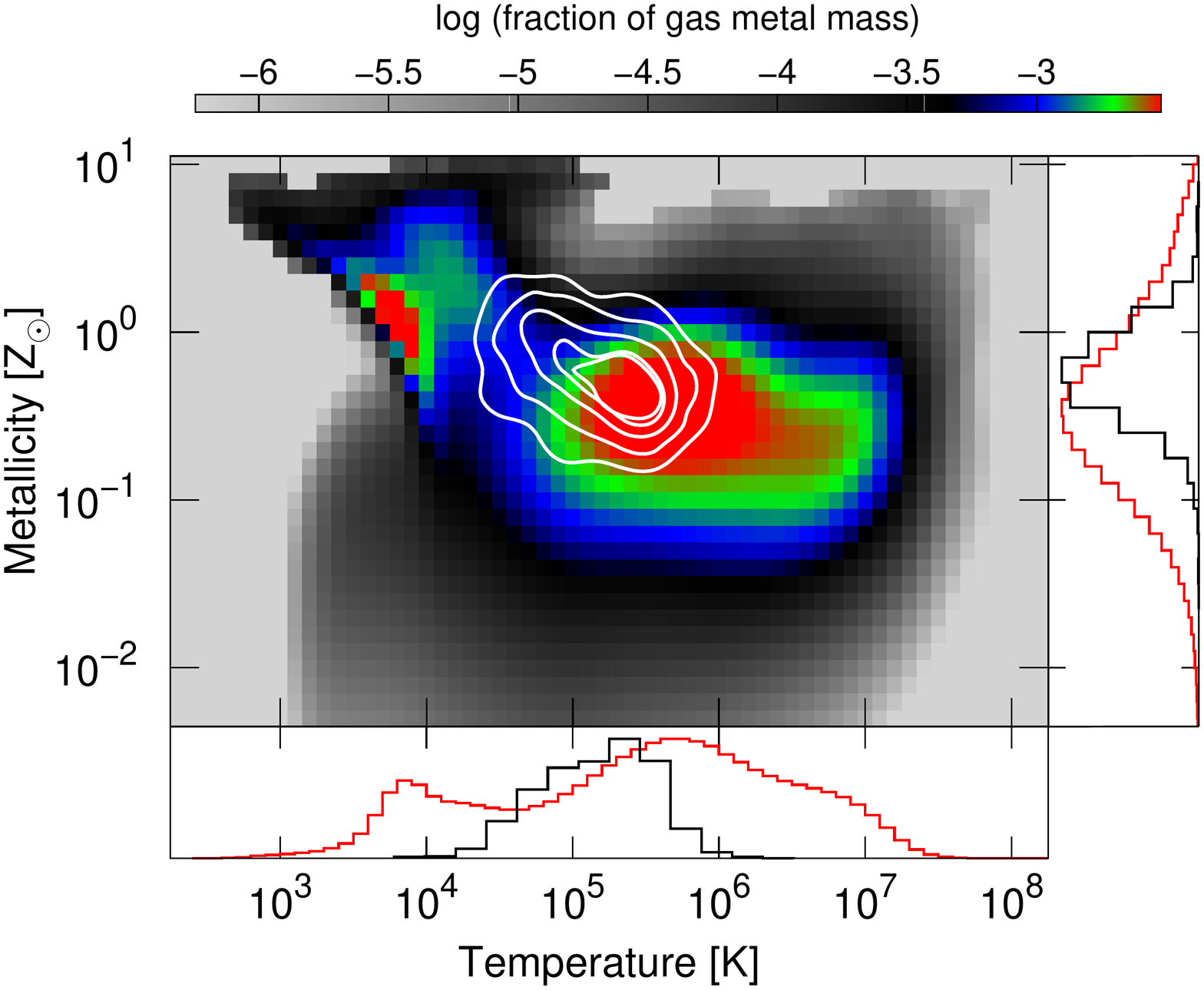}}}
\caption[]{
{\em Top row:} Gas mass distribution (coloured areas) on the overdensity-metallicity (left) and temperature-metallicity (right) planes, together with the distribution of the \OVI\ absorbers (white contours) for our fiducial run {\em REF\_L050N512} at $z = 0.25$. Note that the marginalised overdensity distributions (lower sub-panels) are identical to those in the top panel of Fig.~\ref{fig:ODvsT} (right sub-panel).
{\em  Bottom row:} Gas metal mass distribution on the overdensity-metallicity (left) and temperature-metallicity (right) planes for our fiducial run {\em REF\_L050N512} at $z = 0.25$. White contours are as in the top row. Note that the marginalised temperature distributions (lower sub-panels) are identical to those in the middle panel of Fig.~\ref{fig:ODvsT} (lower sub-panel).
Colour coding and contour levels are as in Fig.~\ref{fig:ODvsT}. We use a solar metallicity of $Z_{\odot} = 0.0127$ to express the absolute metallicities predicted by the simulation in solar units.}
\label{fig:mmf}
\end{figure*}
%--------------------------------------------------------------------------------------------------------------------------------------------------------------------------------

%--------------------------------------------------------------------------------------------------------------------------------------------------------------------------------
\subsubsection{Metallicity} \label{sec:mmf}

In this section, we analyse the local metallicity of the gas traced by the \OVI\ absorbers (in sample 2). We proceed in the same manner as with gas temperatures and overdensities, and estimate optical-depth weighted metallicities $\metaloviw$ for each of our identified \OVI\ absorbers. Note that an accurate measurement of the elemental abundances in the absorbing gas is very important, since this quantity is crucial for an indirect estimate of the baryon fraction in the gas traced by a given absorption species (see Sec.~\ref{sec:bar}).

In our simulation, as is done in all OWLS runs, we trace how stellar evolution alters the abundance of 11 elements (\H, \He, \C, \N, \Ox, \Ne, \Mg, \Si, \Fe, \Ca, \Su) explicitly, while simultaneously following the metallicity of each particle \citep[see][for further details]{wie09b}. We choose to use {\em smoothed} metallicities -- as opposed to {\em particle} metallicities -- for the various processes that are metallicity-dependent, in particular for the calculation of radiative metal cooling. The smoothed metallicity of a particle is defined as the ratio of the SPH estimates of the metal mass density and the total gas mass density at the location of the particle, while the particle metallicity is given by the ratio of the particle's metal mass and its total gas mass. \citet[][]{wie09b} have shown that in low-metallicity gas, smoothed metallicities are generally higher than particle metallicities. The use of smoothed metallicities in part counters (but does not solve) the lack of metal mixing inherent to SPH, which leads to differences when compared to the use of particle metallicities (although these differences decrease with increasing resolution). First, the use of smoothed metallicities results in higher fractions of metals residing in gas both at lower temperatures ($T < 10^{5} \K$ as compared to $T \sim 10^{6} \K$) and lower metallicities. Second, the use of smoothed abundances enhances radiative cooling and increases the predicted SFR. These differences may be important for the comparison of our results to the results from studies employing particle metallicities \citep[\eg][]{opp09b}.

In the top row of Fig.~\ref{fig:mmf}, we show the gas mass distributions $\partial^{2} \, M_{\rm gas} / (M_{\rm gas, tot} \, \partial \log Z \, \partial \log \Delta)$ (left panel; coloured areas) and $\partial^{2} \, M_{\rm gas} / (M_{\rm gas, tot} \, \partial \log Z \, \partial \log T )$ (right panel, coloured areas), together with the distribution of the \OVI\ absorbers in the $(\Delta,~Z)$ and $(T, ~Z)$ planes (white contours), respectively. Again, for the case of the absorbers the axes correspond to the \OVI\ optical-depth weighted temperature $\tempw$, density $\deltaw$, and metallicity $\metaloviw$. The meaning of colours and contours is the same  as in Fig.~\ref{fig:ODvsT}. The histograms in the lower and right sub-panels show the corresponding distributions marginalised over metallicity and overdensity (left panel) or temperature (right panel) normalised to their peak values. The panels in the bottom row show the gas metal mass distributions $\partial^{2} \, M_{\rm Z} / (M_{\rm Z, tot} \, \partial \log Z \, \partial \log \Delta)$ (left panel; coloured areas), and $\partial^{2} \, M_{\rm Z} / (M_{\rm Z, tot} \, \partial \log Z \, \partial \log T)$ (right panel; coloured areas), as well as their corresponding marginalised distributions (histograms in the low and right sub-panels), together with the distribution of the \OVI\ absorbers (white contours). Note that these panels follow the same colour scheme and include identical contours as the corresponding top-row panels, but that the $y$-axis range is smaller.

Several interesting conclusions can be drawn from this figure. First, the top-row panels reveal that a high fraction of the baryonic mass in our simulation is contained in gas with local metallicities $-2 < \log (Z / Z_{\odot}) < -1$. In contrast, \OVI\ absorbers clearly trace over-enriched material with metallicities $\log (\metaloviw / Z_{\odot}) > -1$. The median metallicity of the gas traced by \OVI\ is $\metaloviw \approx 0.6 ~Z_{\odot}$, which is almost an order of magnitude higher than the median gas metallicity at similar overdensities and temperatures. Note that the use of optical-depth weighted metallicities ensures that we track (and average over) the metallicity of all individual particles that contribute to the \OVI\ absorption. Hence, $\metaloviw$ is an estimate of the {\em local}, \ie\  around the spatial resolution limit, metallicity. Note that the overall {\em mean} metallicity of the structures giving rise to the \OVI\ absorption -- as is derived from observations by using the ratio of the \OVI\ column density to the total \HI{} column density in the absorbing gas -- can be much smaller than $\metaloviw$. In our case, this is supported  by the fact that gas at overdensities and temperatures comparable to the centres of mass of the $(\deltaw, \, \metaloviw)$ and $(\tempw, \, \metaloviw)$ distributions display a large range of metallicities. The large scatter in metallicities, in particular at the lowest overdensities, thus suggests that the metal distribution is quite inhomogeneous.

\begin{figure}
\resizebox{\colwidth}{!}{\includegraphics{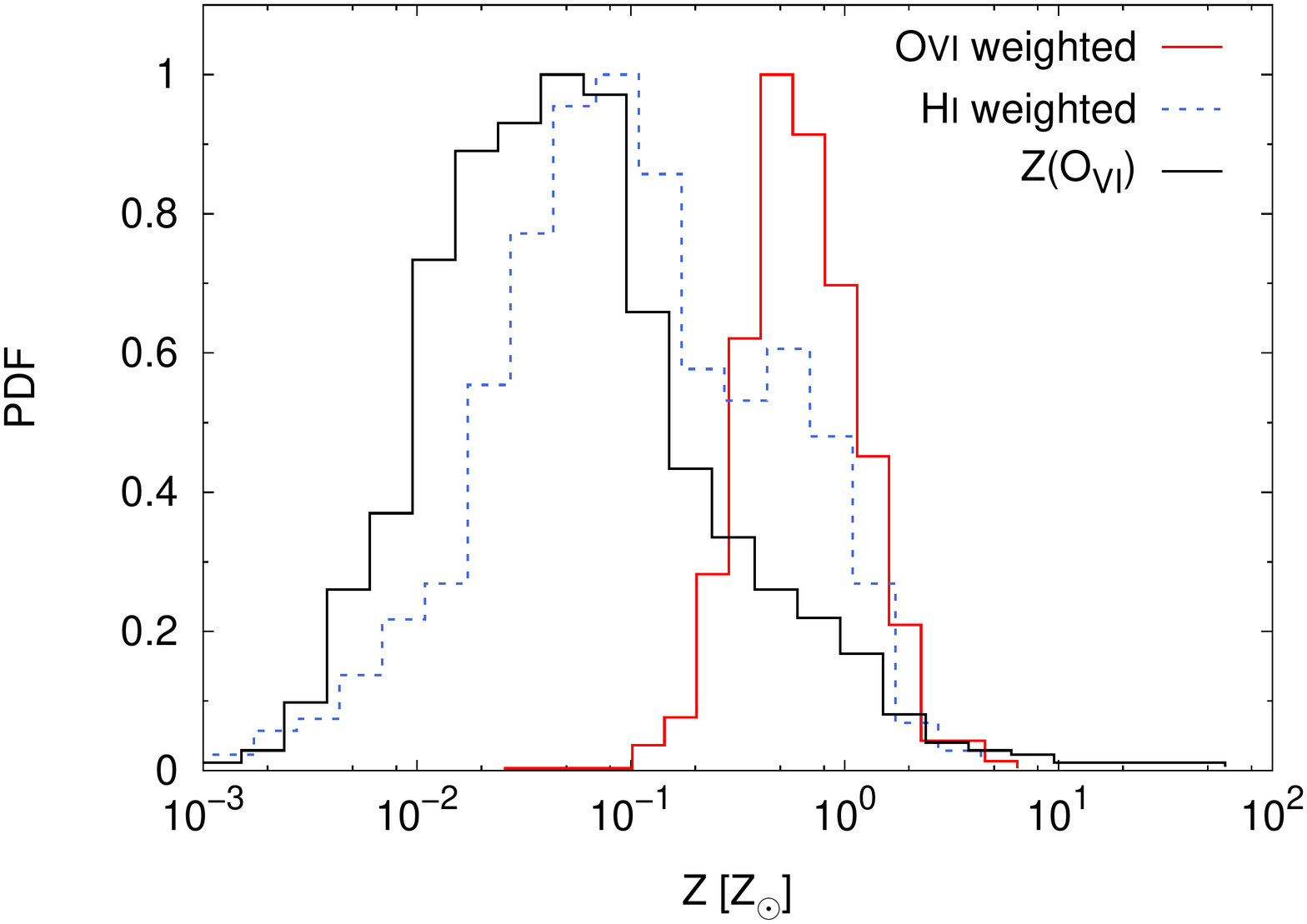}}
\caption[ ]{Metallicity distribution for our sample of {\em well-aligned} \HI\ - \OVI\ absorber pairs. The red (blue) histogram shows the distribution of \OVI\ (\HI) optical-depth weighted metallicities. The distribution of metallicities estimated from the the \OVI\ - \HI\ column density ratio $( \NOVI / \NHI )$ is shown by the black curve (see eq. \ref{eq:metal} and corresponding text for details).}
\label{fig:metal}
\end{figure}

In order to investigate more closely the metal distribution in the structures where the \OVI\ absorption takes place, we proceed as follows: First, we fit all \HI\ \HIstrong\ absorption features along the same spectra from which we obtain our \OVI\ sample 1. Next, we identify every \OVI\ component in the latter sample that is coincident with a given \HI\ component within $\pm 10 \kms$ in velocity space. In this way we obtain a sample of {\em well-aligned} \HI\ - \OVI\ absorber pairs. Assuming that the coincidence in velocity space implies that the \HI\ and \OVI\ absorption arise in the same gas, we estimate the (average) metallicity $Z(\OVI)$ of the structure giving rise to the absorption via
\beq \label{eq:metal}
	Z(\OVI) = \left(\frac{\NOVI}{\NHI}\right) \left(\frac{\fhiw}{\foviw}\right) \left(\frac{\mO}{\mH}\right) \xhoviw \, .
\eeq
We compute $Z(\OVI)$ for each well-aligned absorber pair using the corresponding optical-depth weighted ionisation fractions, $\fhiw$ and $\foviw$, and the \OVI\ optical-depth weighted hydrogen fraction, $\xhoviw$. Also, we compute for each individual \HI- and \OVI-member of a well-aligned absorber pair a  corresponding pair of optical-depth weighted metallicities, $\metalhiw$ and $\metaloviw$, respectively. The distributions of the resulting metallicities for all well-aligned \HI\ and \OVI\ components in our sample are shown in Fig.~\ref{fig:metal}. Clearly, the distribution of the mean metallicity estimated from the \OVI\ -\HI\ column density ratio, $Z(\OVI)$, shows a larger scatter and has a peak value which is lower by an order of magnitude than that of the $\metaloviw$ distribution. In contrast, the $Z(\OVI)$ distribution is comparable in width and range to the distribution of $\metalhiw$, although it peaks at a slightly lower value. Interestingly, there seems to be a sub-population of well-aligned \HI - \OVI\ absorber pairs with metallicities around $~0.6 ~Z_{\odot}$, where the optical-depth weighted metallicity of the \HI-member matches well the optical-depth weighted metallicity of the \OVI-member.

This exercise readily shows that {\em local} gas metal content, as measured by the \OVI\ optical-depth weighted metallicity, can differ substantially from the overall, average metallicity of the structure giving rise to the observed absorption. This in turn strongly suggests that the metallicity distribution is highly in-homogeneous, and that \OVI\ absorbers are biased towards high-metallicity patches embedded in gas with an otherwise lower, average metallicity.

We note that the patchy metal distribution may in part be due to the metal-mixing problem inherent to SPH which is alleviated (but by no means solved) by the use of smoothed abundances, as mentioned above. While the implementation of metal diffusion \citep{gre09a,she09a} could, depending on the choice of parameter values, reduce the scatter in the metallicities, this may in fact be in conflict with observations which indicate that intergalactic metals are poorly mixed on small scales \citep{sch07a}. Our simulations, on the other hand, seem to agree with the latter observational fact, at least in terms of well-aligned \HI\ - \OVI\ absorber pairs. We will present a detailed analysis of the alignment statistics, as well as of the physical properties of aligned pairs, in a companion paper ({\em Tepper-Garc{\'\i}a et al.}, in preparation).

%--------------------------------------------------------------------------------------------------------------------------------------------------------------------------------
% FIGURE: Metallicity histograms
%--------------------------------------------------------------------------------------------------------------------------------------------------------------------------------

In terms of the metal mass distribution, we can see in Fig.~\ref{fig:mmf} that the vast majority of the metals in our simulation reside in gas with metallicities $-1 \lesssim \log (Z / Z_{\odot}) \lesssim 0$, temperatures $T \gtrsim 10^{5} \, \K$, and overdensities $10^{-1} < \Delta \lesssim 10^{3}$. From the overlap of the \OVI\ distribution with the overall metal mass  distribution, we may conclude that the \OVI\ we see in absorption traces the moderate-density, warm, enriched gas relatively well. Nevertheless, despite the coexistence of the \OVI\ absorbers with this gaseous phase on the $(T,~Z)$- and $(\Delta, ~Z)$-planes, the gas traced by \OVI\ contains actually only a vanishingly small (less than 0.1 per cent) fraction of the metals (see Sec~\ref{sec:temp_od}).

To sum up, the results of this section indicate that \OVI\ traces over-enriched gas, which is neither representative for the bulk of the baryons nor for the bulk of metals. The metallicity of the gas traced by \OVI\ is high, with a narrow distribution extending over the range  \mbox{$(0.1,~1) ~Z_{\odot}$}, and a peak at \mbox{$\metaloviw \approx 0.6 ~Z_{\odot}$}, which is consistent with results from other simulations \citep[\eg\ ][]{cen06a,opp09b}, and metallicity measurements in metal-line absorption systems at low redshift \citep[\eg\ ][]{sav02a,pro04a,tri05a,jen05a,coo08b}. In spite of the general agreement with observations, we want to emphasise again that our quoted values refer to {\em local} estimates of the metallicity, while observed values are generally derived from metal-line systems aligned (\ie\  within a given range in velocity space) with \HI{} absorption, which can lead to significantly lower {\em mean} metallicities if the metal distribution is patchy \citep[see][for a discussion]{sch07a}. We showed that this can be indeed the case by comparing optical-depth weighted metallicities with metallicity estimates derived from \OVI\ -\HI\ column density ratios for well-aligned \HI\ and \OVI\ absorber pairs.

%--------------------------------------------------------------------------------------------------------------------------------------------------------------------------------
\subsubsection{Baryon content} \label{sec:bar}

%--------------------------------------------------------------------------------------------------------------------------------------------------------------------------------
% FIGURE: title
\begin{figure}
{\resizebox{1.15\colwidth}{!}{\includegraphics{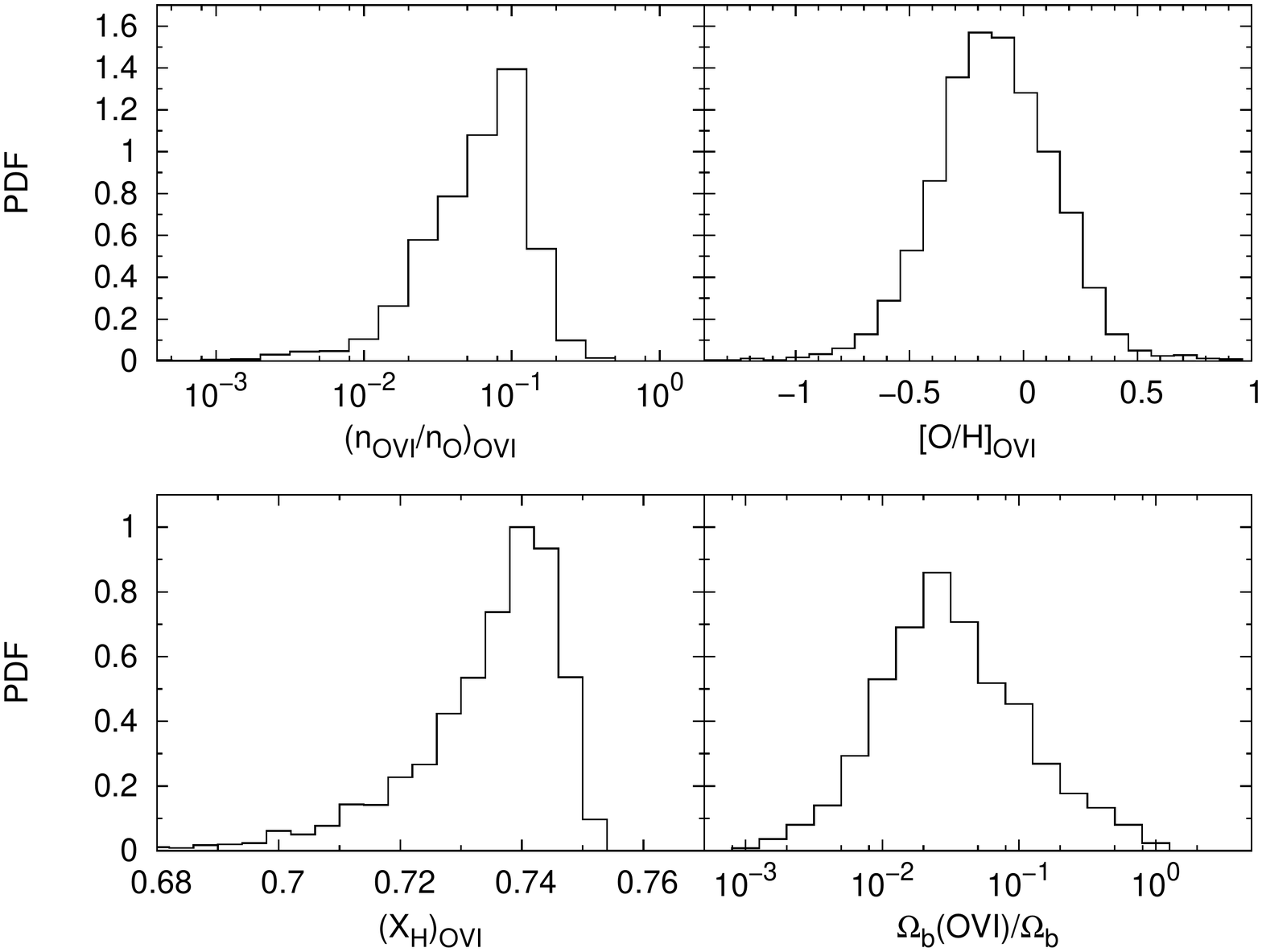}}}
\caption[]{Distribution of optical-depth weighted \OVI\ fraction (top-left), local oxygen abundance (top-right), and local hydrogen mass fraction (bottom-left) for our sample of 3034 \OVI\ absorbers at $z = 0.25$ in spectra with ${\rm S/N} = 50$ (sample 2). The bottom-right panel shows the distribution of the baryon density, relative to the cosmic mean, traced by \OVI\ along individual \loss{}. Note that we use the default \textsc{cloudy} solar oxygen abundance $\left(n_{\rm O}/n_{\rm H}\right)_{\odot} = 4.9\times10^{-4}$ to compute $\rm [O/H]_{\ionsubscript{O}{VI}}$.}
\label{fig:omb}
\end{figure}
%--------------------------------------------------------------------------------------------------------------------------------------------------------------------------------

The baryon content of gas traced by \OVI\ is a quantity of great interest since it has long been anticipated by cosmological simulations that nearly half of the baryons in the Universe \citep[absent in observational inventories, see \eg\ ][]{fuk04a} might be hidden in the WHIM \citep[][]{cen99a,dav01a,ber08a}, and that this gas phase might be detectable through its different emission and absorption signatures, in particular absorption by \OVI\ \citep[][]{cen99a}.

A standard approach used in the literature to indirectly estimate the baryon fraction of \OVI\ bearing gas is to write
\beq \label{eq:omb}
	\Omega_{\rm b}(\OVI) = \frac{\mH }{\rho_{\rm c}} ~\left( \frac{c}{H_0} \sum_{i=1}^{N_{\rm LOS}} \Delta \chi_{i} \right)^{-1} \sum_{i=1}^{N_{\rm LOS}} \sum_{j=1}^{N_{\rm ab s}} \frac{\NOVI_{,ij}}{{\rm X_{H}}~(n_{\ionsubscript{O}{VI}}/n_{\rm O}) ~(n_{\rm O}/n_{\rm H})},
\eeq
where $\mH$ is the hydrogen mass, $\rho_{\rm c}$ the critical density, and $X_{\rm H}$ the (assumed) hydrogen mass fraction. The factor in brackets is the total surveyed physical length with $\Delta \chi_i$ the so-called absorption path length \citep[see \eg\ ][]{sch01a} along an individual \los.

One major caveat of this approach is that the estimated value for $\Omega_{\rm b}(\OVI)$ is highly sensitive to both the ion fraction and the metallicity of the gas and that, owing to the difficulties in measuring these quantities, plausible values need to be assumed. Usually, the \OVI\ ion fraction is set to $(n_{\ionsubscript{O}{VI}}/n_{\rm O}) = 0.2$ -- corresponding to the peak ion fraction of \OVI\ at $T \sim 2-3\times10^{\,5} \, \K$ either in CIE \citep[][]{sut93a} or non-equilibrium \citep[][]{gna07a} --  and the metallicity to $\rm [O/H] = 0.1 ~{\rm dex}$ \citep[see \eg\ ][]{tri00a}.

Here we want to estimate the baryon content of the gas traced by our sample of \OVI\ absorbers, and we want to do this in such a way that we can compare our result to the observed range of values. Therefore, we follow the above approach, \ie, using equation \eqref{eq:omb}, but exploit the advantage of our simulation, which allows us to measure the ion fraction $(n_{\ionsubscript{O}{VI}}/n_{\rm O})$, oxygen abundance $(n_{\rm O}/n_{\rm H})$, and hydrogen mass fraction $\rm X_{H}$ of the gas giving rise to {\em each} \OVI\ absorber. To this end,  and  to be consistent, we again use optical-depth weighted quantities, estimated for the \OVI\ absorbers in sample 2. The resulting distributions of the values for each of these quantities are shown in Fig.~\ref{fig:omb} (top row; bottom-left).

Inserting the values for each of the latter quantities and for each identified \OVI\ absorber in equation \eqref{eq:omb}, we obtain $\Omega_{\rm b}(\OVI)/\Omega_{\rm b} = 0.014$ over 5000 \loss\ spanning a total absorption path $\Delta \chi = 139.3$. The distribution of the baryon content along {\em individual} \loss\ is shown in the bottom-right panel of Fig.~\ref{fig:omb}. Using the same 5000 spectra\footnote{In this case we identify a total of 1044 \OVI\ components rather than 3034.}, this time with ${\rm S/N} = 10$ rather than ${\rm S/N} = 50$, we find $\Omega_{\rm b}(\OVI)/\Omega_{\rm b} = 0.0096$. Either value is lower than values estimated from observations which range from $\Omega_{\rm b}(\OVI)/\Omega_{\rm b} \gtrsim  0.019$ \citep[][]{tri00a}, through $\sim  0.045$ \citep[][]{tho08a}, $\gtrsim  0.054$ \citep[][]{sem04a}, and $0.086\pm0.008$ \citep[][]{dan08b}. 

Two things are noteworthy. First, as can be judged from the bottom-right panel of Fig.~\ref{fig:omb}, the distribution of $\Omega_{\rm b}(\OVI)$ along individual \loss\ shows a large scatter, reflecting the effect of cosmic variance. This indicates that measurements of the baryon content using data taken along different \loss{} might vary drastically, even in case that the ionisation fraction and the metallicity are known to high accuracy, which they are usually not (although note that our \loss\ span $\Delta z = 0.02$ at $z = 0.25$). Second, and related to the previous point, is the fact that the {\em median} optical-depth weighted ionisation fraction resulting from our simulation is $\foviw \approx 0.052$, \ie,  is a factor four smaller than usually assumed, and the median optical-depth weighted metallicity is $\rm [O/H]_{\ionsubscript{O}{VI}} \approx -0.19 ~{\rm dex}$ or about six times larger than the `canonical' 10 per cent solar-value. Also, the distribution of both these  quantities shows a large dispersion (see top panels in Fig.~\ref{fig:omb}). This in turn implies that baryon density measurements from observations which assumed fixed values for the \OVI\ ionisation fraction and the metallicity may be off (most probably overestimated) by non-negligible factors, independent of the effect of cosmic variance. This in turn might explain why we find an average baryon density which is lower than the various measured values.

A second quantity we can easily compute and compare to measured values is the baryon content, relative to the critical density, of  \OVI\ in all identified absorbers. The expression used to compute this quantity is very similar to equation \eqref{eq:omb} and is given by
\beq \label{eq:omi}
	\Omega_{\ionsubscript{O}{VI}} = \frac{\mO }{\rho_{\rm c}} ~\left( \frac{c}{H_0} \sum_{i=1}^{N_{\rm LOS}} \Delta \chi_{i} \right)^{-1} \sum_{i=1}^{N_{\rm LOS}} \sum_{j=1}^{N_{\rm ab s}} \NOVI_{,ij} \, ,
\eeq
where $\mO$ is the oxygen atomic mass. This quantity is directly measurable, and has the advantage that it is independent of the metallicity and ionisation corrections. If we use the {\em recovered}, \ie, fitted \OVI\ column densities, we get \mbox{$\Omega_{\ionsubscript{O}{VI}} = 1.1 \times 10^{-7}$} for \mbox{S/N = 50}, and \mbox{$\Omega_{\ionsubscript{O}{VI}} = 9.3 \times 10^{-8}$} for \mbox{S/N = 10}, down to \mbox{$W_{\,r} = 10 ~{\rm m\AA}$}. If instead we use the {\em true}, total \OVI\ column density along each \los\ we find \mbox{$\Omega_{\ionsubscript{O}{VI}} = 1.2 \times 10^{-7}$}. Note that the latter value exactly matches (neglecting numerical round-off errors) the \OVI\ mass density relative to the critical density obtained by adding up the \OVI\ mass of each SPH particle in the simulation and dividing by the simulation volume. For comparison with our results, \citet[][]{tho08a} report \mbox{$\Omega_{\ionsubscript{O}{VI}} = 1.7 \pm 0.3 \times 10^{-7}$} for \mbox{$W_{\,r} > 30 ~{\rm m\AA}$}., while \citet[][]{dan08b} find \mbox{$\Omega_{\ionsubscript{O}{VI}} = 4.9 \pm 0.4 \times 10^{-7}$} for \mbox{$W_{\,r} > 10 ~{\rm m\AA}$}. Both values are higher than ours, but this is expected since the amplitude of our CDDF is also lower compared to observations.

%--------------------------------------------------------------------------------------------------------------------------------------------------------------------------------
% FIGURE:  Physical conditions as a function of observables
\begin{figure*}
\resizebox{1.1\textwidth}{!}{\includegraphics{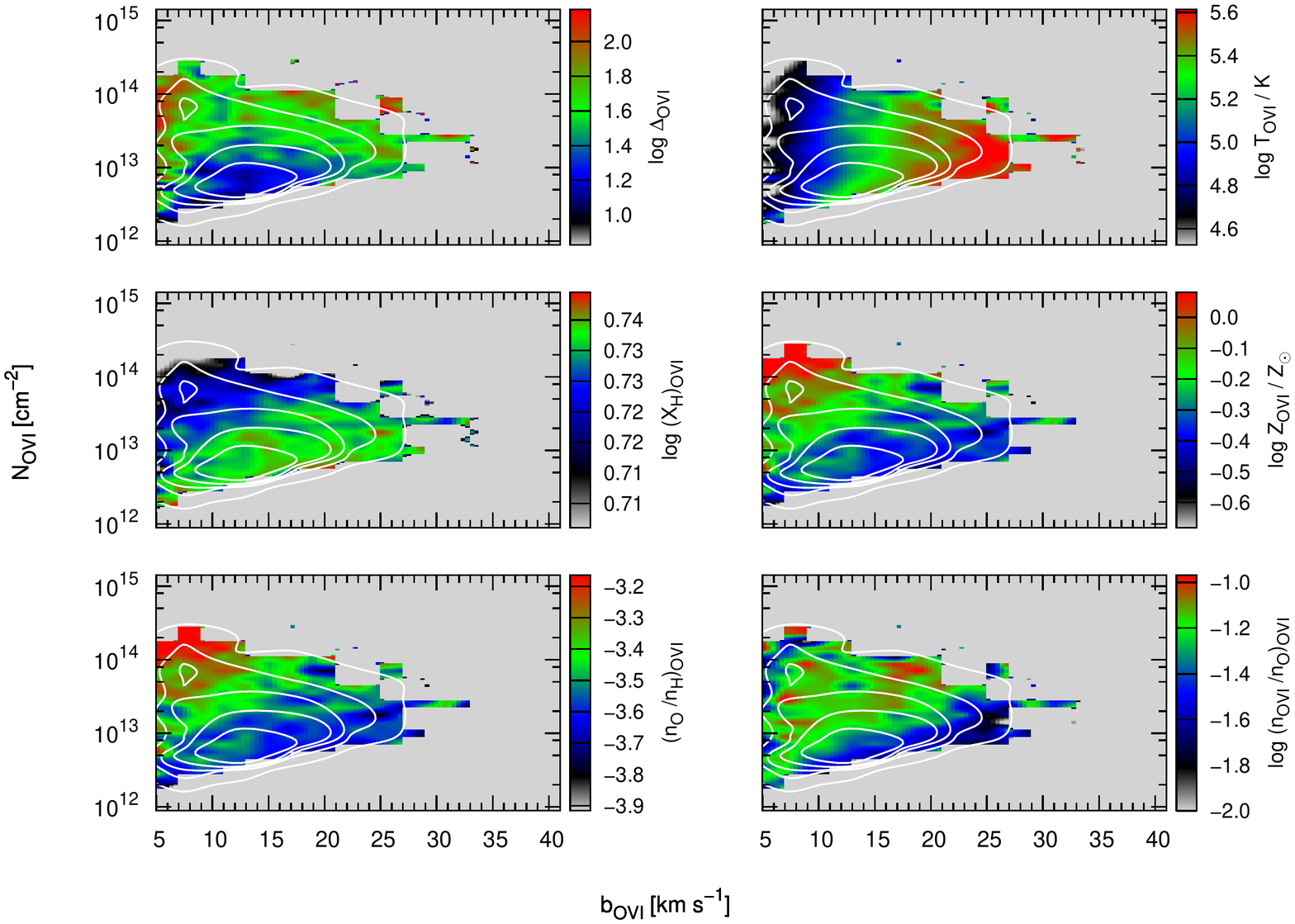}}
\caption[ ]{Physical conditions of the \OVI\ bearing gas as a function of \OVI\ observables, \ie\  column density and Doppler parameter. The colour scale shows, for each $(\bovi,\NOVI)$-cell of size $\Delta \bovi = 2.0 ~\kms$ and $\Delta \log \NOVI = 0.2 ~{\rm dex}$, the median overdensity (top-left), temperature (top-right), hydrogen mass fraction (middle-left), metallicity (middle-right), oxygen abundance (bottom-left), and \OVI\ ionisation fraction (bottom-right). For reference, we have included contours showing the distribution (by number) of \OVI\ absorbers on the $\bovi-\NOVI$ plane. These contours contain, starting from the innermost, 25, 50, 75, 90, and 99 per cent of the identified \OVI\ absorbers and they are the same in all panels.}
\label{fig:phc}
\end{figure*}
%--------------------------------------------------------------------------------------------------------------------------------------------------------------------------------

%--------------------------------------------------------------------------------------------------------------------------------------------------------------------------------
\subsection{Physical conditions of the gas and \OVI\ observables} \label{sec:phc}

Now that we have analysed the physical conditions (density, temperature, etc.) of the gas giving rise to the \OVI\ absorption in our simulation, we want to make a connection between these and the actual \OVI\ observables, \ie\  the \OVI\ column density, $\NOVI$, and the \OVI\ line width as measured by the Doppler parameter, $\bovi$.

For this purpose we bin the set of $(\bovi, ~\NOVI)$ values for our \OVI\ sample 2 using \mbox{$\Delta \bovi = 2.0 ~\kms$} and \mbox{$\Delta \log \NOVI = 0.2 ~{\rm dex}$}, and compute in each cell the median of the physical quantity under consideration, \eg\ density. The result of this exercise is shown in Fig.~\ref{fig:phc}. The panels show the median overdensity (top-left), temperature (top-right), hydrogen mass fraction (middle-left), metallicity (middle-right), oxygen abundance (bottom-left), and \OVI\ ionisation fraction (bottom-right) as a function of $\bovi$ and $\NOVI$. For reference, we have included contours showing the distribution (by number) of \OVI\ absorbers in the $\bovi-\NOVI$ plane. These contours contain, starting from the innermost, 25, 50, 75, 90, and 99 per cent of the identified \OVI\ absorbers. 

Looking at the top-left panel, we can see that there is a mild correlation between column density and overdensity, with no apparent correlation between the line width and overdensity, which is consistent with our missing $\bovi - \NOVI$ correlation (see Fig.~\ref{fig:obs}, bottom-right panel). In contrast, there is a strong correlation between Doppler parameter and the optical-depth weighted gas temperature, as  shown by the top-right panel. Furthermore, the middle-right panel shows a good correlation between column density and gas metallicity, \ie\ a metallicity-density relationship. This is consistent with the results shown in the middle-left and bottom-left panels, where we can see that the hydrogen mass fraction of the gas giving rise to \OVI\ absorption decreases with $\NOVI$, and correspondingly, the oxygen abundance increases. Finally, the bottom-right panel shows that there is no clear trend of the ionisation state of the \OVI\ bearing gas with $\bovi$ nor with $\NOVI$.

All these results together imply that {\em some} information about the physical conditions can be gained from the measured \OVI\ column densities and line widths alone, but only in a statistical sense, given the large scatter in the correlations between directly observable and derived physical quantities. In particular, the bottom panels of Fig.~\ref{fig:phc} reinforce our previous statement that assuming a fixed value for the oxygen abundance and \OVI\ ionisation fraction to estimate the baryon content of \OVI\ bearing gas is dangerous.

%--------------------------------------------------------------------------------------------------------------------------------------------------------------------------------
\subsubsection{Gas temperature, Doppler widths, and line strengths} \label{sec:tvsb}

One of the main purposes of recent observational studies about intergalactic \OVI\ absorption at low redshift has been to determine the temperature, and thus the ionisation state, of the gas giving rise to the observed absorption. Some of these \citep[\eg\ ][]{tho08a,tri08b} have estimated the gas temperature using the line width of  well-aligned, \ie,  within some velocity uncertainty $\sigma(\Delta v)$, \OVI\ - \HI{} absorbers, and have found that the implied gas temperature are typically $T < 10^{5}~\K$, indicating that the associated \OVI\ is mainly photo-ionised. This approach, however, implicitly assumes that \OVI\ and \HI{} absorbers arise in the same gas, which might not be true in general, even for small velocity displacements. As also noted by \citet[][]{tho08a} and \citet[][]{tri08b}, the gas is most probable multi-phase, and a broader \HI{} component related to the \OVI\ absorption might be too weak to be detected, particularly if the metallicity of the component is high. Such a broad component would relax the upper limit on the temperature imposed by the \HI{} line width, leading to higher gas temperatures. Indeed, \citet[][]{tri08b} have  estimated an upper limit on the temperature of the \OVI\ bearing gas using the \OVI\ line widths alone and assuming pure thermal broadening, and they find significantly higher temperatures than those allowed by well-aligned OVI\ - \HI{} absorbers. 

\citet[][]{dan08b} find a strong correlation between the column densities of \OVI\ and \NV{} absorbers, and use this to model the ionisation state of the absorbing gas, concluding that the observed column density ratios are consistent with collisionally ionised gas at $T = 10^{5.3\pm0.1}~\K$, assuming a solar $(\rm N/O)$ abundance. \citet[][]{dan08b} note that although the observed $\NOVI/\NNV$ ratios are also consistent with pure photo-ionisation models, the implied ionisation parameters, the required metallicities, and/or the spectral hardness of the photo-ionising radiation are not compatible with other constraints.

\begin{figure}
\resizebox{1.\colwidth}{!}{\includegraphics{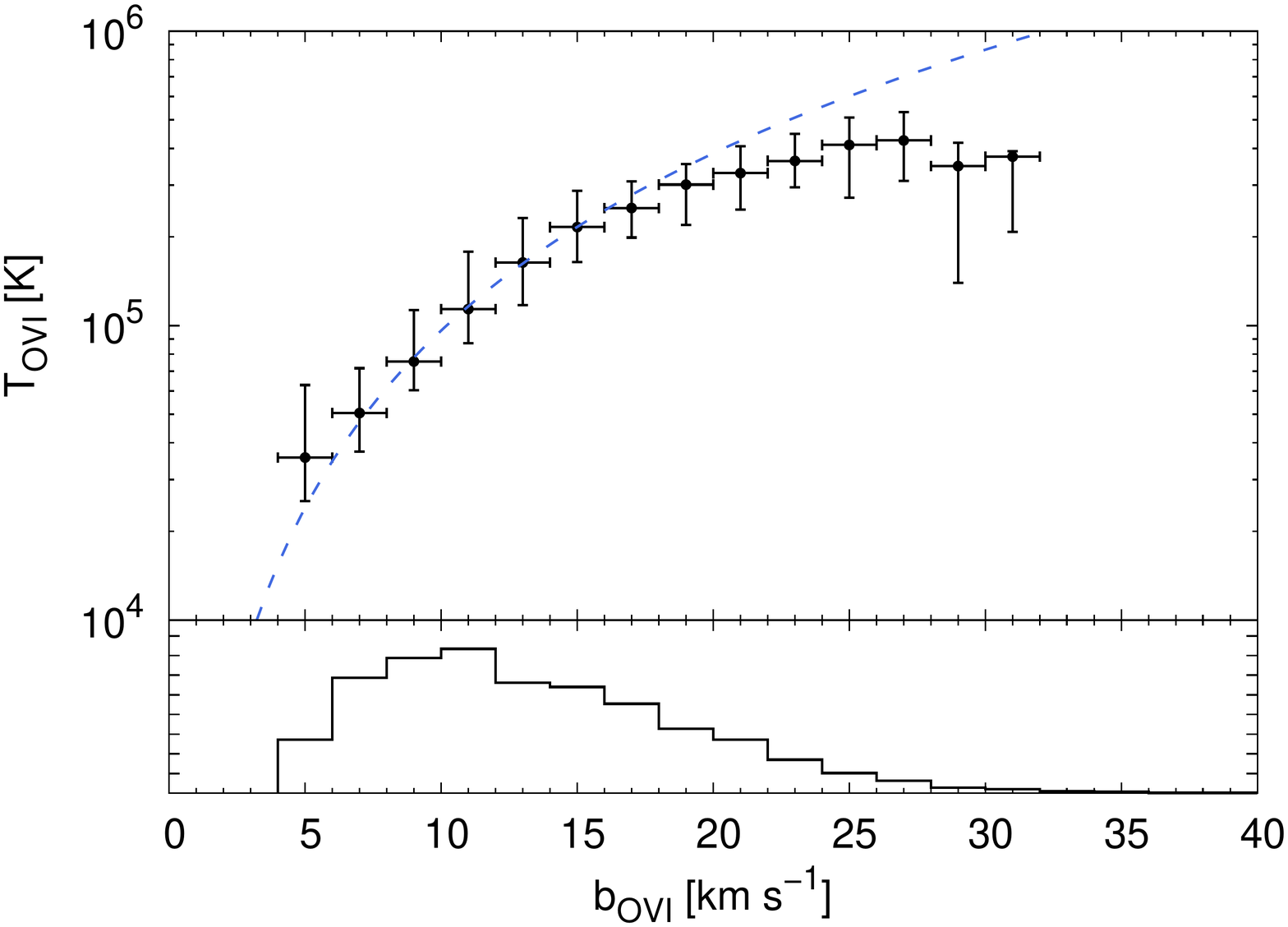}}
\caption[]{Comparison between fitted Doppler parameters and \OVI\ gas temperatures for all components (3034) in sample 2. The filled dots show the median value in each bin of size $\Delta \bovi = 2 \kms$ (indicated by the $x$-bars), while the $y$-bars show the 25th and 75th percentiles, respectively. The blue dashed curve shows the relation between temperature and Doppler width assuming pure thermal broadening (see text for details). The (normalised) histogram in the sub-panel shows the distribution of $\bovi$ values.}
\label{fig:tvsb}
\end{figure}

Given the current disagreement about the temperature, and hence the ionisation state, of the gas traced by the observed \OVI, we inspect the correlation between \OVI\ line width and gas temperature suggested by the top-right panel of Fig.~\ref{fig:phc} in more detail. For this purpose, we bin the $\tempw$ values using bins of size $\Delta \bovi = 2 \kms$ for all identified \OVI\ absorbers with $\bovi < 40 \, \kms$ in sample 2, and compute the median, and 25th and 75th percentiles in each bin. The result is shown in Fig.~\ref{fig:tvsb}. The relation between gas temperature and Doppler parameter assuming pure thermal broadening,
\beq \label{eq:bth}
	\tempw = \frac{m_{\rm O}~\bovi^{\,2}}{2 k} \approx 9.7\times10^4 \,\K \left (\frac{\bovi}{10~\kms}\right )^2,
\eeq
has been included in this plot for reference (blue dashed curve).

%--------------------------------------------------------------------------------------------------------------------------------------------------------------------------------
% FIGURE: Temperature vs. line width
%--------------------------------------------------------------------------------------------------------------------------------------------------------------------------------

Clearly, there is a tendency for the the optical-depth weighted gas temperature to increase with $\bovi$. For \mbox{$\bovi > 15 \kms$} (corresponding to $\log (T/{\rm K}) > 5.3$), non-thermal broadening becomes progressively more important, leading to larger $b$-values for a given temperature than expected for pure thermal broadening. Note that for the bin just above the resolution limit in our synthetic spectra, $b_{\rm min} = 4.2 \, \kms$, the median temperature is larger than expected for pure thermal broadening, as is the case for the 75th percentiles at higher $b$-values, which is unphysical. As discussed in Sec.~\ref{sec:dop}, this is an artefact of our fitting procedure in which weak, narrow components are added to broad, shallow absorption features to improve the fit. Some of these broader components arise in gas with relatively high temperatures and, correspondingly, high optical-depth weighted temperatures are assigned to the related (narrow) components.

From the previous discussion we can conclude that line widths do reflect the temperature of the absorbing gas in our simulation, although in a statistical sense, but that no definite conclusion can be reached for individual absorbers, unless further information, \eg\ other related absorbing ions, are available. Also, it is plausible that the correlation between gas temperature and line widths might be washed out by line-broadening mechanisms such as turbulence, which are not resolved by our simulation, or by a too low instrumental resolution.

Nevertheless, assuming we would estimate an upper limit on the gas temperature from the \OVI\ line widths alone, \eg\  by inverting the relation shown in Fig.~\ref{fig:tvsb}, we would find a good agreement with the upper limits estimated by \citet[][their Table~7]{tri08b} from their \OVI\ line widths, which are, however, much higher than the temperatures implied by their sample of well-aligned \OVI\ - \HI{} absorbers. This, in turn, suggests that gas temperatures derived from $\bhi / \bovi$ ratios alone might be too low. Furthermore, this supports the idea that there are broad \HI{} components which truly are associated with the \OVI\ components, but which are too weak to be detected, as anticipated by \citet[][]{tri08b}.

In addition, \OVI\ detections might be biased towards lower temperatures. This is plausible since absorption strength, as measured by the optical depth at the line centre\footnote{The central optical depth of an absorption line with column density $N$ and Doppler parameter $b$ is given by
\beq
	\tau_{\,0} = \frac{\sqrt{\pi} \, e^{\,2}}{m_{\,e} \, c} \lambda f \frac{N}{b} \approx 2.048~\left(\frac{N/10^{14}\cm^{-2}}{b/10\kms} \right), \notag
\eeq
where $e$ and $m_{\,e}$ are the electron's charge and mass, $\lambda$ and $f$ are the transition's rest wavelength and oscillator strength, respectively, and the last numerical factor is valid for the \OVI\ \OVIstrong\ transition.
}
, $\tau_{\,0}$, is proportional to the ratio of the column density to the Doppler width, $N / b$. Hence, at a fixed column density, line strength is inversely proportional to the Doppler parameter and thus, assuming pure thermal broadening, inversely proportional to the square-root of the gas temperature. Also, at a fixed $b$-value, line strength is directly proportional to column density, and given the correlation between column density and metallicity presented in Sec.~\ref{sec:mmf}, also directly proportional to the gas metallicity. Since metallicity enhances cooling, it is thus natural to expect that stronger \OVI\ absorption traces gas at lower temperatures.

\begin{figure}
\resizebox{\colwidth}{!}{\includegraphics{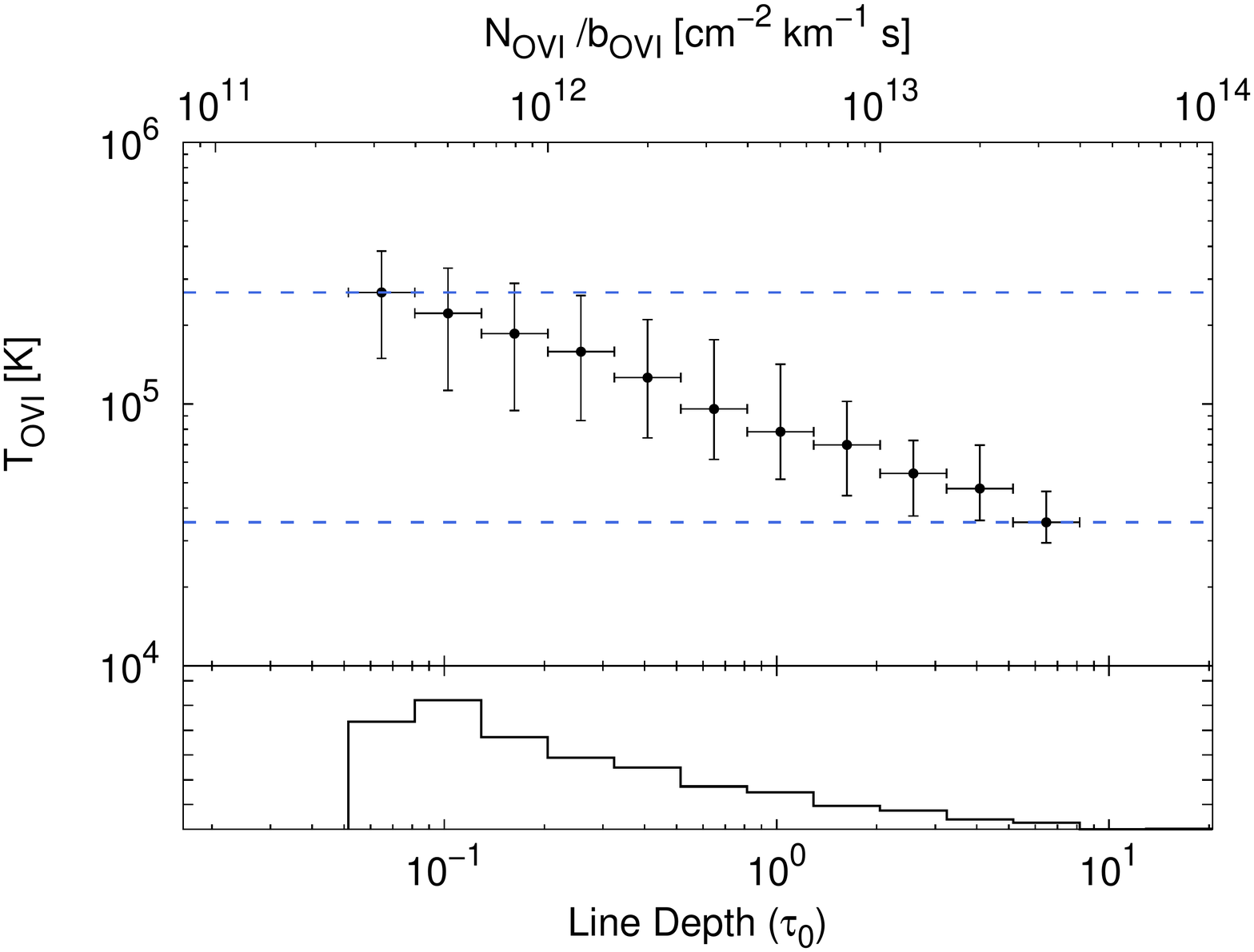}}
\caption[]{Optical depth-weighted gas temperature of the \OVI\ absorbing gas as a function of central optical depth $\tau_{\,0}$ (or line strength $\NOVI/\bovi$) for all components identified in 5000  synthetic spectra with ${\rm S/N} = 50$ at $z = 0.25$. The blue dashed lines indicate the median gas temperature at the smallest and largest line depths. The (normalised) histogram in the lower panel shows the distribution of \OVI\ components contributing to each bin.}
\label{fig:tvsbovern}
\end{figure}

In order to test this, we bin the distribution of gas temperatures $\tempw$ as as function of the line depth (central optical depth) for our sample 2, and compute the median temperature and the 25th and 75th percentiles in each bin. The result is shown in Fig.~\ref{fig:tvsbovern}. As we anticipated, we see a clear anti-correlation between the median gas temperature and absorption strength. We note that \citet[][]{tri08b} estimated their detection threshold to be $\tau_{0} \approx 0.1$. We can see from the histogram in the sub-panel -- which shows the distribution of lines as a function of line strength -- that there is a non-negligible  number of weak \OVI\ absorption features arising in warm-hot gas, which would not have been detected by \citet{tri08b}. Qualitatively, this conclusion is neither affected by the assumed ${\rm S/N}$ (the same behaviour is found for spectra with ${\rm S/N} = 10$), nor by the fact that in our synthetic spectra there is no apparent correlation between $b$ and $\NOVI$. Incidentally, the latter facilitates the interpretation of the anti-correlation between gas temperature and line strength shown in Fig.~\ref{fig:tvsbovern} as a signature of enhanced cooling in enriched gas, which is discussed in the next section. The bottom line of this exercise is that observations, even if the gas temperature is measured accurately, may be biased towards absorption systems tracing gas at somewhat lower temperatures than expected for collisionally ionised gas, due to the limited ${\rm S/N}$ of the data.

%--------------------------------------------------------------------------------------------------------------------------------------------------------------------------------
% FIGURE: Line strength; overdensity vs. column density
%--------------------------------------------------------------------------------------------------------------------------------------------------------------------------------

%--------------------------------------------------------------------------------------------------------------------------------------------------------------------------------
\subsection{Cooling times} \label{sec:cool}

%--------------------------------------------------------------------------------------------------------------------------------------------------------------------------------
% FIGURE: Temperature-overdensity phase diagram: cooling times
\begin{figure}
\resizebox{1.\colwidth}{!}{\includegraphics{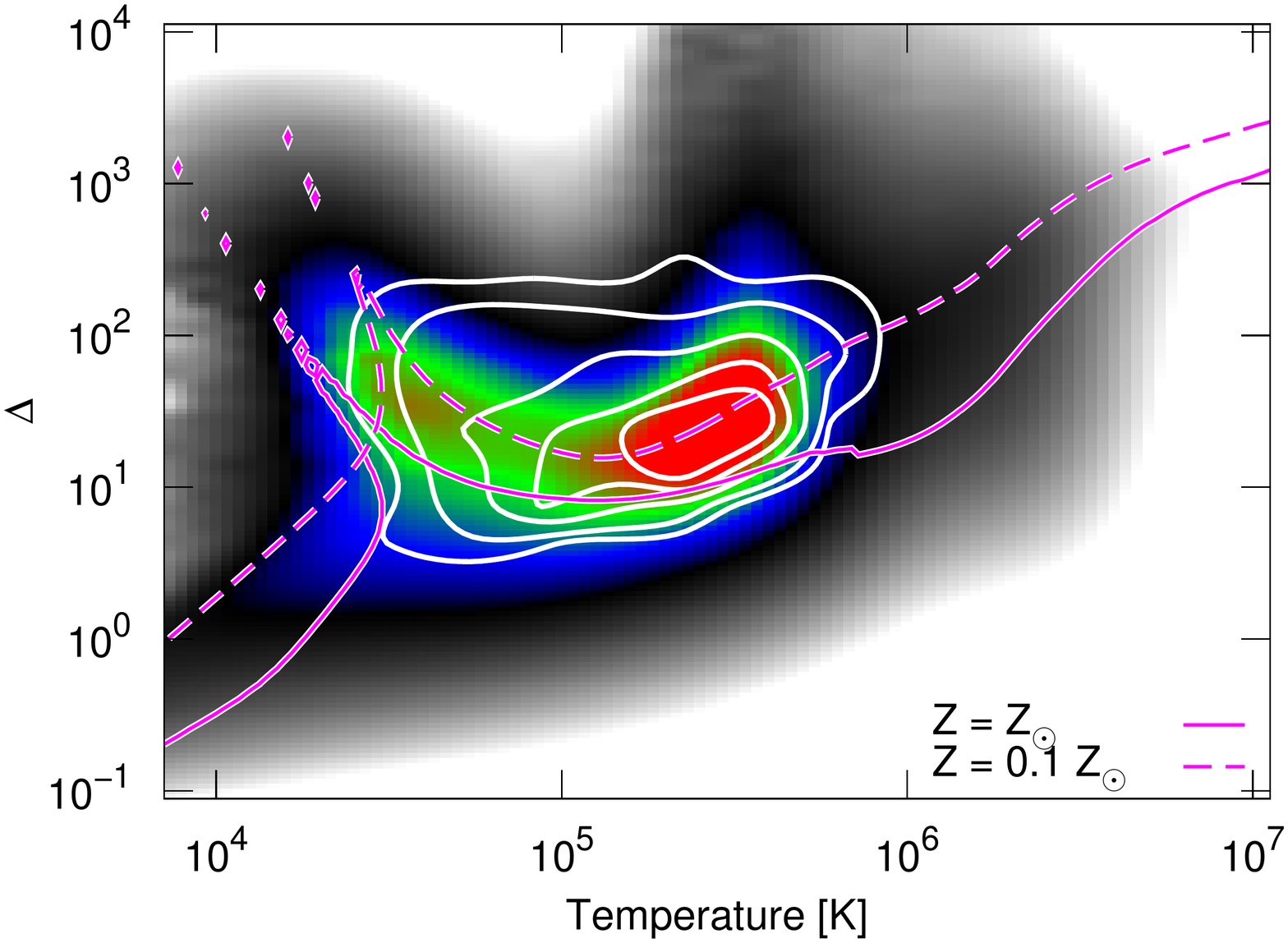}}
\caption{Distribution of \OVI\ mass (coloured regions) and distribution of \OVI\ absorbers (white contours) in the temperature-overdensity plane. Note that these distributions are identical to those shown in the bottom panel of Fig.~\ref{fig:ODvsT}. Cooling time contours (magenta lines) for two different metallicities (solar, solid; 10 per cent solar, dashed) split the plane into regions corresponding to temperatures and overdensities for which the cooling time is shorter (upper part) and longer (lower part) than a Hubble time.
}
\label{fig:cool_1}
\end{figure}
%--------------------------------------------------------------------------------------------------------------------------------------------------------------------------------

We presented in Sec.~\ref{sec:temp_od} our key result that simulated \OVI\ absorbers preferentially trace gas at temperatures in the range \mbox{$T \sim 10^{\,5} - 10^{\,6} \K$}, and thus trace gas in the low temperature regime of the WHIM. This is in marked contrast with the results of \citet[][their Fig.~14]{opp09b}, who find that the majority of \OVI\ absorbers arise in gas at temperatures \mbox{$10^{\,3.8} \lesssim T \lesssim 10^{\,4.8} \K$} with a peak at \mbox{$T \approx 10^{\,4.2 \pm 0.2} \K$}. The purpose of this section is to address the source of this discrepancy.

\citet[][]{opp09b} find that in their simulations \OVI\ traces over-enriched (by factors of four to six) regions with a clumpy metallicity distribution, and they argue that these regions are thus subject to enhanced metal-line cooling, such that \OVI\ bearing gas which was initially shock-heated to temperatures $T > 10^{\,6} \, \K$ is able to cool to photo-ionised temperatures well within a Hubble time. Our results are consistent with \citet[][]{opp09b} inasmuch as we also find that \OVI\ traces in-homogeneously enriched gas with relatively high metallicities, but at significantly higher temperatures.  In other words, our results indicate that, in our simulation, the vast majority of the gas traced by \OVI\ that was shock-heated to temperatures $T > 10^5 \, \K$ has not yet cooled down to photo-ionisation temperatures, {\em in spite of} its enhanced metallicity.

%--------------------------------------------------------------------------------------------------------------------------------------------------------------------------------
% FIGURE: Temperature-overdensity phase diagram: cooling rates
\begin{figure*}
\resizebox{1.\colwidth}{!}{\includegraphics{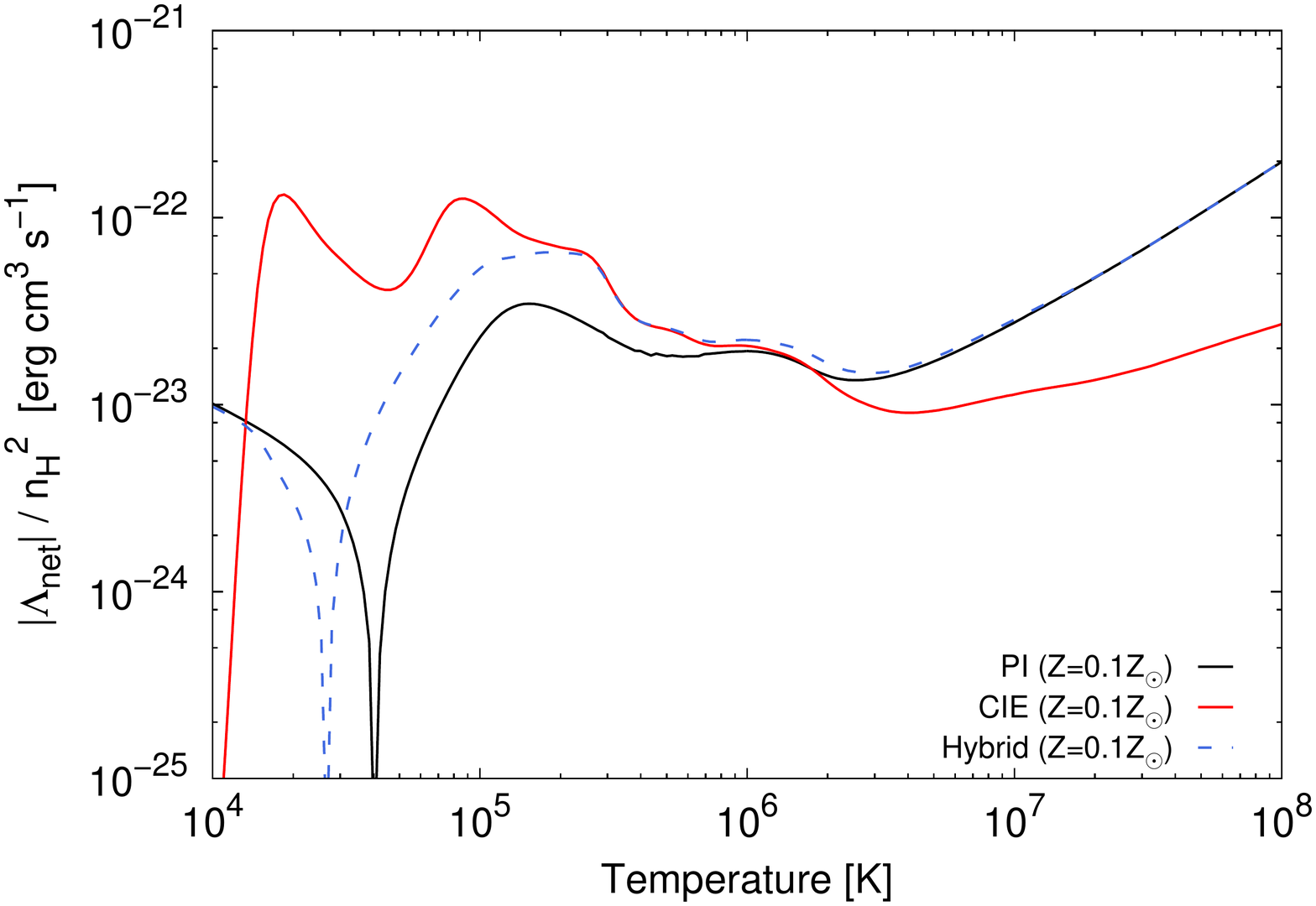}}
\resizebox{1.\colwidth}{!}{\includegraphics{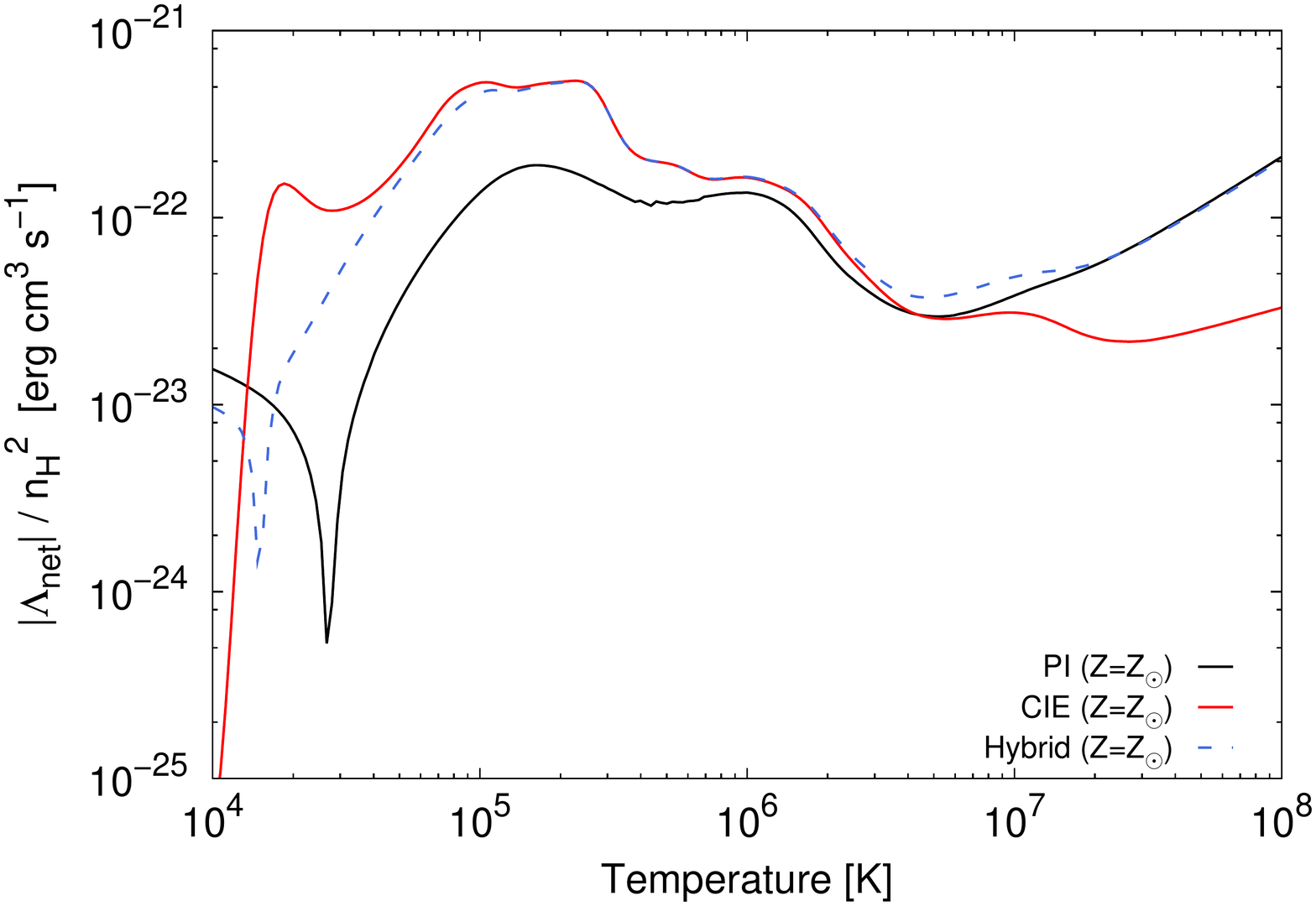}}
\caption{Normalised, absolute net cooling rates at $z = 0.25$ for gas with ${\rm n_{H}} = 10^{\,-5} \cm^{\,-3}$, for gas with metallicity of 10 per cent solar (left panel) and solar (right panel), computed assuming collisional ionisation equilibrium (CIE; red), including photoionisation \citep[][PI; black]{wie09a}, and including photoionisation only for gas of primordial composition while assuming CIE for metal-line cooling (Hybrid; blue dashed). See text for details.
}
\label{fig:cool_2}
\end{figure*}
%--------------------------------------------------------------------------------------------------------------------------------------------------------------------------------

This fact is demonstrated in Fig.~\ref{fig:cool_1}, where we reproduce the bottom panel of Fig.~\ref{fig:ODvsT} showing the \OVI\ mass distribution (colour shading) and the distribution of \OVI\ absorbers (white contours) in the $T - \Delta$ phase diagram, together with the locus of $(T,~\Delta)$-values for which the cooling time equals the Hubble time for gas of solar (solid) and ten per cent solar (dashed) metallicity. Note that the metallicity values used roughly span the range of gas metallicities traced by our \OVI\ absorbers (see Fig.~\ref{fig:mmf}). Here, the cooling time is defined as
\beq \label{eq:cool}
	t_{\rm cool} \equiv \frac{T}{\diff T / dt} = \frac{3}{2} \frac{n k T}{\left| \Lambda_{\rm net}\right|},
\eeq
where $\Lambda_{\rm net}$ is the {\em net} (\ie, cooling minus heating) radiative cooling rate at a fixed density. Note that the contour segments to the left from $T \approx 10^{4.5}~\K$ correspond actually to {\em heating} times. The point where heating and cooling time contours merge defines the equilibrium temperature, \ie\  the temperature at which the {\em net} cooling rate vanishes and the cooling time becomes effectively infinite (see also Fig.~\ref{fig:cool_2}). From Fig.~\ref{fig:cool_1}, it becomes apparent that, while virtually all \OVI\ absorbers in our simulation at $z = 0.25$  will cool down to the equilibrium temperature in less than a Hubble time\footnote{For reference, the Hubble time $\tau_{\rm Hubble} \equiv t_{\rm cosmic}(z=0) = 13.8~\Gyr$ and $t_{\rm cosmic}(z=0.25) = 10.9~\Gyr$ for our adopted cosmology.} (in the absence of further shock-heating), most will still be hot by $z = 0$. Note that, as anticipated in Sec.~\ref{sec:temp_od}, the low-temperature \OVI\ bearing gas has nearly reached the equilibrium temperature \mbox{$T \sim 10^{4.5} \K$},  typical for photo-ionised \OVI.

The above result strongly suggests that an important factor leading to the notable difference between the results of \citet[][]{opp09b} and our findings regarding the mean temperature of gas traced by \OVI\ is the way the radiative cooling of gas is treated in the simulations. As described in \citet[][]{opp09b}, metal-line cooling is included in their simulations using the models by \citet[][]{sut93a}, which assume collisional ionisation equilibrium and fixed relative abundances. They only included photo-ionisation for hydrogen and helium. In contrast, as described in detail in \citet[][]{sch10a}, the OWLS runs include radiative cooling implemented according to the method by \citet[][]{wie09a}, who compute cooling rates on an element-by-element basis in the presence of the cosmic microwave background (CMB) and an ionising (\ie, UV/X-Ray) radiation field as modelled by \citet[][]{haa96a}. \citet[][]{wie09a} showed that including photo-ionisation by the meta-galactic UV/X-Ray background not only drastically decreases the cooling rates for hydrogen and helium, but also for heavy elements.  Moreover, they showed that photo-ionisation of heavy elements shifts the equilibrium temperatures to higher values. In other words, photo-ionisation can drastically affect the radiative cooling rates of enriched gas and calculations neglecting photo-ionisation will thus underestimate the temperature of the gas.

A comparison of the cooling rates computed according to each method is shown in Fig.~\ref{fig:cool_2}. In each case, the gas is exposed to the CMB and the UV/X-Ray background at $z = 0.25$, and has a hydrogen density \mbox{$\nH = 10^{-5} \, \cm^{-3}$}  (\mbox{$\Delta \approx 50$}), a hydrogen mass fraction $X_{\H} = 0.752$ -- which roughly correspond to the median values of $\nhoviw$ and $\xhoviw$ in sample 2 -- and a metallicity of 10 per cent solar (left panel) and solar (right panel). The black curve in either panel shows the net cooling rate including photo-ionisation of heavy elements (PI); the red curve, the net cooling rate assuming pure collisional ionisation equilibrium (CIE) and fixed (solar) abundances, while the blue curve shows the net cooling rate assuming CIE, but including photo-ionisation for hydrogen and helium (Hybrid). The dip apparent in each cooling curve shows the equilibrium temperature. Note, in particular, the difference in equilibrium temperatures and overall amplitude between the black and blue curves, which correspond to the cooling rates implemented in the OWLS runs and the approach adopted by \eg, \citet[][]{opp09b}, respectively. It is clear that simulations using the latter prescription overestimate the cooling rates. The difference in the overall amplitude and in the equilibrium temperature is more severe at lower densities -- over an order of magnitude at $\nH = 10^{\,-6} \cm^{\,-3}$ -- and higher metallicities (see right panel).

The reduction in the cooling rates and the corresponding increase in the cooling times, together with the shift of the equilibrium temperature to higher values, may explain the difference between the results of \citet[][]{opp09b} and ours regarding the temperature of gas traced by simulated \OVI\ absorbers. This difference is even more striking considering the fact that \citet[][]{opp09b} use particle abundances while we use smoothed metallicities throughout the simulation to compute the metal cooling rates. As previously mentioned and discussed in detail by \citet[][]{wie09b}, the use of smoothed abundance results in increased cooling for most of the enriched gas, although the cooling rates are reduced for gas particles that have metal mass fractions higher than those of their neighbours. One would thus naively expect to find exactly the opposite result, namely that {\em our} simulated \OVI\ absorbers trace gas at lower temperatures. That this is not the case indicates that the reduction of the radiative cooling rates due to photo-ionisation has a large impact on the thermal state of the gas.

One should keep in mind that our fiducial run is not directly comparable to the set of simulations by \citet[][]{opp09b}, since they differ in various aspects, in particular in the assumed cosmology (WMAP-5 vs. our assumed WMAP-3), and in the sub-grid prescriptions to model, \eg, SN feedback in the form of galactic winds. Also, recall that we use different methods to ascribe physical properties to absorption features identified in synthetic spectra. However, we have carried out runs in which we have varied the cosmology, the prescriptions for SN feedback, and even have implemented the method by \citet[][]{opp09b} to estimate the gas temperature and density related to \OVI\ absorption features, and our results remain practically unaltered. We will report on such variations to our reference model in a future paper where we analyse the physical conditions of \OVI\ absorbers in several of the OWLS runs, and we will show that the fact that \OVI\ traces gas at temperatures \mbox{$T \sim 10^{\,5} - 10^{\,6} \K$} and overdensities \mbox{$1 \ll \Delta < 10^{3}$} is a robust prediction of these simulations. This demonstrates the capital importance of a correct treatment of radiative cooling in simulations that attempt to shed light on the much debated physical state of absorbing gas giving rise to absorption by highly ionised species, currently observed in QSO/AGN spectra. We should therefore keep in mind that, although our implementation of radiative cooling is more advanced than the ones used before in that we include photo-ionisation of heavy elements by the UV background and in that we do not assume fixed relative abundances of different elements, we are still assuming ionisation equilibrium and we are still neglecting local sources of ionising radiation.

%--------------------------------------------------------------------------------------------------------------------------------------------------------------------------------
\section{Conclusions and discussion} \label{sec:dis}

It is now well established that the inventory of observed baryons in the low-redshift Universe falls short of the baryonic mass measured at high redshift and expected from big bang nucleosynthesis calculations. However, numerical simulations of structure formation predict that at low redshift a significant amount of baryons are hidden in a hard-to-observe, diffuse, hot gas phase termed the Warm-Hot Intergalactic Medium (WHIM), thus providing a potential solution for the mismatch between observed and expected baryons. As a consequence, a large observational campaign has been launched with the primary goal of detecting this gas phase and estimating its baryon content, mainly through the observation of \OVI\ seen in absorption in the spectra of QSO and AGN.

Although statistically significant samples of low-redshift \OVI\ absorption systems have been obtained by way of UV absorption spectroscopy, the nature and the physical state of these absorbers remains controversial. Due to the complexity of the absorbing gas in terms of its density, temperature, and ionisation state, no secure estimate of the amount of baryons contained in this gas has been achieved. Nor is it clear in which gaseous phase these absorbers arise and whether they are predominantly collisionally or photo-ionised. Observations thus need to be supported by theoretical work and, given the intricate interplay between the various physical processes governing the formation and evolution of metal absorption systems, the use of numerical simulations has become a powerful tool to complement (and/or reinforce) the results inferred from observational data.

In this paper, we have made use of a set of cosmological hydrodynamical simulations of the `reference model' from the OverWhelmingly Large Simulations (OWLS) project \citep{sch10a} to study the physical properties and baryon content of intervening \OVI\ absorbers at low redshift. The main advantage of these simulations compared to previous ones is the implementation of important physical processes that have been largely ignored in earlier studies. The delayed release of mass by massive stars, intermediate mass stars, and supernovae of types Ia and II is followed for all the 11 elements that contribute significantly to the radiative cooling rates. Radiative cooling is computed element-by-element and the effect of photo-ionisation by the UV/X-ray background is taken into account not only for hydrogen and helium, but also for the heavy elements. \citet[][]{wie09a} showed that the suppression of cooling by photo-ionisation of heavy elements, which was ignored in previous studies, is important for the WHIM and we confirmed that it has large consequences for the gas that is responsible for the \OVI\ absorption. 

We have generated a large number of synthetic spectra that closely match the properties (S/N, resolution, etc.) of observed spectra. We have identified and fitted individual \OVI\ absorption components, and have measured their column density, Doppler parameter, and rest equivalent width, and their corresponding distributions. We have shown that we can reliably identify lines with \OVI\ column densities above a given threshold that increases with decreasing signal-to-noise ratio, even though we still miss a fraction of the (weak) absorption lines present in the spectra. Moreover, we showed that we are able to measure rest equivalent widths quite accurately, and that the fitted Doppler parameters are reliable for moderate signal-to-noise ratios. Finally, we showed that the line parameter distributions are converged with respect to box size and resolution. 

Our simulations reproduce the shape of the observed column density distribution function, although with a lower (by a factor $\sim2$) amplitude. Similarly, the predicted cumulative equivalent width distribution has a lower amplitude, a lack of strong lines, and an excess of low equivalent width lines. Furthermore, our simulations do not show the positive correlation between $\bovi$ and $\NOVI$ that appears to be observed for \OVI\ absorbing systems in a variety of environments, although the observations are uncertain and even contradictory in this respect. We attribute the lack of strong (\ie, high equivalent width) lines in our simulation to a lack of lines with high \OVI\ column density {\em and} large Doppler parameter values. 

A similar disagreement between the predicted and the observed \OVI\ cumulative line-number density and $\bovi-\NOVI$ correlation has been reported by \citet[][]{opp09b}, who argued that small-scale turbulence is the crucial mechanism leading to the large $b$-values measured in IGM \OVI\ systems, and solved the discrepancy between their simulations and observations by adding sub-resolution turbulent broadening to the Doppler parameters of their simulated absorption lines. Exploring the effects of such a post-run variation on the physical properties of the absorbing gas is beyond the scope of the present work, but we will do so in a future paper. 

We stated that even with simulations it is not trivial to associate physical properties, such as a density, temperature, chemical composition and ionisation balance, to a particular absorber because gas parcels with a range of properties will generally contribute to a single absorption line. Therefore, we chose to assign physical properties to each pixel in velocity space by weighting all gas parcels by their contributions to the \OVI\ optical depth in that pixel. The resulting optical-depth weighted quantities were then in turn averaged over the line profile, weighted by the optical depth of each pixel. The use of optical-depth weighted quantities allows the direct association of an absorption feature with the gas where the absorption takes place, even in the presence of redshift-space distortions resulting from peculiar velocity gradients and thermal broadening.  

In terms of the physical state and gas (metal) mass content of the \OVI\ bearing gas, we have found that:

\ben
\item Detectable \OVI\ absorbers trace gas at temperatures \mbox{$10^{4.5} \K \lesssim T < 10^{6} \K$} with a peak at \mbox{$T \approx10^{5.3} \K$}, and overdensities \mbox{$1 \ll \Delta \lesssim 10^{2}$}.

\item The bulk of the \OVI\ mass in our simulation is found in gas at moderate overdensities $\Delta \sim 10^{0.5} - 10^{2.5}$, and 
temperatures between $T = 10^{4} \K$ and $10^{6} \K$. While approximately 30 per cent of the \OVI\  seen in absorption arises in gas at $T < 10^{5} ~\K$, the rest is found to trace a the hotter gaseous phase at temperatures around $T \approx 10^{5.3} \K$.

\item The metallicity distribution of the gas traced by \OVI\ is narrow, extending over the range  \mbox{$(0.1, \, 1) ~Z_{\odot}$}, 
with a peak at \mbox{$\metaloviw \approx 0.6 ~Z_{\odot}$}, \ie, almost an order of magnitude higher than the median metallicity of gas at comparable overdensities and temperatures and also about an order of magnitude higher than would be inferred from the column density ratio of \OVI\ to \HI\ for aligned \OVI-\HI\ pairs. 

\item The \OVI\ bearing gas contains a baryon fraction \mbox{$\Omega_{\rm b}(\OVI)/\Omega_{\rm b} \approx 0.014$}, and contains less than one per cent of the total metal mass in our simulation.
\een

The low baryon fraction associated with the \OVI\ absorbers is mainly a consequence of the fact that the metal distribution is patchy. The \OVI\ absorption arises mostly in high-metallicity regions that are small compared with the scales on which the density and the temperature fluctuate. Hence, if we were to recompute the synthetic spectra after smoothing the metal distribution, we would probably find that \OVI\ traces a larger fraction of the baryons. The small-scale inhomogeneity of the metal distribution therefore makes the question of which baryons to associate with the \OVI\ gas somewhat ill defined. This is particularly relevant because the lack of small-scale metal mixing may be partly numerical. However, there is in fact strong observational evidence that the metals in the real Universe are also poorly mixed on small scales \citep{sch07a} and that the metal distribution is indeed patchy on the scales over which the \HI\ absorption is uniform.

The common approach of comparing \OVI\ and \HI\ column densities to estimate the physical conditions in intervening absorbers from QSO observations may thus be misleading, as most of the \HI\ (and most of the gas) is not physically connected with the high-metallicity patches that give rise to the \OVI\ absorption. A possible solution to this problem is to rely only on upper limits for the absorption by \HI\ (and other species) at the velocity of \OVI, as was done for \CIV\ absorbers in \citet{sch07a}.

The physical conditions in the absorbing gas can in principle be inferred from the measured \OVI\ column densities and line widths, but only in a statistical sense, given the large scatter in the correlations between directly observable and derived physical quantities. In particular, the gas temperature can be recovered reasonably well from the \OVI\ line widths for $b$-parameters $<15~\kms$ ($\log(T/{\rm K})<5.3$), but for larger line widths non-thermal broadening becomes increasingly important. 

The conclusion that \OVI\ absorption preferentially traces gas at \mbox{$T \sim 10^{5.3\pm0.5} \K$}, typical of collisionally ionised gas, is in marked contrast to the results of \citet[][]{opp09b}, who find that the overwhelming majority of the \OVI\ absorbers in their simulations trace gas at temperatures \mbox{$T \approx 10^{\,4.2 \pm 0.2} \K$}, which are typical of photo-ionised gas. We showed that this drastic difference is most probably a consequence of the different treatment of radiative cooling of enriched gas. In particular, while \citet{opp09b} assumed collisional ionisation equilibrium for the heavy elements when computing cooling rates, we included the effect of ionisation by the UV/X-ray background radiation, which strongly suppresses the cooling rates for low-density gas.

While the mismatch between our predicted and observed line parameter distributions is certainly unsatisfying, we think that it may not undermine our conclusions about the physical properties of the \OVI\ absorbing gas. As \citet[][]{opp09b} showed, it is possible to alleviate the discrepancy with observations by using a physically motivated {\em post-run} modification, namely the addition of turbulence on scales that are unresolved by the simulation. We do not expect such a modification to significantly affect the physical properties derived for the absorbing gas, precisely because it is done in a post-run fashion. 

It should also be stressed that it would a priori be surprising if we matched the column density distributions very well, given that the nucleosynthetic yields are uncertain at the factor of a two level, even for a fixed stellar initial mass function \cite[e.g.][]{wie09b}. Some rescaling of the abundances in post-processing would thus be justified, although this would break the self-consistency of the simulation if it were to change the cooling rates significantly.

We postpone a detailed presentation and discussion of a series of modifications to our reference run (\ie, simulation runs with different parameters) as well as post-run variations to a future paper.

The results presented in this work clearly show that \OVI\ absorbers trace neither the main metal reservoirs nor the bulk of the baryons in our simulation. Instead, they trace over-enriched, shock-heated gas at moderate overdensities. Hence, our study indicates that \OVI\ - albeit collisionally ionised - is a poor tracer for the bulk of the baryons in the WHIM, as most of the baryons reside in shock-heated gas that has much lower metallicities. Instead, our simulations indicate that \OVI\ absorbers may trace cooling metal-ejecta from galaxies and can thus be used to study the circulation of metal-enriched gas in the intergalactic environment of galaxies.

%--------------------------------------------------------------------------------------------------------------------------------------------------------------------------------
\section*{Acknowledgments}
We are very grateful to Volker Springel for invaluable help with the simulations. We also want to thank Ben Oppenheimer for a careful reading of the manuscript. We also thank the referee, Todd Tripp, for useful comments and suggestions which helped improving the presentation of our results. The simulations presented here were run on Stella, the LOFAR Blue Gene/L system in Groningen and on the Cosmology Machine at the Institute for Computational Cosmology in Durham as part of the Virgo Consortium research programme. This work was sponsored by the National Computing Facilities Foundation (NCF) for the use of supercomputer facilities, with financial support from the Netherlands Organisation for Scientific Research (NWO), an NWO VIDI grant, and the \emph{Deutsche Forschungsgemeinschaft} (DFG) through Grant DFG-GZ: Ri 1124/5-1.\\

%--------------------------------------------------------------------------------------------------------------------------------------------------------------------------------
\bibliographystyle{mn2e_eprint} % style mn2e_eprint.bst, allows eprint field to be included
%\bibliographystyle{mn2e} % style mn2e.bst
%\bibliography{/Users/tepper/references/complete} %references list

\input{tepper_etal_OVI.bbl}
%--------------------------------------------------------------------------------------------------------------------------------------------------------------------------------

%--------------------------------------------------------------------------------------------------------------------------------------------------------------------------------
\label{lastpage}
%--------------------------------------------------------------------------------------------------------------------------------------------------------------------------------

\clearpage %to force LaTeX to output all floating objects processed before this line
%

%--------------------------------------------------------------------------------------------------------------------------------------------------------------------------------
% Appendix
%--------------------------------------------------------------------------------------------------------------------------------------------------------------------------------
\input{appendix}
%--------------------------------DOCUMENT END-----------------------------------
\end{document}

%% file: newcommands_mnras.tex
\newlength{\colwidth}
\newcommand{\cm}{{\rm cm}}
\newcommand{\cmsq}{{\rm cm^{-2}}}

\newcommand{\kms}{{\, \rm km}\,{\rm s}^{-1}}
\newcommand{\K}{{\, \rm K}}

\newcommand{\hkpc}{h^{-1}\,{\rm kpc}}

\newcommand{\hMpc}{h^{-1}\,{\rm Mpc}}

\newcommand{\hMsun}{{h^{-1}\,{\rm M}_\odot}}

\newcommand{\nH}{{{\rm n}_{\rm H}}}
\newcommand{\mH}{{{m}_{\rm H}}}
\newcommand{\nHs}{{{\rm n}_{\rm H}^*}}

\newcommand{\Gyr}{{\rm Gyr}}

\newcommand{\ion}[2]{\hbox{#1\,{\sc #2}}}
\newcommand{\ionsubscript}[2]{\hbox{\scriptsize #1\,{\tiny #2}}}
\renewcommand{\H}{{\rm H}}
\newcommand{\He}{{\rm He}}
\newcommand{\C}{{\rm C}}
\newcommand{\N}{{\rm N}}
\newcommand{\Ox}{{\rm O}}
\newcommand{\Ne}{{\rm Ne}}
\newcommand{\Mg}{{\rm Mg}}
\newcommand{\Si}{{\rm Si}}
\newcommand{\Fe}{{\rm Fe}}
\newcommand{\Ca}{{\rm Ca}}
\newcommand{\Su}{{\rm S}}
\newcommand{\HI}{\ion{H}{i}}

\newcommand{\CIV}{\ion{C}{iv}}

\newcommand{\NV}{\ion{N}{v}}

\newcommand{\OVI}{\ion{O}{vi}}
\newcommand{\OVII}{\ion{O}{vii}}
\newcommand{\OVIII}{\ion{O}{viii}}

\newcommand{\lya}{Ly$\alpha$}

\newcommand{\ie}{\textit{i.e.}}

\newcommand{\eg}{\textit{e.g.}}
\newcommand{\bea}{\begin{eqnarray}}
\newcommand{\eea}{\end{eqnarray}}
\newcommand{\beq}{\begin{equation}}
\newcommand{\eeq}{\end{equation}}
\newcommand{\bit}{\begin{itemize}}
\newcommand{\eit}{\end{itemize}}
\newcommand{\ben}{\begin{enumerate}}
\newcommand{\een}{\end{enumerate}}
\newcommand{\los}{sightline}
\newcommand{\loss}{sightlines}

\newcommand{\OVIstrong}{\mbox{$\lambda 1031.93$ \AA}}
\newcommand{\OVIweak}{\mbox{$\lambda 1037.62$ \AA}}

\newcommand{\HIstrong}{\mbox{$\lambda 1215.67$ \AA}}

\newcommand{\mO}{{{m}_{\rm O}}}

\newcommand{\NOVI}{{\rm N_{\ionsubscript{O}{VI}}}}

\newcommand{\NHI}{{\rm N_{\ionsubscript{H}{I}}}}

\newcommand{\NNV}{{\rm N_{\ionsubscript{N}{V}}}}

\newcommand{\deltaw}{\Delta_{\ionsubscript{O}{VI}}}

\newcommand{\tempw}{T_{\ionsubscript{O}{VI}}}

\newcommand{\metaloviw}{Z_{\ionsubscript{O}{VI}}}

\newcommand{\metalhiw}{Z_{\ionsubscript{H}{I}}}

\newcommand{\foviw}{(n_{\ionsubscript{O}{VI}}/n_{\rm O})_{\ionsubscript{O}{VI}}}

\newcommand{\fhiw}{(n_{\ionsubscript{H}{I}}/n_{\rm H})_{\ionsubscript{H}{I}}}

\newcommand{\nhoviw}{\left({\rm n_H}\right)_{\ionsubscript{O}{VI}}}

\newcommand{\bovi}{b_{\ionsubscript{O}{VI}}}

\newcommand{\bhi}{b_{\ionsubscript{H}{I}}}

\newcommand{\xhoviw}{({\rm X_ H})_{\ionsubscript{O}{VI}}}

\newcommand{\diff}{\rm d}

%% file: tables/refsim_list.tex
\begin{table*} 
\begin{center}
\caption{List of simulations --which assume the reference model of the OWLS project-- used in this study. Our fiducial run \emph{REF\_L050N512} (shown in bold) is used for the comparison to observations and our general analysis, whilst all other simulations are used in the Appendix when addressing the convergence of our results with respect to box size and resolution. The columns show$^{\rm a}$: simulation identifier$^{\rm b}$; comoving box size ($L$); number of dark matter particles ($N$; there are equally many baryonic particles); baryonic particle mass ($m_{\rm b}$); dark matter particle mass ($m_{\rm dm}$); comoving (Plummer-equivalent) gravitational softening ($\epsilon_{\rm com}$); maximum physical softening ($\epsilon_{\rm prop}$); final redshift ($z_{\rm end}$).} 
\label{tab:ref_sims}
\begin{tabular}{lrrrlrrl}
\hline
Simulation & $L$ & $N$ & $m_{\rm b}$ & $m_{\rm dm}$ & $\epsilon_{\rm com}$ & $\epsilon_{\rm prop}$ & $z_{\rm end}$\\  
& $(\hMpc)$ & & $(\hMsun)$ & $(\hMsun)$ & $(\hkpc)$ & $(\hkpc)$ & \\
\hline 
\emph{REF\_L025N128} &  25.00 & $128^3$ & $8.7 \times 10^7$ & $ 4.1 \times 10^8$ & 7.81 & 2.00 & 0 \\
\emph{REF\_L050N128} &  50.00 & $128^3$ & $6.9 \times 10^8$ & $ 3.3 \times 10^9$ & 15.62 & 4.00 & 0 \\
\emph{REF\_L050N256} &  50.00 & $256^3$ & $8.7 \times 10^7$ & $ 4.1 \times 10^8$ & 7.81 & 2.00 & 0 \\
{\bf \emph{REF\_L050N512}} &  {\bf 50.00} & $\mathbf{512^3}$ & $\mathbf{1.1 \times 10^7}$ & $\mathbf{5.1 \times 10^7}$ & {\bf 3.91} & {\bf 1.00} & {\bf 0} \\
\emph{REF\_L100N512} & 100.00 & $512^3$ & $8.7 \times 10^7$ & $ 4.1 \times 10^8$ & 7.81 & 2.00 & 0 \\
\hline
\end{tabular}
\end{center}
\begin{list}{}{}
\item[$^{\,\rm a}$] For a complete list of the OWLS runs as well as the detailed description of the corresponding model see \citet[][]{sch10a}.
\item[$^{\,\rm b}$] The name convention is {\em [model]\_LxxxNyyy}, where '{\em xxx}' and '{\em yyy}' are, respectively, the box size in comoving $\hMpc$ and cube root of the number of particles per species (dark matter or baryonic).
\end{list}
\end{table*}

%% file: appendix.tex
%--------------------------------------------------------------------------------------------------------------------------------------------------------------------------------
\appendix

%--------------------------------------------------------------------------------------------------------------------------------------------------------------------------------
\section{Absorption line parameters} \label{sec:spec_quant}

In this section we explore the effect of the signal-to-noise ratio (S/N) on the distributions of \OVI\ line parameters (column density, Doppler parameter, rest equivalent width) obtained from the analysis of 5000 synthetic spectra in the range $z = 0.0 - 0.5$ which span a total redshift path $\Delta z = 98.5$ (sample 1; see Sec.~\ref{sec:syn_spec}). We analyse each of these distributions in detail to assess the strengths and weaknesses of our spectral analysis.

%--------------------------------------------------------------------------------------------------------------------------------------------------------------------------------
\subsection{Column density distribution} \label{sec:cddf}

\begin{figure}
{\resizebox{1.\colwidth}{!}{\includegraphics{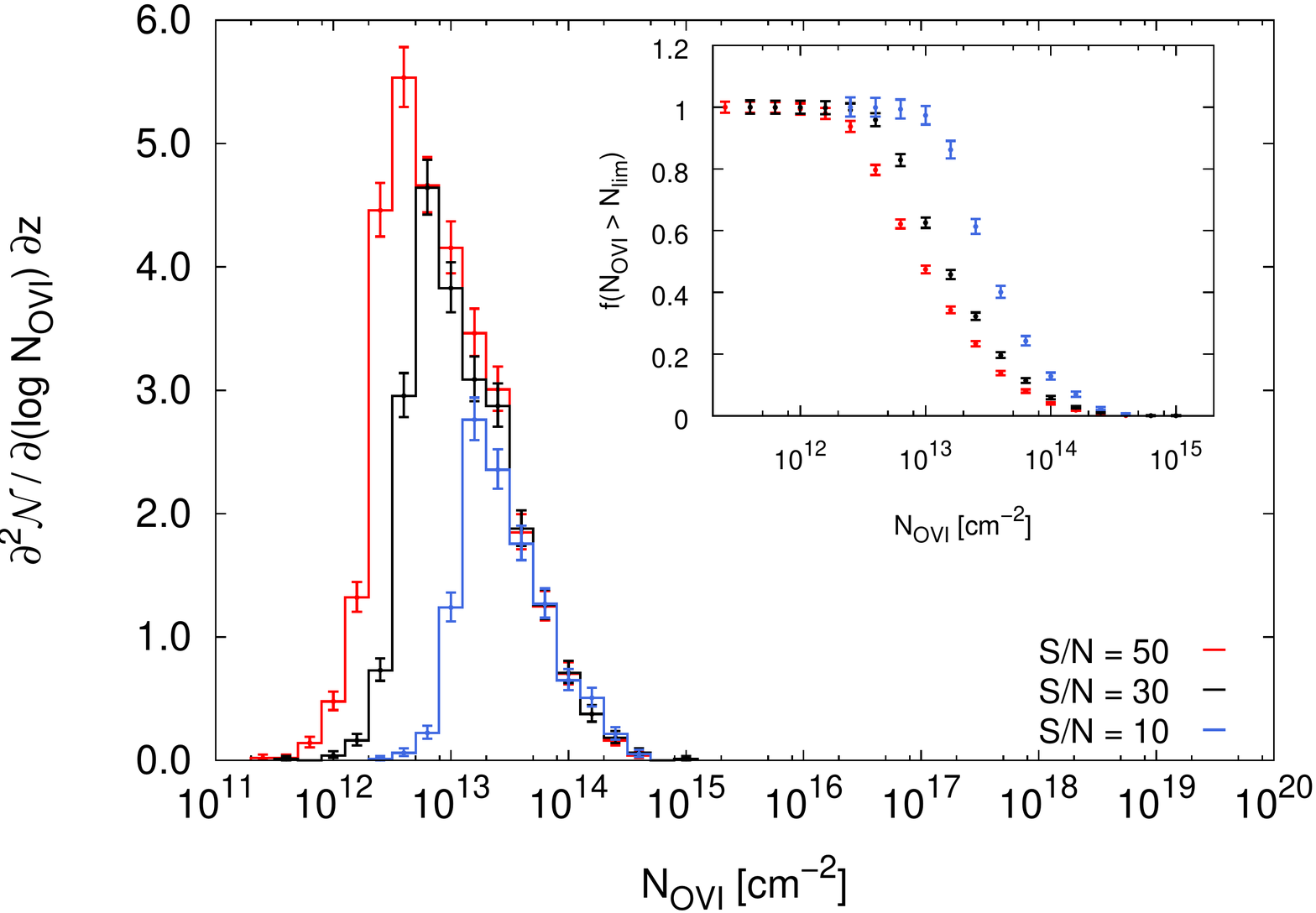}}}
{\resizebox{1.\colwidth}{!}{\includegraphics{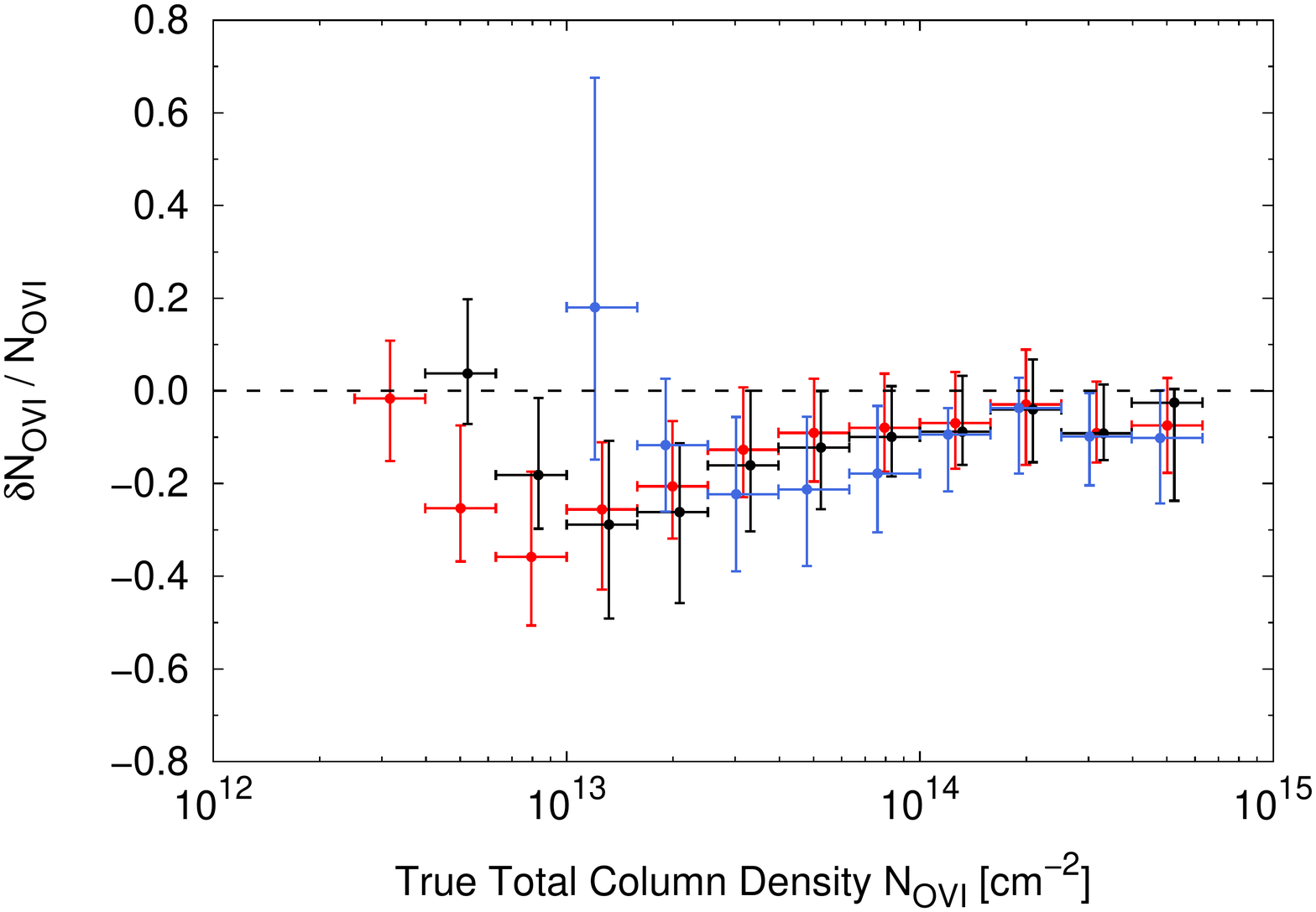}}}
\caption[]{{\em Top} Differential column-density distribution $\partial^{\,2} \mathcal{N} /(\partial \log \NOVI \, \partial z)$ (histograms) for different S/N values. The error bars show Poisson single-sided $1\sigma$ confidence limits, computed according to the tables by \citet[][]{geh86a}. The inset displays the corresponding normalised cumulative distributions, \ie, the number of lines above a given column density.
{\em Bottom:} Comparison between the integrated column density along each physical \los\ (the {\em true} column density) and the sum of the column densities of the fitted lines in the corresponding spectrum (the {\em recovered} column density). The $y$-axis displays the relative difference between true and recovered column densities, $\delta \NOVI~/~\NOVI$. We have binned the results using \mbox{$\Delta \log \NOVI = 0.2 \, {\rm dex}$} (as indicated by the $x$ error-bars); the dots and $y$ error-bars show the median value in each bin, and 25 and 75th percentiles, respectively. The black dashed line shows a one-to-one correspondence. Colours as in the top panel (see legend).}
\label{fig:cddf}
\end{figure}

We first consider the distribution of \OVI\ column densities, obtained by binning the set of fitted \OVI\ column densities for each adopted S/N in logarithmic bins of size $\Delta \log \NOVI = 0.2 ~{\rm dex}$, and normalising by the total redshift path surveyed. In each bin, we compute Poisson single-sided $1\sigma$ confidence limits using the tables by \citet[][]{geh86a}. Note that this uncertainty does not include the fitting error in the column density, which is typically around 10 per cent. The histograms thus obtained are displayed in the top panel of Fig.~\ref{fig:cddf}. This plot clearly shows the effect of the S/N on the detection limit in terms of column density: First, a decreasing S/N leads to a smaller number of identified components (3115 at  S/N = 50 vs. 1093 at S/N = 10). Second, it increases the column density threshold -- $\log (\NOVI/\cm^{-2}) \approx 12.3$ at  S/N = 50 vs. $\log (\NOVI / \cm^{-2}) \approx 13.2$ at  S/N = 10 -- above which lines can be reliably detected. The column density detection threshold for each S/N can be best appreciated in the inset displaying the normalised cumulative distribution, \ie, the number of lines with column densities above a given value. Note that a column density threshold $\log (\NOVI/\cm^{-2}) \approx 13.2$ for S/N = 10 is typical for \OVI\ absorbers identified in FUSE and HST/STIS data at a similar S/N \citep[\eg{}][]{tri00a,dan05a,tho08a}.

In order to assess the completeness of our line sample and to check the quality of our fits, we compare the integrated \OVI\ column density along each physical \los\ in the simulation (the {\em true} total \OVI\ column density) to the sum of the column densities of all \OVI\ absorption components identified in the corresponding spectrum (the {\em recovered} total \OVI\ column density). The result is shown in the bottom panel of Fig.~\ref{fig:cddf}. We bin the relative difference $\delta \NOVI/\NOVI$ of each \los\ (spectrum) as a function of the true column density using \mbox{$\Delta \log \NOVI = 0.2~{\rm dex}$} (as displayed by the $x$ error-bars), and compute the 25th and 75th percentile in each bin (corresponding to the lower and upper $y$ error-bar, respectively). Comparison to a perfect match given by the black dashed line shows that the agreement between the true and the recovered column densities is good, although with a significant scatter, and with the true amount of \OVI\ along each physical \los\ being underestimated at all column densities but the lowest. In particular, we can see that we miss $\sim 30$ per cent of the \OVI\ column density along \loss\ with a total column density around \mbox{$\NOVI = 10^{13} ~\cmsq$}. This is due to the fact that along some of these \loss\ ($\sim 20$ per cent), more than one absorption component (usually two or three) are identified. Hence, some of the absorption components along these \loss\ have column densities that fall below the detection threshold. Of course, this is true in general for all \loss, which explains why we systematically underestimate the total \OVI\ column density. Note that the particular column density at which the recovery is worst depends on the adopted S/N. Moreover, at high column densities, saturation effects may also play a role. Indeed, since \OVI\ saturates at \mbox{$\NOVI \approx 2\times10^{\,14} \,\cm^{\,-2}$} for $\bovi = 4.2 \, \kms$ (\ie, at our chosen spectral resolution limit; see next section), it is possible that lines with true column densities above this value are fitted using components with somewhat lower column densities. The latter effect is, however, negligible, given the small number of components with $\NOVI> 2\times10^{\,14} \,\cm^{\,-2}$ (see inset in top panel of Fig.\ref{fig:cddf}).

%--------------------------------------------------------------------------------------------------------------------------------------------------------------------------------
% FIGURE: Column density distribution function and integrated column density for S/N=10, 30, 50
%--------------------------------------------------------------------------------------------------------------------------------------------------------------------------------

%--------------------------------------------------------------------------------------------------------------------------------------------------------------------------------
\subsection{Doppler parameter distribution} \label{sec:dop}

Since the STIS resolution limit of \mbox{${\rm FWHM} = 7 \kms$} translates into a minimum Doppler parameter $b_{\rm min} = 4.2 \kms$, we impose the condition that our fitting algorithm discards any identified features with fitted $b$-values smaller than this to constrain the detection of spurious features. However, this does not have a large impact on our resulting line width statistics, given the small value of the threshold relative to the typical \OVI\ line widths ($\bovi \sim 12 \kms$; see below). Note that for other instruments with lower resolution such as {\em COS} (\mbox{${\rm FWHM} \approx 17 \kms$}), imposing a minimum allowed $b$-value in the fitting procedure might have a noticeable effect on the line width distribution.

\begin{figure}
\resizebox{1\colwidth}{!}{\includegraphics{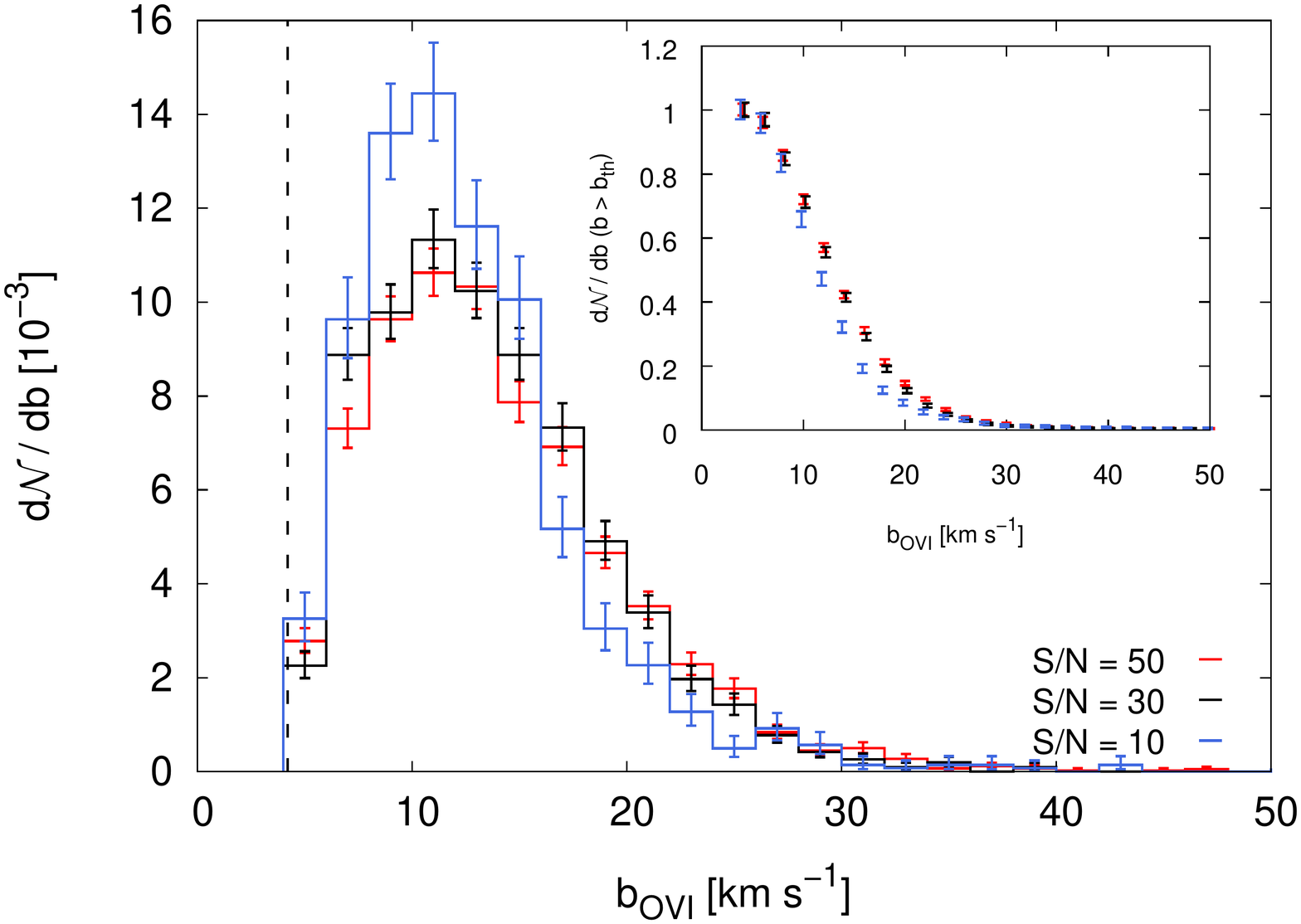}}
\caption{Distribution of Doppler parameters (histograms). Note that the distributions have been normalized to unit area. The inset shows the normalised cumulative distribution for each S/N. The black, dashed line shows the STIS resolution limit at \mbox{${\rm FWHM} = 7 \kms$}, which translates into a minimum Doppler parameter $b_{\rm min} = 4.2 \kms$. The cut-off at this value is due to the condition imposed on our fitting algorithm that any line with fitted $b$-value smaller than $b_{\rm min}$ be discarded.
}
\label{fig:bdf}
\end{figure}

The distributions of Doppler parameters obtained from our spectra for S/N = 10, 30, and 50 are shown in Fig.~\ref{fig:bdf}. As can be appreciated, all three distributions are similar in width, with peaks at Doppler values in the range \mbox{$\bovi \approx 10 - 12 \kms$}, and all show an extended tail towards $b$-values as large as $50 \, \kms$. The median Doppler parameter is \mbox{$b_{\,\rm med}$ = 12.9, 12.7, and 11.7 $\kms$} for \mbox{S/N = 50, 30, and 10}, respectively, and typical relative errors in the fitted Doppler parameters are around 10 per cent, independent of the adopted S/N. The error bars display Poisson uncertainties. We find that a low S/N leads to a slight decrease in the number of  components with \mbox{$\bovi > 10 \, \kms$}, as can be seen in the inset displaying the normalised cumulative distribution for each S/N. This is expected, since broader components are generally shallower (for a fixed equivalent width) and get progressively lost in the noise with decreasing S/N. To a lesser extent, it is also true that with decreasing S/N, intrinsically broad lines are occasionally contaminated by strong noise spikes and may be thus be fitted with multiple narrow components, thus effectively improving the fit. This can have important (and unwanted) consequences for the correlation between the temperature assigned to an absorption line (\ie, to the gas where the absorption takes place) and its corresponding Doppler parameter. Note on the other hand that, irrespective of the S/N, a high fraction (more than 70 per cent) of the fitted components have Doppler widths \mbox{$10 \kms < \bovi < 40 ~\kms$}, which for oxygen correspond to temperatures \mbox{$10^{5} ~\K  \lesssim T \lesssim 10^{6} ~\K$} (assuming pure thermal broadening; see equation \ref{eq:bth}). The relation between Doppler parameter and gas temperature is discussed in detail in Sec.~\ref{sec:tvsb}. The similarity between the Doppler parameter distributions for \mbox{S/N = 30} and \mbox{S/N = 50} suggests that our fitted Doppler parameter are reliable for ${\rm S/N} \gtrsim 30$.

%--------------------------------------------------------------------------------------------------------------------------------------------------------------------------------
% FIGURE: b-value (differential and cumulative) distribution
%--------------------------------------------------------------------------------------------------------------------------------------------------------------------------------

%--------------------------------------------------------------------------------------------------------------------------------------------------------------------------------
\subsection{Equivalent width distribution} \label{sec:dndw}

%--------------------------------------------------------------------------------------------------------------------------------------------------------------------------------
% FIGURE: Cumulative line number density for different signal-to-noises; curve-of-growth
\begin{figure*}
\resizebox{1.\colwidth}{!}{\includegraphics{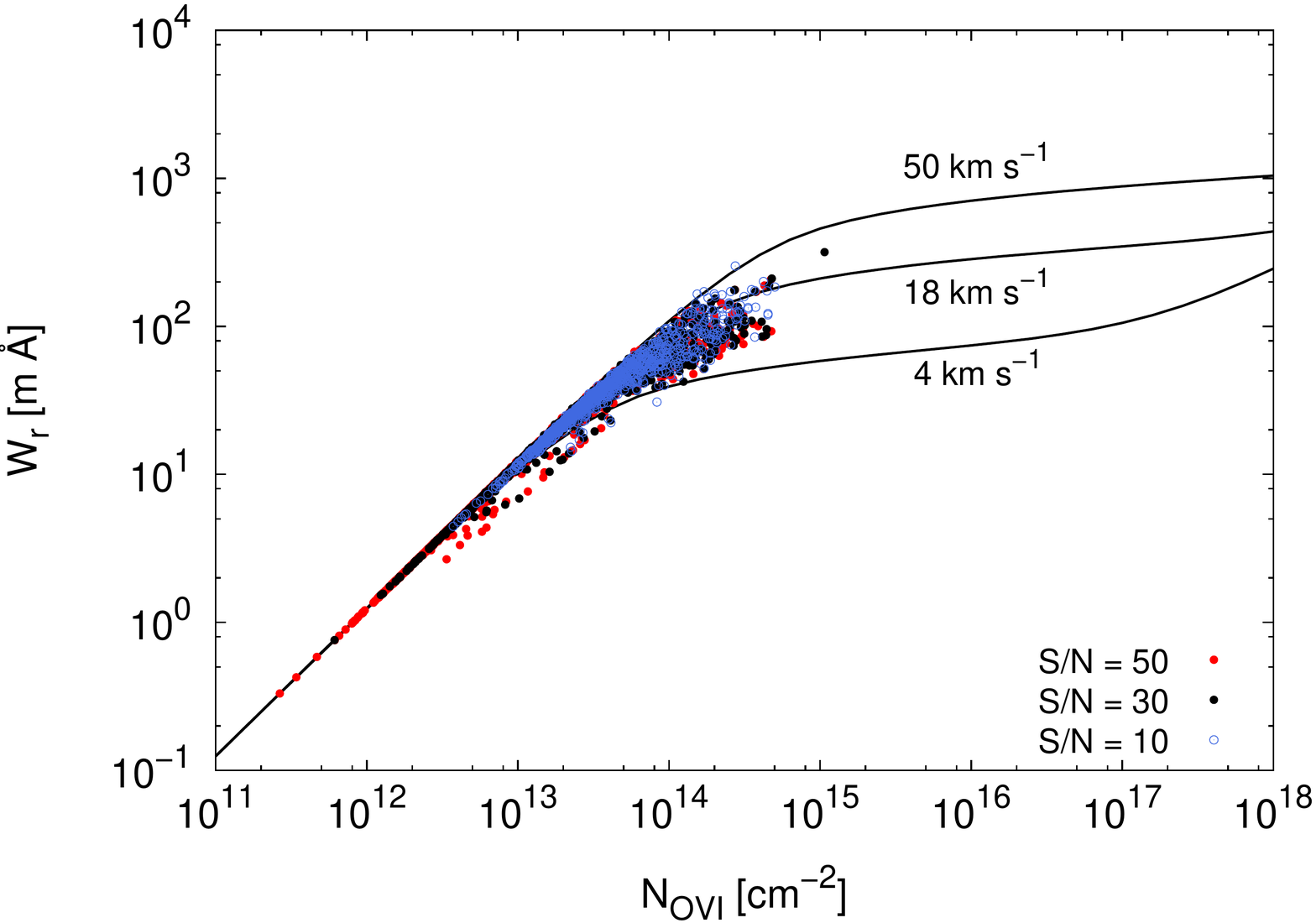}}
\resizebox{1.\colwidth}{!}{\includegraphics{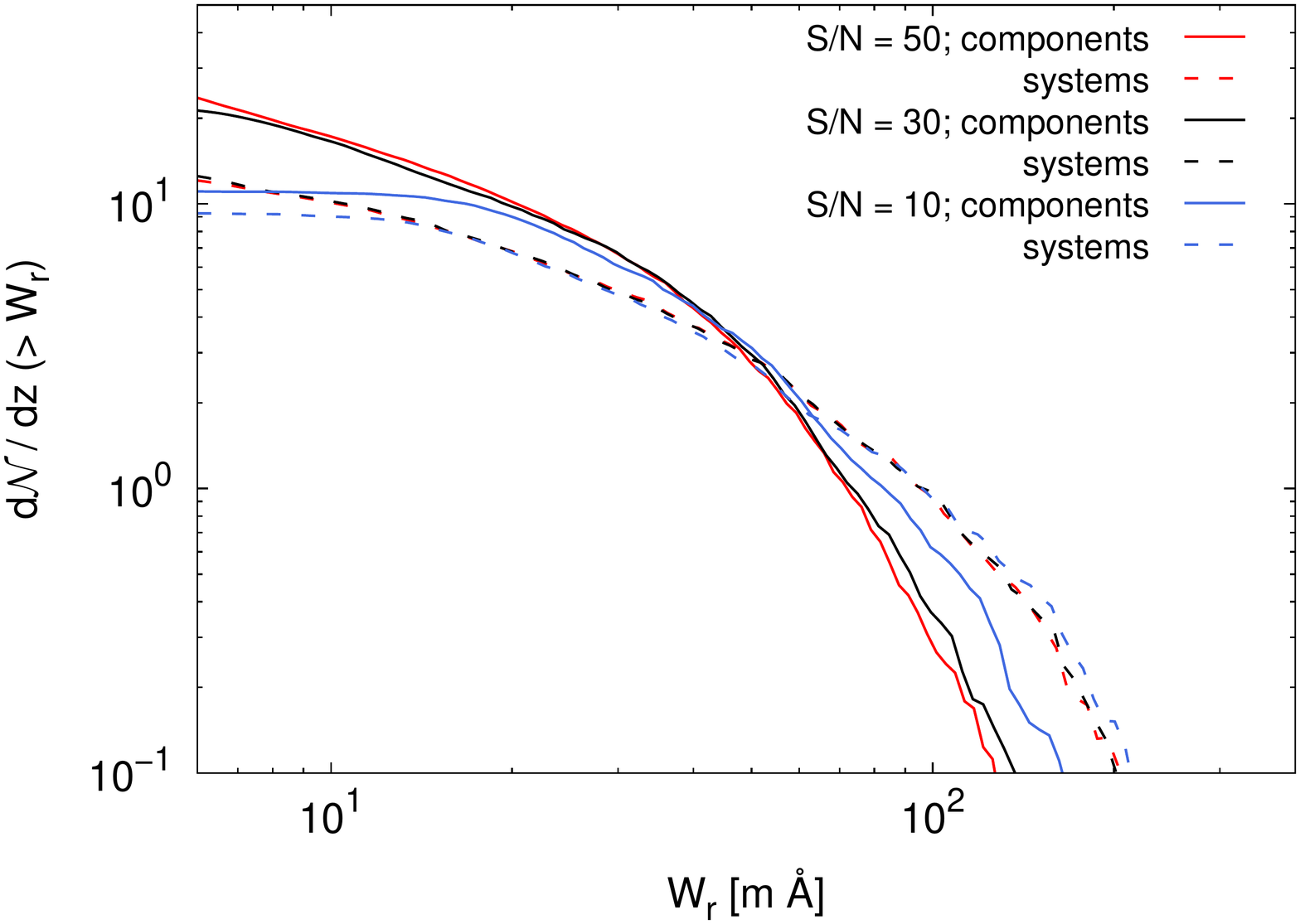}}
\resizebox{1.\colwidth}{!}{\includegraphics{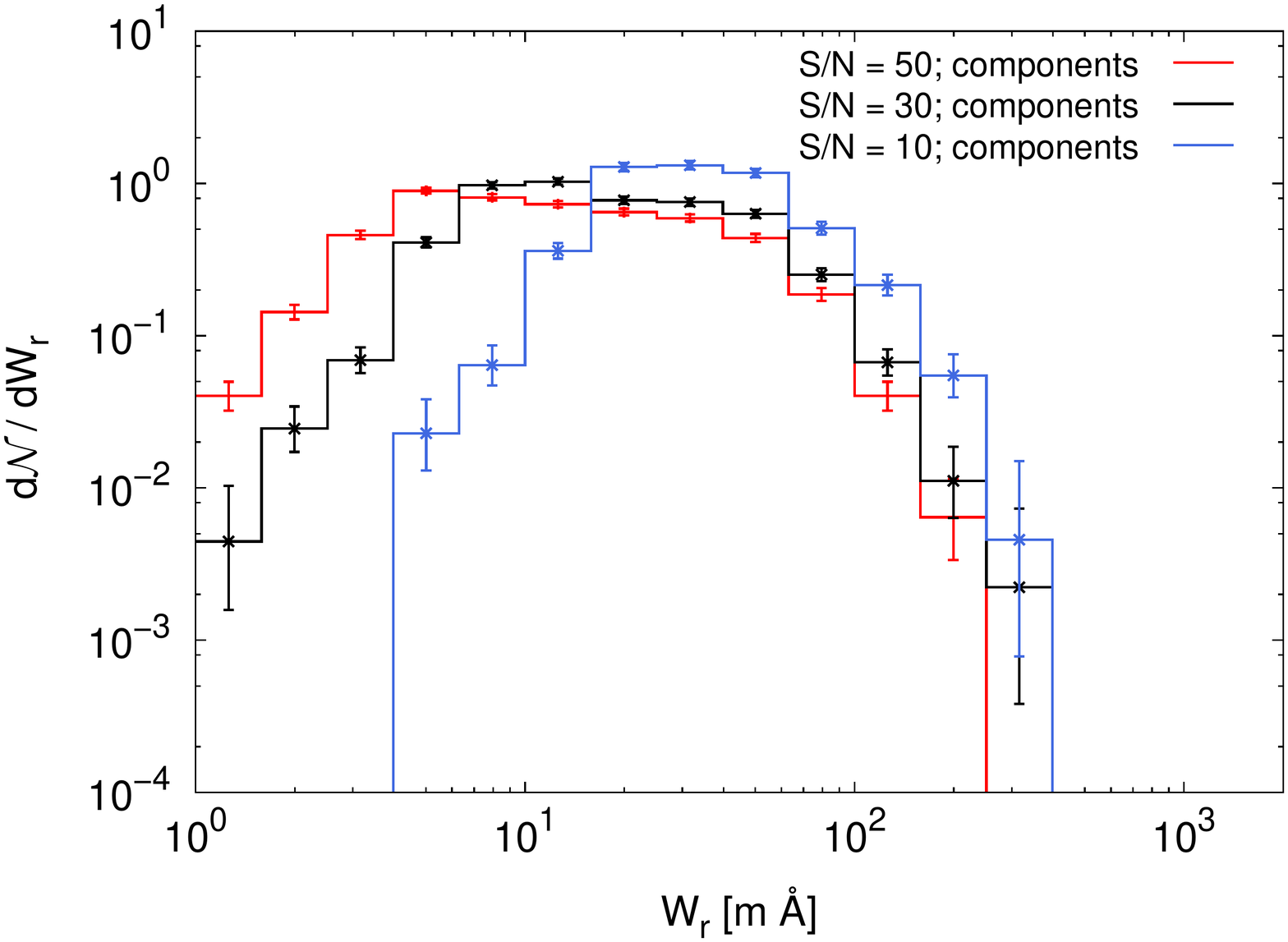}}
\resizebox{1.\colwidth}{!}{\includegraphics{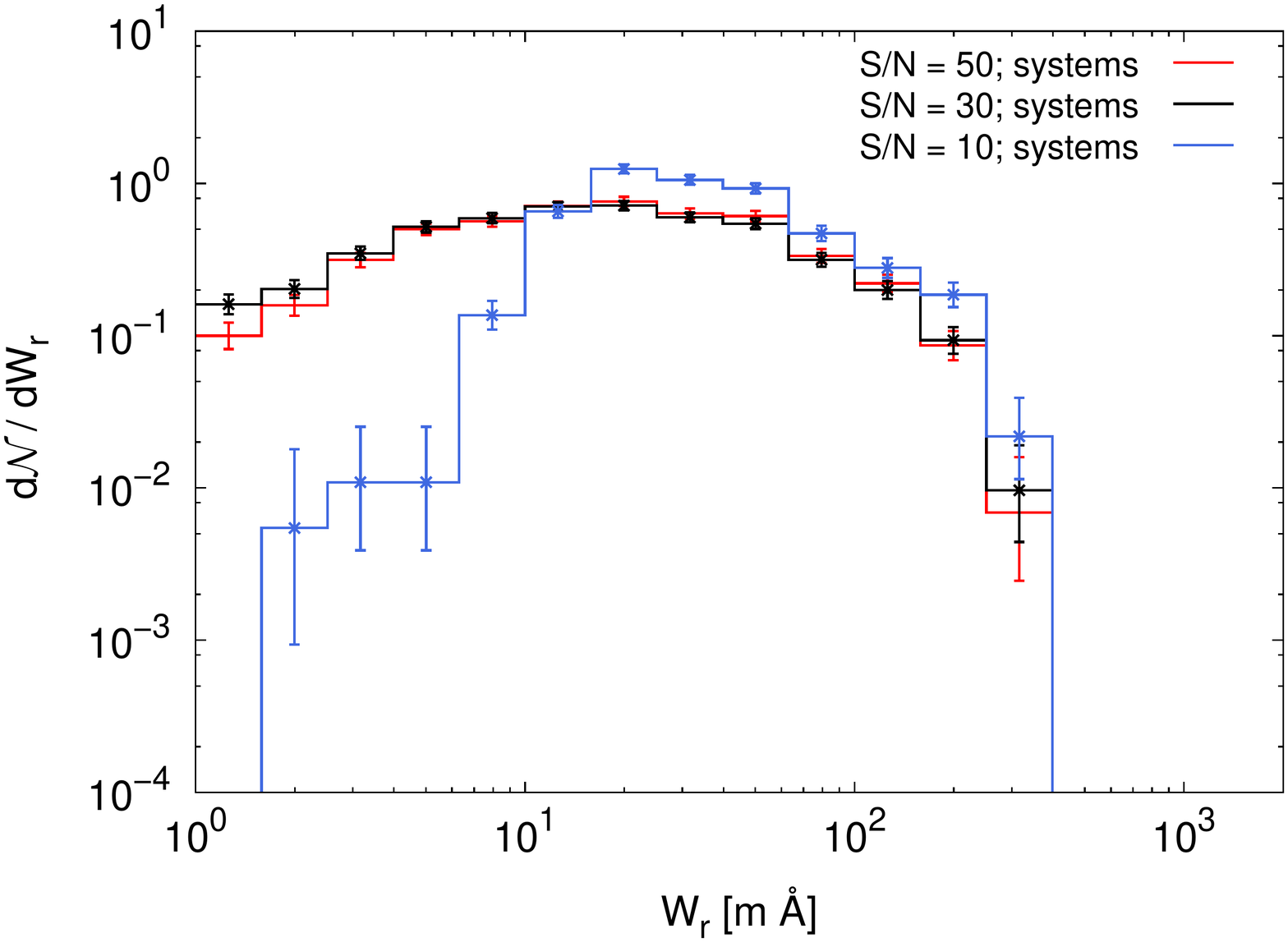}}
\caption[ ]{{\em Top left:} Rest equivalent width $W_{\rm r}$ as a function of \OVI\ column density for each \OVI\ component and for different S/N (coloured symbols). The solid curves show theoretical curves-of-growth (CoG); the $b$-value used to compute each CoG is indicated along the corresponding curve.
{\em Top right:} Cumulative equivalent width distribution, \ie, number of lines per unit redshift above a given rest equivalent width for {\em components} (solid) and {\em systems} (dashed) identified in our synthetic spectra for different S/N.
{\em Bottom left:} Frequency of absorption lines per equivalent width bin for individual components and different S/N. Error bars denote Poisson uncertainties.
{\em Bottom right:} Same as bottom-left panel, but for absorption {\em systems}.}
\label{fig:dndw_sn}
\end{figure*}
%--------------------------------------------------------------------------------------------------------------------------------------------------------------------------------

Here we further explore the quality of our fits by analysing the measured rest equivalent widths $W_{\rm r}$. The rest equivalent width of an \OVI\ absorption line is measured directly by integrating the flux depression over the entire spectrum, and correcting for the line broadening due to cosmological expansion. In the top-left panel of Fig.~\ref{fig:dndw_sn}, we compare the rest equivalent width as a function of column density for each \OVI\ component identified in spectra with different S/N to theoretical curves-of-growth (CoG) for a range of Doppler parameters which roughly corresponds to the width of the distributions shown in Fig.~\ref{fig:bdf}. It is remarkable that the overwhelmingly majority of the identified components lie in the range allowed by the curves-of-growth. The exception are a few absorption lines for which either the measured equivalent width is systematically underestimated (assuming the column density has been accurately measured), or the column density is slightly and systematically overestimated. The good correspondence between rest equivalent width and column density as compared to the CoG is even more notable if one considers that the rest equivalent width of a given line has not been determined from a CoG analysis, \ie, its measurement is completely independent of the corresponding fitted \OVI\ column density and the fitted $b$-value. This demonstrates further the accuracy of the line fitting algorithm we use. We note that most of the absorption components, regardless of the S/N, lie on the linear part of the CoG. For these lines, the rest equivalent width scales linearly with the column density, and is (nearly) independent of the Doppler parameter. Unfortunately, this degeneracy makes it difficult to assess the goodness of the Doppler parameter fitted to a particular line. On the other hand, this explains why all identified \OVI\ components seem to have $(\NOVI, \, W_{r})$-values consistent with CoG for $\bovi = 4 - 50 ~\kms$, even though we do find a few lines with (possibly artificial) $b$-values as high as $\bovi = 60 ~\kms$.

Given the measured rest equivalent widths $W_{\,r}$, we compute the cumulative equivalent width distribution, \ie, the frequency of absorption lines per unit redshift above a given equivalent width threshold, for S/N = 10, 30, and 50. As already discussed in Sec.~\ref{sec:obs}, we explicitly distinguish between individual absorption {\em components} and {\em systems} in order to have a better comparison to different sets of observations. The cumulative equivalent width distributions for individual components and systems adopting different S/N are shown in the top-right panel of Fig.~\ref{fig:dndw_sn}.  A lower S/N leads to a smaller number of lines at rest equivalent widths $W_{\rm r} \lesssim 30 {\rm m\AA}$. The reason for this is two-fold: First, weak absorption features simply get lost in the noise. Second, with decreasing S/N, otherwise resolved narrow components are occasionally fitted by a single, broader absorption feature. Both these effects reduce the relative number of components at low equivalent widths and increase their number at high equivalent widths. The cumulative equivalent width distribution for systems shows qualitative trends which are similar to those of the distribution of  individual components at low equivalent widths. However, an important difference is that the number of identified systems seems to be nearly independent of the adopted S/N at the high equivalent-width end. This can be understood in terms of the definition of systems together with the fact that noise has little effect on strong absorption features. Note also that there are more strong (\ie, at $W_{\,r} \gtrsim 50 {\rm m \AA}$) systems than strong components, as expected, since the latter are a subset of the former.

While a {\em cumulative} distribution gives an idea about the {\em amplitude} of the overall equivalent width distribution, a {\em differential} distribution gives insight into the {\em shape} of the distribution. It is thus of interest to take a look at the differential distribution of rest equivalent widths both of components and systems at different S/N values. The bottom-left panel of Fig.~\ref{fig:dndw_sn} shows the distribution of components (normalised to unit area). Here we see again that low $W_{\,r}$ lines get systematically lost with decreasing S/N, while the opposite is true for high $W_{\,r}$ lines, owing to the fact that individual features are occasionally merged in noisy spectra. This suggests that some of the high $W_{\,r}$ lines identified in real QSO spectra could be artificial. The bottom-right panel shows the differential $W_{\rm r}$ distribution of systems. Again, weak systems seem to get progressively lost in the noise\footnote{Recall that the detection threshold $\tau_{\rm th}$ used to identify absorption systems depends on S/N.}, as is the case for single components, even though the effect is almost not present for ${\rm S/N} > 30$, as can be judged by comparing the distributions for S/N = 30 and S/N = 50 which are nearly indistinguishable. This indicates that, above a certain S/N value, the identification of absorption systems is less sensitive to the adopted S/N as compared to single components.

%--------------------------------------------------------------------------------------------------------------------------------------------------------------------------------
\section{Convergence Tests} \label{sec:conv}

In this section we address the numerical convergence of our results with respect to box size and resolution, focusing on the predicted \OVI\ line parameter (column density, Doppler parameter, differential equivalent width) distributions. To this end, we generate and analyse 5000 synthetic spectra at $z =0.25$, adopting a standard S/N of 50, for different runs including our reference model. Note that similar convergence tests in terms of the predicted star-formation history (SFH) and cosmic metal distribution have previously been presented and discussed in detail by \citet[][their section 3.2]{sch10a} and \citet[][their appendices B and C]{wie09b}, respectively, using the same simulation runs that we use here. For this reason, we will restrict ourselves to the presentation of our conclusions and kindly refer the reader to the studies mentioned above for a thorough discussion.

%--------------------------------------------------------------------------------------------------------------------------------------------------------------------------------
\subsection{Box size}

%--------------------------------------------------------------------------------------------------------------------------------------------------------------------------------
% FIGURE: title
\begin{figure*}
{\resizebox{0.49\textwidth}{!}{\includegraphics{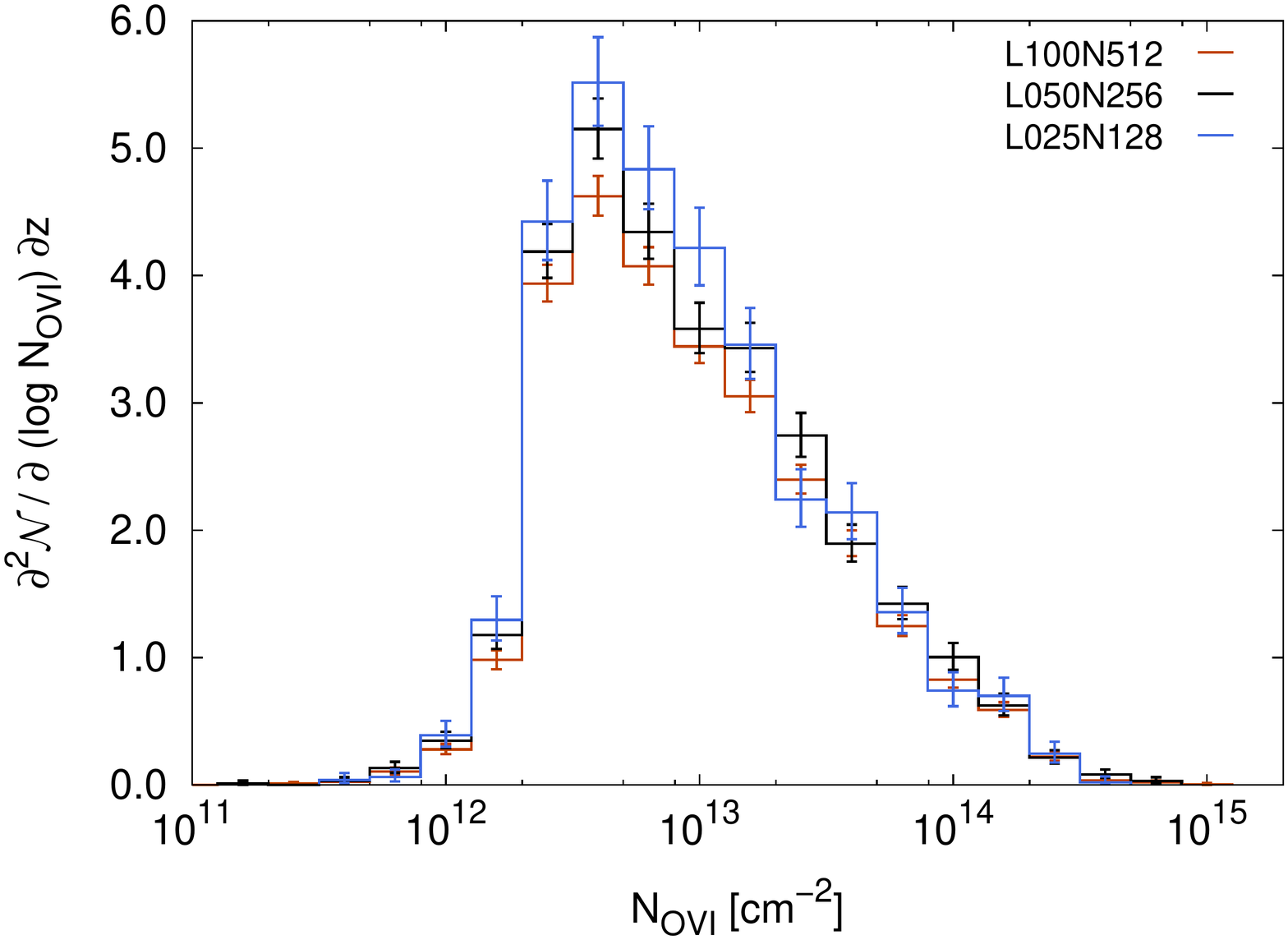}}}
{\resizebox{0.49\textwidth}{!}{\includegraphics{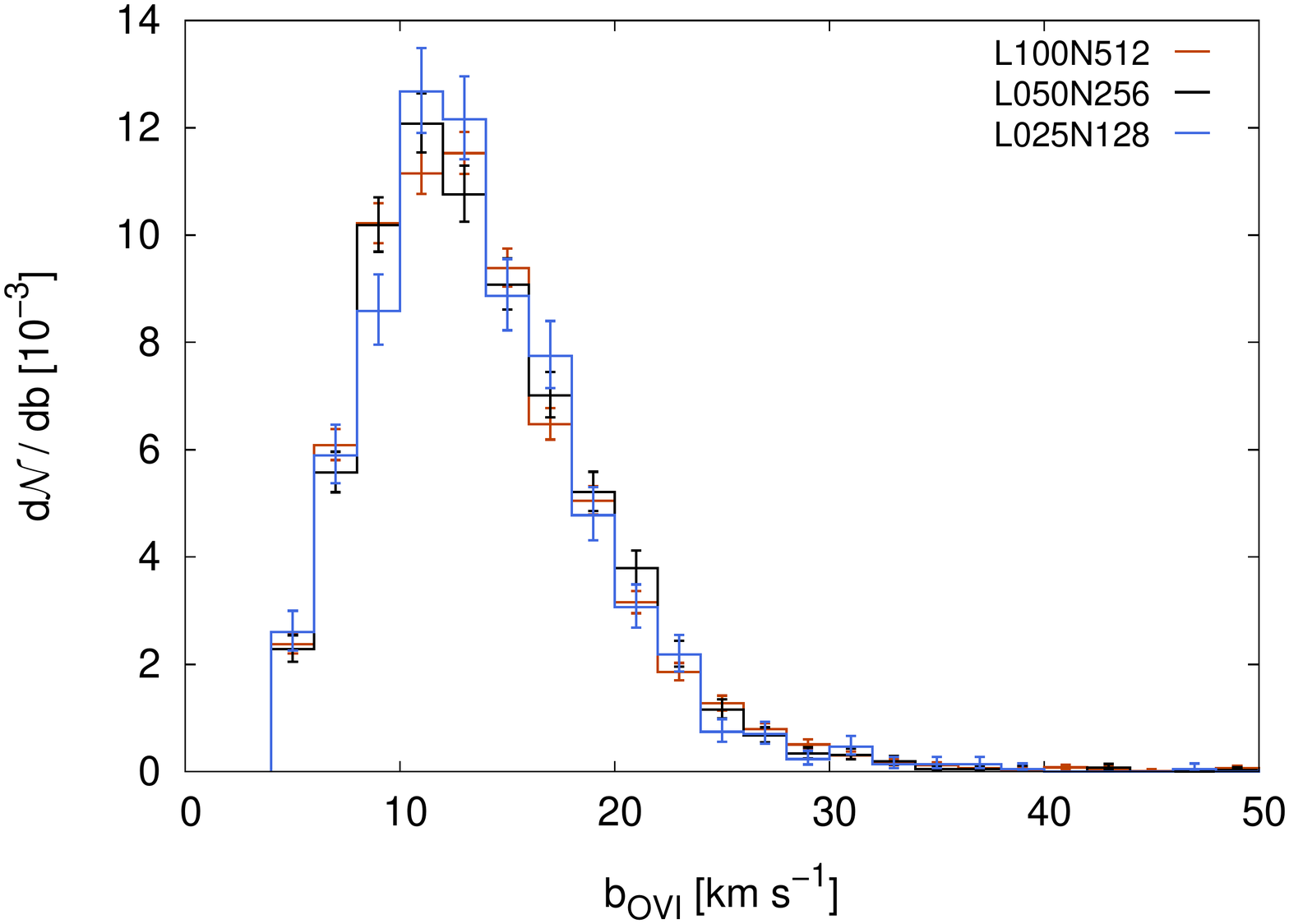}}}
{\resizebox{0.49\textwidth}{!}{\includegraphics{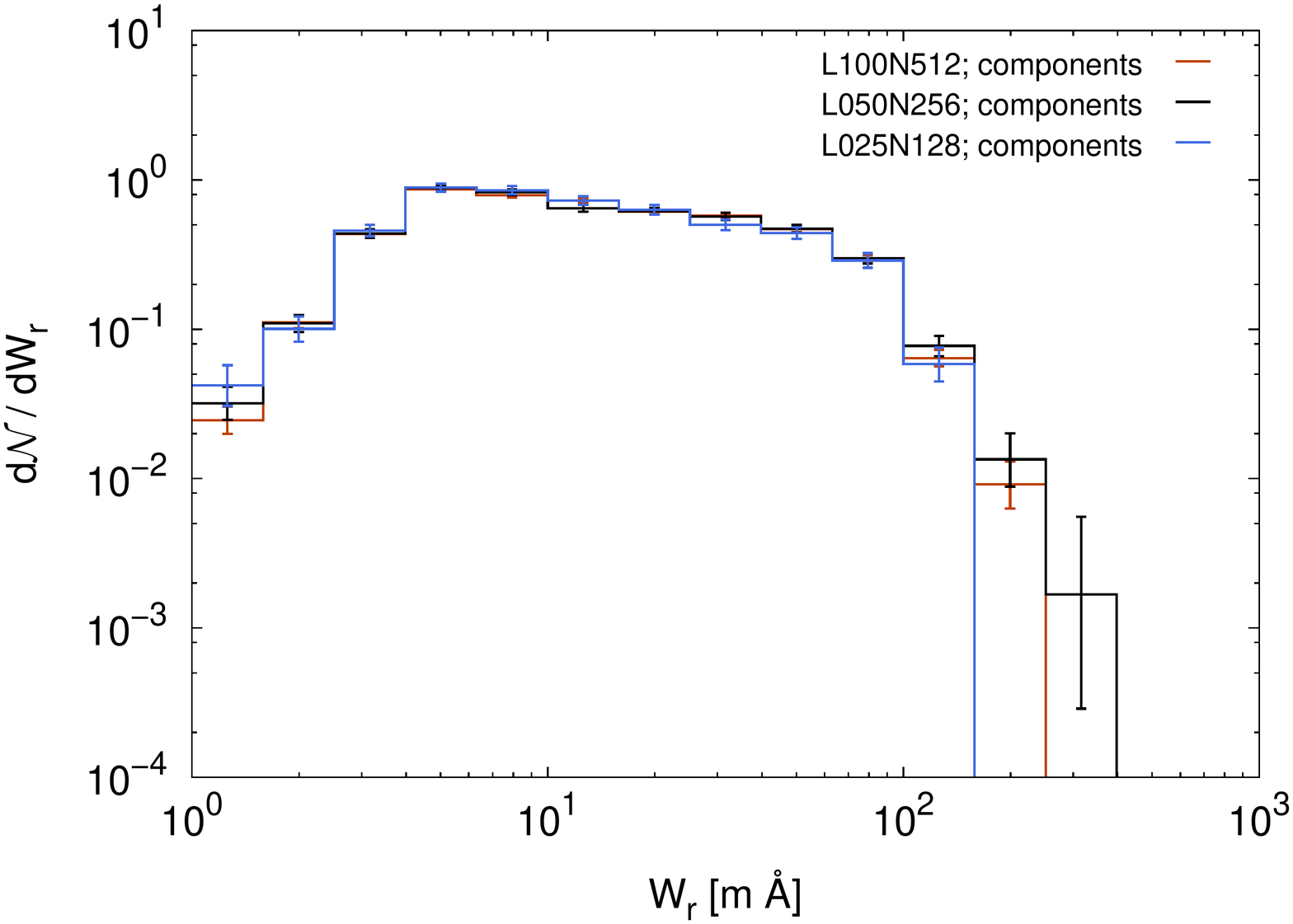}}}
{\resizebox{0.49\textwidth}{!}{\includegraphics{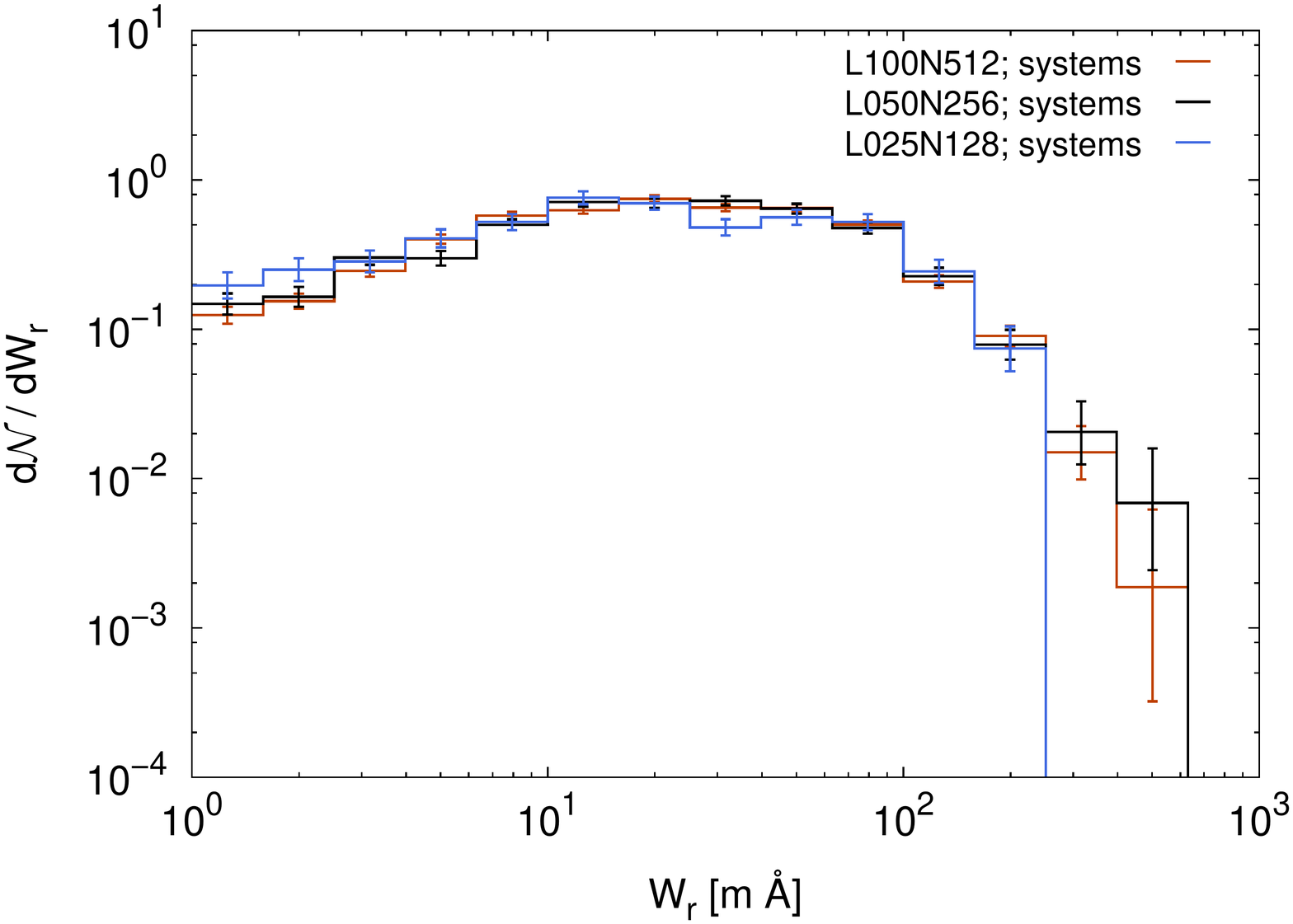}}}
\caption[]{Convergence of our reference model with respect to simulation box size (at fixed spatial and mass resolution; see Table \ref{tab:ref_sims} for details).
{\em Top-left:} \OVI\ column density distribution.
{\em Top-right:} \OVI\ Doppler parameter distribution.
{\em Bottom-left:} \OVI\ equivalent width distribution for single components.
{\em Bottom-right:} \OVI\ equivalent width distribution for systems.
}
\label{fig:conv1}
\end{figure*}
%--------------------------------------------------------------------------------------------------------------------------------------------------------------------------------

Fig.~\ref{fig:conv1} shows the \OVI{} column density distribution function (top-left), Doppler parameter distribution (top-right), and differential equivalent width distribution for single components (bottom-left) and systems (bottom-right), obtained from spectra synthesised  using three different simulations: {\em  REF\_L025N128}, {\em  REF\_L050N256}, and {\em  REF\_L100N512}. All these simulations have the same spatial and mass resolution (see Table \ref{tab:ref_sims}), but the sizes of the simulation boxes differ by factors of two. This plot shows that the prediction from our reference run for the \OVI{} line parameter distributions is robust with respect to the size of the simulation volume. A box of $L = 25~\hMpc$ on a side is sufficiently large to compute a reliable prediction for all three observables for lines with $W_{\rm r} < 200~{\rm m \AA}$. Note, however, that we require a box of at least $50~\hMpc$ to correctly sample the rarer strong (\ie, $W_{\rm r} > 500~{\rm m \AA}$) absorbers.

%--------------------------------------------------------------------------------------------------------------------------------------------------------------------------------
\subsection{Resolution}

%--------------------------------------------------------------------------------------------------------------------------------------------------------------------------------
% FIGURE: title
\begin{figure*}
{\resizebox{0.49\textwidth}{!}{\includegraphics{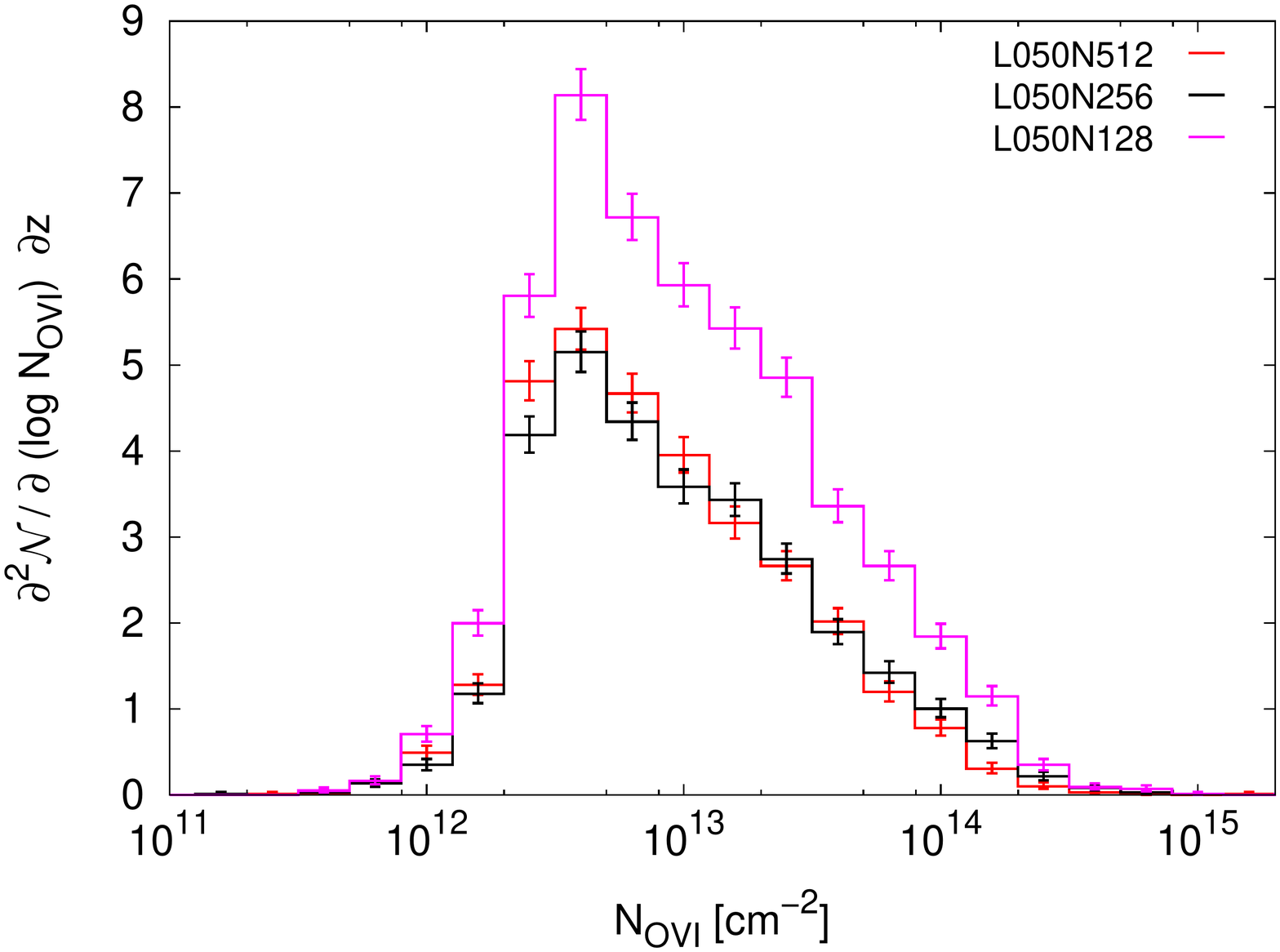}}}
{\resizebox{0.49\textwidth}{!}{\includegraphics{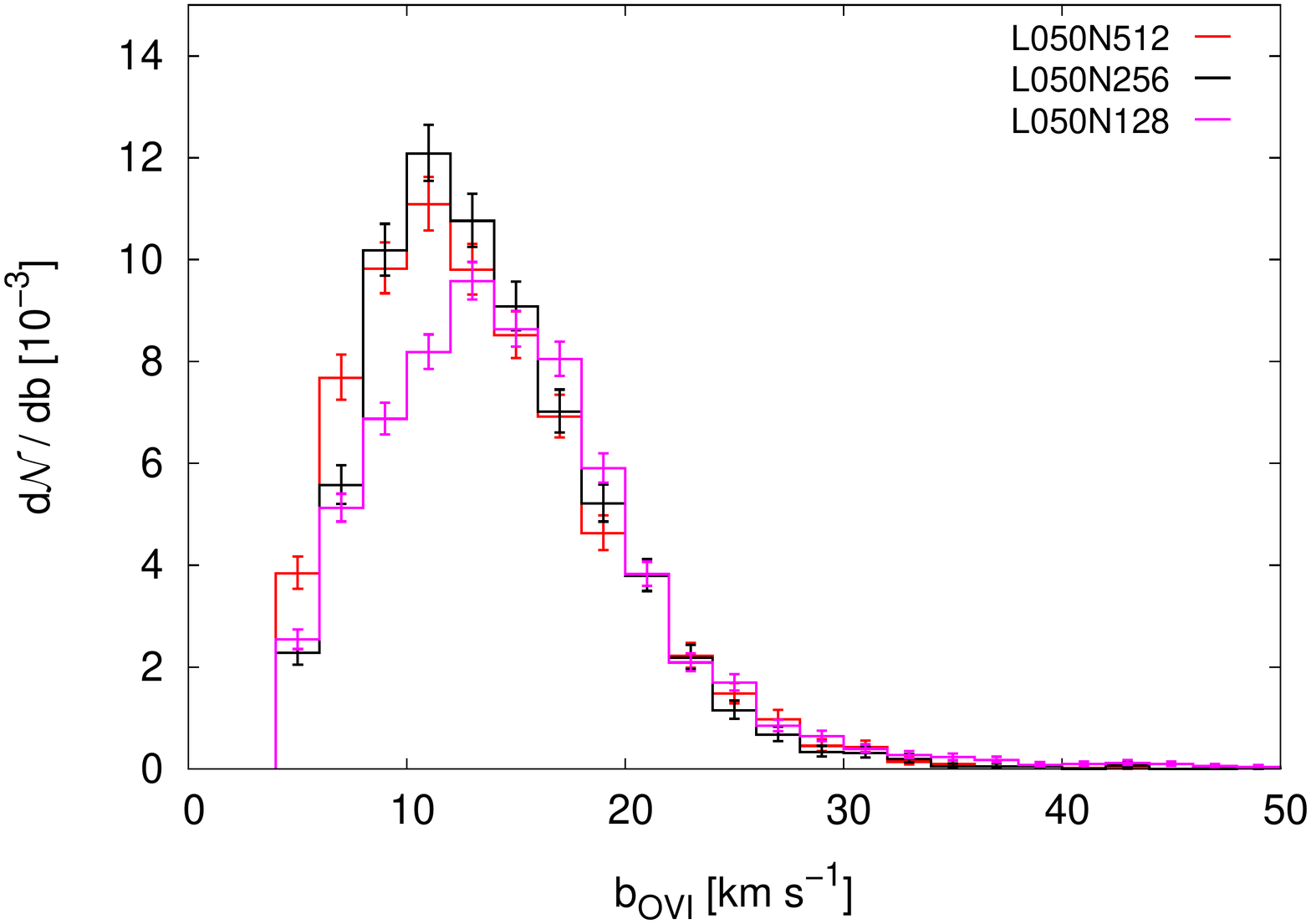}}}
{\resizebox{0.49\textwidth}{!}{\includegraphics{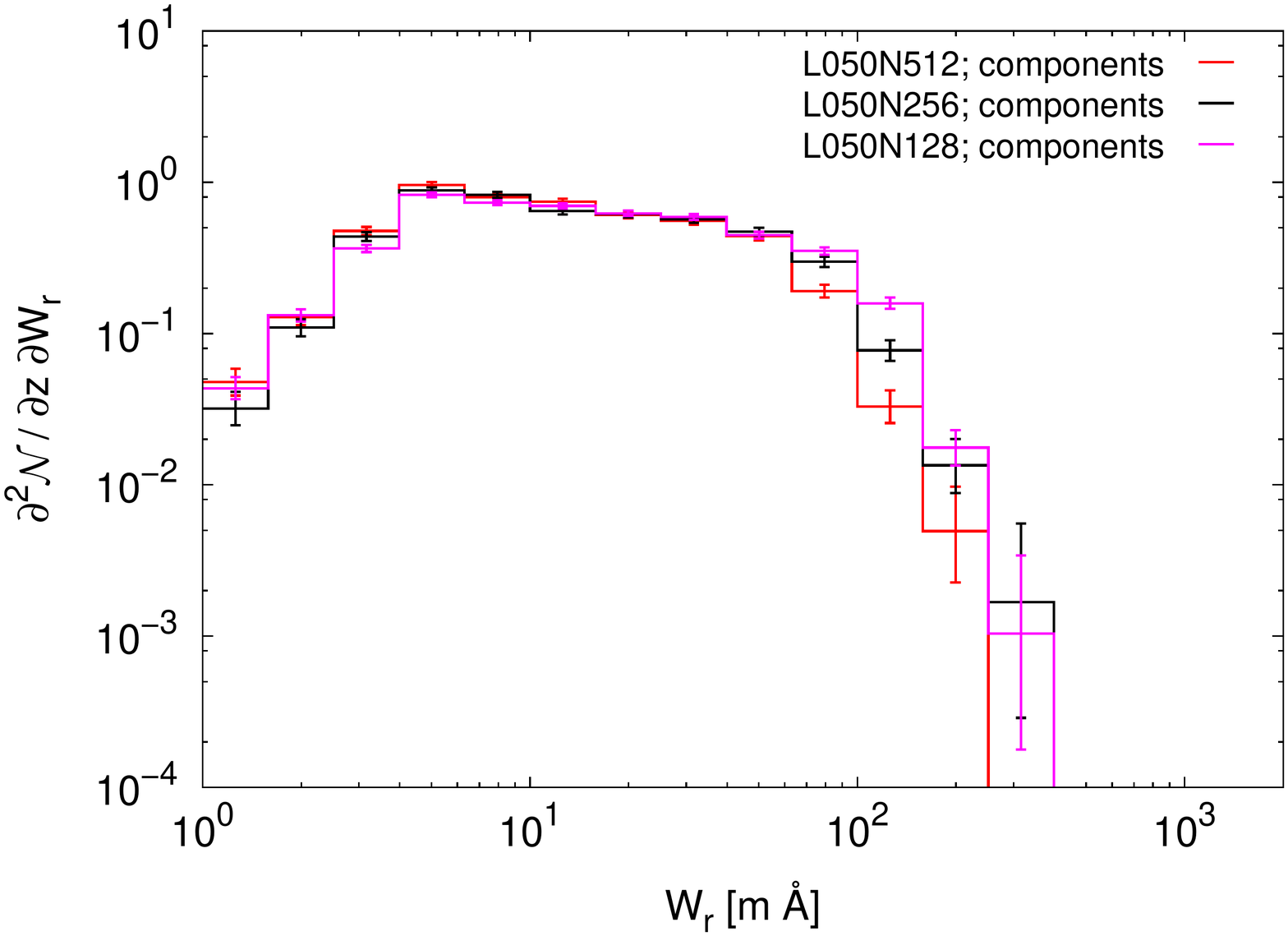}}}
{\resizebox{0.49\textwidth}{!}{\includegraphics{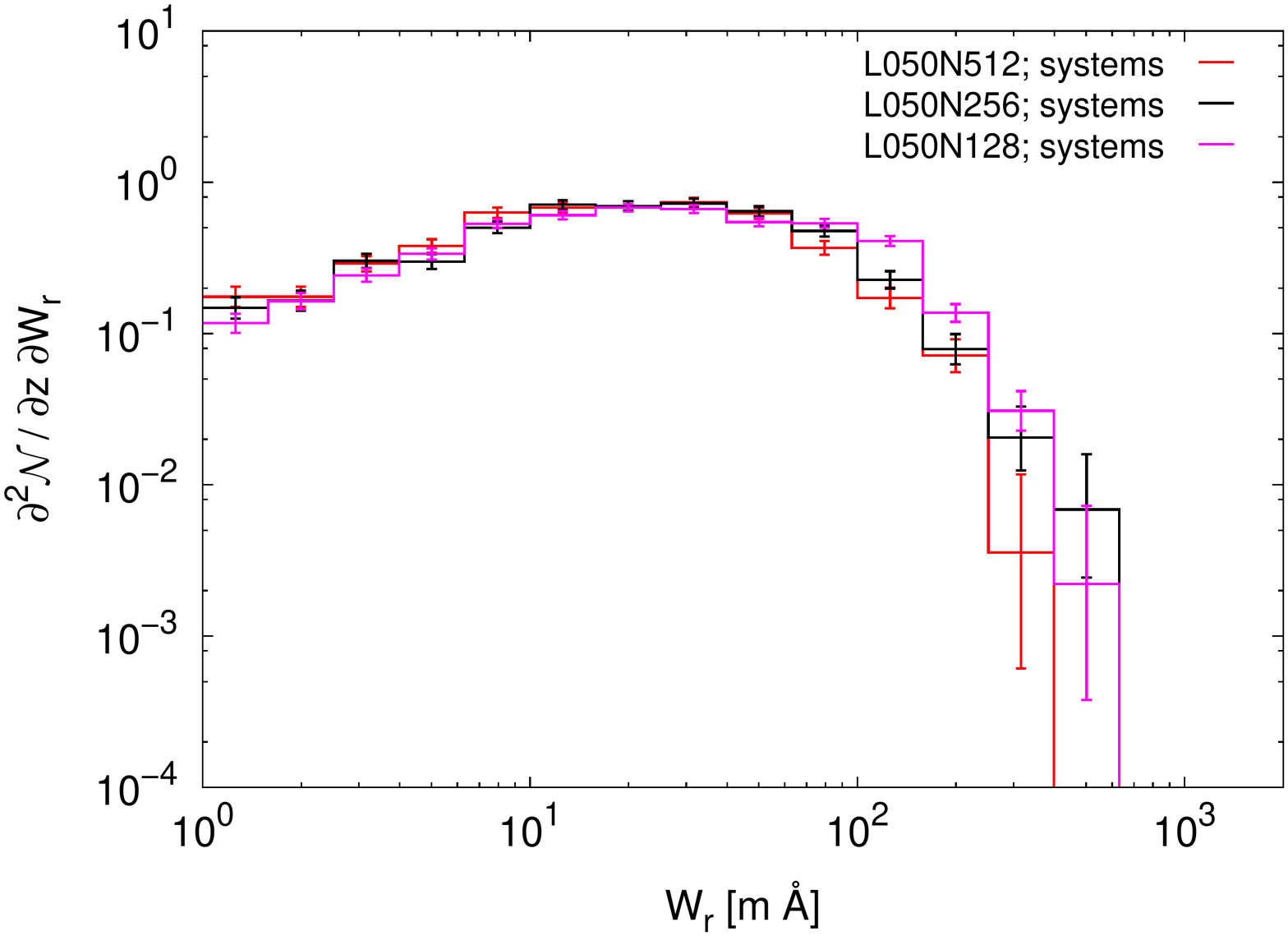}}}
\caption[]{Convergence of our reference model (and fiducial run) with respect to resolution (at fixed box size; see Table \ref{tab:ref_sims} for details).
{\em Top-left:} \OVI\ column density distribution function.
{\em Top-right:} \OVI\ Doppler parameter distribution.
{\em Bottom-left:} \OVI\ equivalent width distribution for single components.
{\em Bottom-right:} \OVI\ equivalent width distribution for systems.
}
\label{fig:conv2}
\end{figure*}
%--------------------------------------------------------------------------------------------------------------------------------------------------------------------------------

We now explore the effect of varying resolution on the predicted \OVI{} observables. For this purpose, Fig.~\ref{fig:conv2} compares a low resolution run {\em  REF\_L050N128} and a moderate resolution run {\em  REF\_L050N256} to our fiducial high-resolution simulation {\em  REF\_L050N512}. Note that we keep the box size fixed in this comparison. While the low-resolution run is clearly different, the intermediate resolution model is nearly identical to our fiducial run, suggesting that the results are converged with respect to the numerical resolution, except perhaps for the strongest absorbers with $W_{\rm r} > 200~{\rm m \AA}$.

%--------------------------------------------------------------------------------------------------------------------------------------------------------------------------------

%% file: tepper_etal_OVI.bbl
\begin{thebibliography}{}

\bibitem[\protect\citeauthoryear{{Bertone}, {Schaye}, {Booth}, {Dalla Vecchia},
  {Theuns} \& {Wiersma}}{{Bertone} et~al.}{2010}]{ber10a}
{Bertone} S.,  {Schaye} J.,  {Booth} C.~M.,  {Dalla Vecchia} C.,  {Theuns} T.,
    {Wiersma} R.~P.~C.,  2010, ArXiv e-prints, \eprint{1002.3393}

\bibitem[\protect\citeauthoryear{{Bertone}, {Schaye}, {Dalla Vecchia}, {Booth},
  {Theuns} \& {Wiersma}}{{Bertone} et~al.}{2009}]{ber09a}
{Bertone} S.,  {Schaye} J.,  {Dalla Vecchia} C.,  {Booth} C.~M.,  {Theuns} T.,
    {Wiersma} R.~P.~C.,  2009, ArXiv e-prints, \eprint{0910.5723}

\bibitem[\protect\citeauthoryear{Bertone, Schaye \& Dolag}{Bertone
  et~al.}{2008}]{ber08a}
Bertone S.,  Schaye J.,    Dolag K.,  2008, Space Science Reviews, 134, 295

\bibitem[\protect\citeauthoryear{{Cen} \& {Chisari}}{{Cen} \&
  {Chisari}}{2010}]{cen10a}
{Cen} R.,  {Chisari} N.~E.,  2010, ArXiv e-prints, \eprint{1005.1451}

\bibitem[\protect\citeauthoryear{{Cen} \& {Ostriker}}{{Cen} \&
  {Ostriker}}{1999}]{cen99a}
{Cen} R.,  {Ostriker} J.~P.,  1999, \apj, 514, 1, \eprint{9806281}

\bibitem[\protect\citeauthoryear{{Cen} \& {Ostriker}}{{Cen} \&
  {Ostriker}}{2006}]{cen06a}
{Cen} R.,  {Ostriker} J.~P.,  2006, \apj, 650, 560, \eprint{astro-ph/0601008}

\bibitem[\protect\citeauthoryear{{Cooksey}, {Prochaska}, {Chen}, {Mulchaey} \&
  {Weiner}}{{Cooksey} et~al.}{2008}]{coo08b}
{Cooksey} K.~L.,  {Prochaska} J.~X.,  {Chen} H.,  {Mulchaey} J.~S.,    {Weiner}
  B.~J.,  2008, \apj, 676, 262, \eprint{0706.1285}

\bibitem[\protect\citeauthoryear{{Cooksey}, {Thom}, {Prochaska} \&
  {Chen}}{{Cooksey} et~al.}{2010}]{coo10a}
{Cooksey} K.~L.,  {Thom} C.,  {Prochaska} J.~X.,    {Chen} H.,  2010, \apj,
  708, 868, \eprint{0906.3347}

\bibitem[\protect\citeauthoryear{{Dalla Vecchia} \& {Schaye}}{{Dalla Vecchia}
  \& {Schaye}}{2008}]{dal08b}
{Dalla Vecchia} C.,  {Schaye} J.,  2008, \mnras, 387, 1431, \eprint{0801.2770}

\bibitem[\protect\citeauthoryear{{Danforth} \& {Shull}}{{Danforth} \&
  {Shull}}{2005}]{dan05a}
{Danforth} C.~W.,  {Shull} J.~M.,  2005, \apj, 624, 555,
  \eprint{astro-ph/0501054}

\bibitem[\protect\citeauthoryear{{Danforth} \& {Shull}}{{Danforth} \&
  {Shull}}{2008}]{dan08b}
{Danforth} C.~W.,  {Shull} J.~M.,  2008, \apj, 679, 194, \eprint{0709.4030}

\bibitem[\protect\citeauthoryear{{Danforth}, {Shull}, {Rosenberg} \&
  {Stocke}}{{Danforth} et~al.}{2006}]{dan06a}
{Danforth} C.~W.,  {Shull} J.~M.,  {Rosenberg} J.~L.,    {Stocke} J.~T.,  2006,
  \apj, 640, 716, \eprint{astro-ph/0508656}

\bibitem[\protect\citeauthoryear{{Dav{\'e}}, {Cen}, {Ostriker}, {Bryan},
  {Hernquist}, {Katz}, {Weinberg}, {Norman} \& {O'Shea}}{{Dav{\'e}}
  et~al.}{2001}]{dav01a}
{Dav{\'e}} R.,  {Cen} R.,  {Ostriker} J.~P.,  {Bryan} G.~L.,  {Hernquist} L.,
  {Katz} N.,  {Weinberg} D.~H.,  {Norman} M.~L.,    {O'Shea} B.,  2001, \apj,
  552, 473, \eprint{astro-ph/0007217}

\bibitem[\protect\citeauthoryear{{Dav{\'e}}, {Hernquist}, {Weinberg} \&
  {Katz}}{{Dav{\'e}} et~al.}{1997}]{dav97a}
{Dav{\'e}} R.,  {Hernquist} L.,  {Weinberg} D.~H.,    {Katz} N.,  1997, \apj,
  477, 21, \eprint{astro-ph/9609115}

\bibitem[\protect\citeauthoryear{{Ferland}, {Korista}, {Verner}, {Ferguson},
  {Kingdon} \& {Verner}}{{Ferland} et~al.}{1998}]{fer98a}
{Ferland} G.~J.,  {Korista} K.~T.,  {Verner} D.~A.,  {Ferguson} J.~W.,
  {Kingdon} J.~B.,    {Verner} E.~M.,  1998, \pasp, 110, 761

\bibitem[\protect\citeauthoryear{{Fukugita}}{{Fukugita}}{2004}]{fuk04a}
{Fukugita} M.,  2004, in {Ryder} S.,  {Pisano} D.,  {Walker} M.,   {Freeman}
  K.,  eds, Dark Matter in Galaxies Vol.~220 of IAU Symposium, {Cosmic Matter
  Distribution: Cosmic Baryon Budget Revisited}.
pp 227--+

\bibitem[\protect\citeauthoryear{{Furlanetto}, {Schaye}, {Springel} \&
  {Hernquist}}{{Furlanetto} et~al.}{2004}]{fur04a}
{Furlanetto} S.~R.,  {Schaye} J.,  {Springel} V.,    {Hernquist} L.,  2004,
  \apj, 606, 221, \eprint{astro-ph/0309736}

\bibitem[\protect\citeauthoryear{{Gehrels}}{{Gehrels}}{1986}]{geh86a}
{Gehrels} N.,  1986, \apj, 303, 336

\bibitem[\protect\citeauthoryear{{Gnat} \& {Sternberg}}{{Gnat} \&
  {Sternberg}}{2007}]{gna07a}
{Gnat} O.,  {Sternberg} A.,  2007, \apjs, 168, 213, \eprint{astro-ph/0608181}

\bibitem[\protect\citeauthoryear{{Greif}, {Glover}, {Bromm} \&
  {Klessen}}{{Greif} et~al.}{2009}]{gre09a}
{Greif} T.~H.,  {Glover} S.~C.~O.,  {Bromm} V.,    {Klessen} R.~S.,  2009,
  \mnras, 392, 1381, \eprint{0808.0843}

\bibitem[\protect\citeauthoryear{{Haardt} \& {Madau}}{{Haardt} \&
  {Madau}}{1996}]{haa96a}
{Haardt} F.,  {Madau} P.,  1996, \apj, 461, 20, \eprint{astro-ph/9509093}

\bibitem[\protect\citeauthoryear{{Haardt} \& {Madau}}{{Haardt} \&
  {Madau}}{2001}]{haa01a}
{Haardt} F.,  {Madau} P.,  2001, in {D.~M.~Neumann \& J.~T.~V.~Tran} ed.,
  Clusters of Galaxies and the High Redshift Universe Observed in X-rays
  {Modelling the UV/X-ray cosmic background with CUBA}

\bibitem[\protect\citeauthoryear{{Heckman}, {Norman}, {Strickland} \&
  {Sembach}}{{Heckman} et~al.}{2002}]{hec02a}
{Heckman} T.~M.,  {Norman} C.~A.,  {Strickland} D.~K.,    {Sembach} K.~R.,
  2002, \apj, 577, 691, \eprint{0205556}

\bibitem[\protect\citeauthoryear{{Hui} \& {Gnedin}}{{Hui} \&
  {Gnedin}}{1997}]{hui97a}
{Hui} L.,  {Gnedin} N.~Y.,  1997, \mnras, 292, 27, \eprint{9612232}

\bibitem[\protect\citeauthoryear{{Jarosik}, {Bennett}, {Dunkley}, {Gold},
  {Greason}, {Halpern}, {Hill} \& {Hinshaw}}{{Jarosik} et~al.}{2010}]{jar10a}
{Jarosik} N.,  {Bennett} C.~L.,  {Dunkley} J.,  {Gold} B.,  {Greason} M.~R.,
  {Halpern} M.,  {Hill} R.~S.,    {Hinshaw} G. e.~a.,  2010, ArXiv e-prints,
  \eprint{1001.4744}

\bibitem[\protect\citeauthoryear{{Jenkins}, {Bowen}, {Tripp} \&
  {Sembach}}{{Jenkins} et~al.}{2005}]{jen05a}
{Jenkins} E.~B.,  {Bowen} D.~V.,  {Tripp} T.~M.,    {Sembach} K.~R.,  2005,
  \apj, 623, 767, \eprint{astro-ph/0501475}

\bibitem[\protect\citeauthoryear{Lehner, Savage, Wakker, Sembach,  \&
  Tripp}{Lehner et~al.}{2006}]{leh06a}
Lehner N.,  Savage B.~D.,  Wakker B.~P.,  Sembach K.~R.,     Tripp T.~M.,
  2006, The Astrophysical Journal Supplement Series, 164, 1

\bibitem[\protect\citeauthoryear{{Oppenheimer} \& {Dav{\'e}}}{{Oppenheimer} \&
  {Dav{\'e}}}{2009}]{opp09b}
{Oppenheimer} B.~D.,  {Dav{\'e}} R.,  2009, \mnras, pp 571--+,
  \eprint{0806.2866}

\bibitem[\protect\citeauthoryear{{Penton}, {Stocke} \& {Shull}}{{Penton}
  et~al.}{2004}]{pen04a}
{Penton} S.~V.,  {Stocke} J.~T.,    {Shull} J.~M.,  2004, \apjs, 152, 29,
  \eprint{astro-ph/0401036}

\bibitem[\protect\citeauthoryear{{Prochaska}, {Chen}, {Howk}, {Weiner} \&
  {Mulchaey}}{{Prochaska} et~al.}{2004}]{pro04a}
{Prochaska} J.~X.,  {Chen} H.,  {Howk} J.~C.,  {Weiner} B.~J.,    {Mulchaey}
  J.,  2004, \apj, 617, 718, \eprint{astro-ph/0408294}

\bibitem[\protect\citeauthoryear{{Rauch}, {Haehnelt} \& {Steinmetz}}{{Rauch}
  et~al.}{1997}]{rau97a}
{Rauch} M.,  {Haehnelt} M.~G.,    {Steinmetz} M.,  1997, \apj, 481, 601,
  \eprint{astro-ph/9609083}

\bibitem[\protect\citeauthoryear{{Richter}, {Fang} \& {Bryan}}{{Richter}
  et~al.}{2006}]{ric06a}
{Richter} P.,  {Fang} T.,    {Bryan} G.~L.,  2006, \aap, 451, 767,
  \eprint{astro-ph/0511609}

\bibitem[\protect\citeauthoryear{{Richter}, {Paerels} \& {Kaastra}}{{Richter}
  et~al.}{2008}]{ric08b}
{Richter} P.,  {Paerels} F.~B.~S.,    {Kaastra} J.~S.,  2008, Space Science
  Reviews, 134, 25, \eprint{0801.0975}

\bibitem[\protect\citeauthoryear{{Richter}, {Savage}, {Tripp} \&
  {Sembach}}{{Richter} et~al.}{2004}]{ric04a}
{Richter} P.,  {Savage} B.~D.,  {Tripp} T.~M.,    {Sembach} K.~R.,  2004,
  \apjs, 153, 165, \eprint{0403513}

\bibitem[\protect\citeauthoryear{{Savage}, {Sembach}, {Tripp} \&
  {Richter}}{{Savage} et~al.}{2002}]{sav02a}
{Savage} B.~D.,  {Sembach} K.~R.,  {Tripp} T.~M.,    {Richter} P.,  2002, \apj,
  564, 631

\bibitem[\protect\citeauthoryear{{Schaye}}{{Schaye}}{2001}]{sch01a}
{Schaye} J.,  2001, \apj, 559, 507, \eprint{astro-ph/0104272}

\bibitem[\protect\citeauthoryear{{Schaye}, {Carswell} \& {Kim}}{{Schaye}
  et~al.}{2007}]{sch07a}
{Schaye} J.,  {Carswell} R.~F.,    {Kim} T.,  2007, \mnras, 379, 1169,
  \eprint{astro-ph/0701761}

\bibitem[\protect\citeauthoryear{{Schaye} \& {Dalla Vecchia}}{{Schaye} \&
  {Dalla Vecchia}}{2008}]{sch08e}
{Schaye} J.,  {Dalla Vecchia} C.,  2008, \mnras, 383, 1210, \eprint{0709.0292}

\bibitem[\protect\citeauthoryear{{Schaye}, {Dalla Vecchia}, {Booth}, {Wiersma},
  {Theuns}, {Haas}, {Bertone}, {Duffy}, {McCarthy} \& {van de Voort}}{{Schaye}
  et~al.}{2010}]{sch10a}
{Schaye} J.,  {Dalla Vecchia} C.,  {Booth} C.~M.,  {Wiersma} R.~P.~C.,
  {Theuns} T.,  {Haas} M.~R.,  {Bertone} S.,  {Duffy} A.~R.,  {McCarthy} I.~G.,
     {van de Voort} F.,  2010, \mnras, 402, 1536, \eprint{0909.5196}

\bibitem[\protect\citeauthoryear{{Schaye}, {Theuns}, {Leonard} \&
  {Efstathiou}}{{Schaye} et~al.}{1999}]{sch99a}
{Schaye} J.,  {Theuns} T.,  {Leonard} A.,    {Efstathiou} G.,  1999, \mnras,
  310, 57, \eprint{astro-ph/9906271}

\bibitem[\protect\citeauthoryear{{Seljak} \& {Zaldarriaga}}{{Seljak} \&
  {Zaldarriaga}}{1996}]{sel96a}
{Seljak} U.,  {Zaldarriaga} M.,  1996, \apj, 469, 437,
  \eprint{astro-ph/9603033}

\bibitem[\protect\citeauthoryear{{Sembach}, {Tripp}, {Savage} \&
  {Richter}}{{Sembach} et~al.}{2004}]{sem04a}
{Sembach} K.~R.,  {Tripp} T.~M.,  {Savage} B.~D.,    {Richter} P.,  2004,
  \apjs, 155, 351, \eprint{astro-ph/0407549}

\bibitem[\protect\citeauthoryear{{Shen}, {Wadsley} \& {Stinson}}{{Shen}
  et~al.}{2009}]{she09a}
{Shen} S.,  {Wadsley} J.,    {Stinson} G.,  2009, ArXiv e-prints,
  \eprint{0910.5956}

\bibitem[\protect\citeauthoryear{{Spergel}, {Bean}, {Dor{\'e}}, {Nolta},
  {Bennett}, {Dunkley}, {Hinshaw} \& {Jarosik}}{{Spergel}
  et~al.}{2007}]{spe07a}
{Spergel} D.~N.,  {Bean} R.,  {Dor{\'e}} O.,  {Nolta} M.~R.,  {Bennett} C.~L.,
  {Dunkley} J.,  {Hinshaw} G.,    {Jarosik} N. e.~a.,  2007, \apjs, 170, 377,
  \eprint{astro-ph/0603449}

\bibitem[\protect\citeauthoryear{{Springel}}{{Springel}}{2005}]{spr05b}
{Springel} V.,  2005, \mnras, 364, 1105, \eprint{astro-ph/0505010}

\bibitem[\protect\citeauthoryear{{Sutherland} \& {Dopita}}{{Sutherland} \&
  {Dopita}}{1993}]{sut93a}
{Sutherland} R.~S.,  {Dopita} M.~A.,  1993, \apjs, 88, 253

\bibitem[\protect\citeauthoryear{{Tepper-Garc{\'\i}a}}{{Tepper-Garc{\'\i}a}}{2006}]{tep06a}
{Tepper-Garc{\'\i}a} T.,  2006, \mnras, 369, 2025, \eprint{astro-ph/0602124}

\bibitem[\protect\citeauthoryear{{Theuns}, {Leonard}, {Efstathiou}, {Pearce} \&
  {Thomas}}{{Theuns} et~al.}{1998}]{the98b}
{Theuns} T.,  {Leonard} A.,  {Efstathiou} G.,  {Pearce} F.~R.,    {Thomas}
  P.~A.,  1998, \mnras, 301, 478, \eprint{astro-ph/9805119}

\bibitem[\protect\citeauthoryear{{Thom} \& {Chen}}{{Thom} \&
  {Chen}}{2008a}]{tho08b}
{Thom} C.,  {Chen} H.-W.,  2008a, \apjs, 179, 37, \eprint{0801.2381}

\bibitem[\protect\citeauthoryear{{Thom} \& {Chen}}{{Thom} \&
  {Chen}}{2008b}]{tho08a}
{Thom} C.,  {Chen} H.-W.,  2008b, \apj, 683, 22, \eprint{0801.2380}

\bibitem[\protect\citeauthoryear{{Tornatore}, {Borgani}, {Viel} \&
  {Springel}}{{Tornatore} et~al.}{2009}]{tor09a}
{Tornatore} L.,  {Borgani} S.,  {Viel} M.,    {Springel} V.,  2009, ArXiv
  e-prints, \eprint{0911.0699}

\bibitem[\protect\citeauthoryear{{Tripp}, {Jenkins}, {Bowen}, {Prochaska},
  {Aracil} \& {Ganguly}}{{Tripp} et~al.}{2005}]{tri05a}
{Tripp} T.~M.,  {Jenkins} E.~B.,  {Bowen} D.~V.,  {Prochaska} J.~X.,  {Aracil}
  B.,    {Ganguly} R.,  2005, \apj, 619, 714, \eprint{astro-ph/0407465}

\bibitem[\protect\citeauthoryear{Tripp, Savage,  \& Jenkins}{Tripp
  et~al.}{2000}]{tri00a}
Tripp T.~M.,  Savage B.~D.,     Jenkins E.~B.,  2000, The Astrophysical Journal
  Letters, 534, L1

\bibitem[\protect\citeauthoryear{{Tripp}, {Sembach}, {Bowen}, {Savage},
  {Jenkins}, {Lehner} \& {Richter}}{{Tripp} et~al.}{2008}]{tri08b}
{Tripp} T.~M.,  {Sembach} K.~R.,  {Bowen} D.~V.,  {Savage} B.~D.,  {Jenkins}
  E.~B.,  {Lehner} N.,    {Richter} P.,  2008, \apjs, 177, 39,
  \eprint{0706.1214}

\bibitem[\protect\citeauthoryear{{Wakker} \& {Savage}}{{Wakker} \&
  {Savage}}{2009}]{wak09a}
{Wakker} B.~P.,  {Savage} B.~D.,  2009, \apjs, 182, 378, \eprint{0903.2259}

\bibitem[\protect\citeauthoryear{{Weinberg}, {Miralda-Escude}, {Hernquist} \&
  {Katz}}{{Weinberg} et~al.}{1997}]{wei97b}
{Weinberg} D.~H.,  {Miralda-Escude} J.,  {Hernquist} L.,    {Katz} N.,  1997,
  \apj, 490, 564, \eprint{astro-ph/9701012}

\bibitem[\protect\citeauthoryear{{White}}{{White}}{1996}]{whi96a}
{White} S.~D.~M.,  1996, in {R.~Schaeffer, J.~Silk, M.~Spiro \& J.~Zinn-Justin}
  ed., Cosmology and Large Scale Structure {Formation and Evolution of
  Galaxies}.
pp 349--+

\bibitem[\protect\citeauthoryear{{Wiersma}, {Schaye} \& {Smith}}{{Wiersma}
  et~al.}{2009a}]{wie09a}
{Wiersma} R.~P.~C.,  {Schaye} J.,    {Smith} B.~D.,  2009a, \mnras, 393, 99,
  \eprint{0807.3748}

\bibitem[\protect\citeauthoryear{{Wiersma}, {Schaye}, {Theuns}, {Dalla Vecchia}
  \& {Tornatore}}{{Wiersma} et~al.}{2009b}]{wie09b}
{Wiersma} R.~P.~C.,  {Schaye} J.,  {Theuns} T.,  {Dalla Vecchia} C.,
  {Tornatore} L.,  2009b, \mnras, 399, 574, \eprint{0902.1535}

\bibitem[\protect\citeauthoryear{{Williger}, {Carswell}, {Weymann}, {Jenkins},
  {Sembach}, {Tripp}, {Dav{\'e}}, {Haberzettl} \& {Heap}}{{Williger}
  et~al.}{2010}]{wil10a}
{Williger} G.~M.,  {Carswell} R.~F.,  {Weymann} R.~J.,  {Jenkins} E.~B.,
  {Sembach} K.~R.,  {Tripp} T.~M.,  {Dav{\'e}} R.,  {Haberzettl} L.,    {Heap}
  S.~R.,  2010, \mnras, 405, 1736, \eprint{1002.3401}

\bibitem[\protect\citeauthoryear{{Zel'Dovich}}{{Zel'Dovich}}{1970}]{zel70a}
{Zel'Dovich} Y.~B.,  1970, \aap, 5, 84

\end{thebibliography}
